\documentclass[aps,prd,reprint,superscriptaddress,longbibliography,showpacs]{revtex4-2}

\usepackage{graphicx}
\usepackage{dcolumn}
\usepackage{bm}

\usepackage[figurename=Figure]{caption}
\usepackage{float}    
\usepackage{verbatim} 
\usepackage{amsmath}  
\usepackage{upgreek} 
\usepackage{amssymb}  
\usepackage{listings} 
\usepackage{graphics}
\usepackage{enumitem} 
\usepackage{latexsym,epsfig}
\usepackage[position=top]{subcaption}
\usepackage{physics}
\usepackage{stackrel}
\usepackage{soul}

\usepackage{placeins} 

\usepackage[colorlinks=true,breaklinks=true,allcolors=blue]{hyperref}
\usepackage{lipsum}
\usepackage{dsfont}

\usepackage[scr=boondox]{mathalfa}

\def\Oo{\ensuremath{{\cal O}}} 
\def\Ee{\ensuremath{{\cal E}}} 
\def\Jj{\ensuremath{{\cal J}}} 

\def\Hh{\ensuremath{{\cal H}}} 

\def\Vv{\ensuremath{{\cal V}}}

\def\Cc{\ensuremath{{\cal C}}}
\def\Dd{\ensuremath{{\cal D}}}

\def\Kk{\ensuremath{{\cal K}}}
\def\Gg{\ensuremath{{\cal G}}} 
\def\Zz{\ensuremath{{\cal Z}}}
\def\Pp{\ensuremath{{\cal P}}}

\usepackage[scr=boondox]{mathalfa}

\def\i{\ensuremath{{\imath}}}

\def\Im{\ensuremath{{\operatorname{Im}}}}

\newcommand{\be}{\begin{equation}}
	\newcommand{\ee}{\end{equation}}
\newcommand{\bea}{\begin{equation}\begin{aligned}}
		\newcommand{\eea}{\end{aligned}\end{equation}}
\newcommand{\ben}{\begin{enumerate}}
	\newcommand{\een}{\end{enumerate}}

\DeclareDocumentCommand{\nint}{ O{} O{} m }{\ensuremath{ \int_{\mbox{\scriptsize $#1$}}^{\mbox{\scriptsize$#2$}}\!\!\! \mbox{\small $\,\mathrm{d}#3$\! }}}

\newcommand{\cj}{\mathcal{J}}
\newcommand{\cC}{\mathcal{C}}
\newcommand{\cE}{\mathcal{E}}
\newcommand{\cV}{\mathcal{V}}
\newcommand{\cK}{\mathcal{K}}
\newcommand{\Tau}{\mathcal{T}}

\usepackage{tikz,bm} 
\usepackage{circuitikz}
\usetikzlibrary{decorations.markings}
\usetikzlibrary{decorations.pathreplacing,calligraphy}

\definecolor{mycolor}{rgb}{1,0.2,0.3}
\definecolor{brightgreen}{rgb}{0.4, 1.0, 0.0}
\definecolor{britishracinggreen}{rgb}{0.0, 0.26, 0.15}
\definecolor{cadmiumgreen}{rgb}{0.0, 0.42, 0.24}
\definecolor{ceruleanblue}{rgb}{0.16, 0.32, 0.75}
\definecolor{darkelectricblue}{rgb}{0.33, 0.41, 0.47}
\definecolor{darkpowderblue}{rgb}{0.0, 0.2, 0.6}
\definecolor{darktangerine}{rgb}{1.0, 0.66, 0.07}
\definecolor{emerald}{rgb}{0.31, 0.78, 0.47}
\definecolor{palatinatepurple}{rgb}{0.41, 0.16, 0.38}
\definecolor{pastelviolet}{rgb}{0.8, 0.6, 0.79}

\begin{document}
	
	\preprint{APS/123-QED}
	
	\title{Thermalization of a Closed Sachdev-Ye-Kitaev System in the Thermodynamic Limit}
	
	\author{Santiago Salazar Jaramillo}
	\email{s.salazarjaramillo@stud.uni-goettingen.de}

 \author{Rishabh Jha}
	\email{rishabh.jha@uni-goettingen.de}

 \author{Stefan Kehrein}
	\email{stefan.kehrein@theorie.physik.uni-goettingen.de}
	\affiliation{%
		Institute for Theoretical Physics, Georg-August-Universit\"{a}t G\"{o}ttingen, Friedrich-Hund-Platz 1, 37077 G\"ottingen, Germany
	}%
	
	
	
	
	

\begin{abstract}
	The question of thermalization of a closed quantum system is of central interest in non-equilibrium quantum many-body physics. Here we present one such study analyzing the dynamics of a closed coupled Majorana SYK system. We have a large-$q$ SYK model prepared initially at equilibrium quenched by introducing a random hopping term, thus leading to non-equilibrium dynamics. We find that the final stationary state reaches thermal equilibrium with respect to the Green's functions and energy. Accordingly, the final state is characterized by calculating its final temperature and the thermalization rate. We provide a detailed review of analytical methods and derive the required Kadanoff-Baym equations, which are then solved using the algorithm developed in this work. Our results display rich thermalization dynamics in a closed quantum system in the thermodynamic limit. 
\end{abstract}
	
	\maketitle
	

\section{Introduction}
\label{sec. introduction}

 The Sachdev-Ye-Kitaev (SYK) model, originally proposed by Sachdev and Ye for spins \cite{sachdev1993}, and later modified by Kitaev for fermions \cite{KitaevCaltech}, has garnered significant attention since its conception due to its unusual properties. Condensed matter theory is replete with models containing quasiparticles \cite{Mattuck1992Jun}, however it has been shown that the SYK model doesn't have such a description \cite{chowdhury2022}. At the same time, the model is non-integrable and maximally chaotic \cite{Maldacena2016Nov} (in the sense that it saturates the Maldacena-Shenker-Stanford bound of chaos \cite{MSS_bound}) while being exactly solvable in the large-$N$ limit, where $N$ is the number of sites. This means that the saddle point equations, namely the Schwinger-Dyson equations, have a closed form for its self-energy, allowing for full treatment of the diagrammatic expansion \textit{in the thermodynamic limit} \cite{Rosenhaus2019Jul}. 
 Furthermore, its Green's function exhibits conformal symmetry in the low-temperature (and strong coupling) limit which allows for a holographic mapping to the gravity dual \cite{Sarosi_2018}. 
 
 The standard SYK model is written for Majorana fermions, including the one considered in this work, but has also been generalized to complex Dirac fermions which at half-filling reduces to the Majorana case \cite{Gu2020Feb}. Usually, two-body interactions are considered but the model can be generalized to $q/2$ body interactions \cite{Maldacena2016Nov}, where the large-$q$ limit allows for further analytical tractability in the solutions of the general Schwinger-Dyson equations. This limit, is justified as it has been shown for transport properties in Ref. \cite{Zanoci2022Jun} that the large-$q$ limit results qualitatively matches with finite-$q$ ones. Due to its all-to-all random coupling, the model is effectively a zero spatial dimensional model, however lattices can be constructed out of the SYK dots in higher spatial dimensions \cite{Gu2017May, Song2017Nov, Jha2023Jun}.

This work addresses the question of thermalization in a closed SYK system in the thermodynamic limit. Thermalization in closed quantum systems has been an area of intense research which tries to address the question of how thermal equilibrium, in accordance with the laws of statistical physics, is achieved during quantum unitary time evolution \cite{Deutsch2018Jul}. One of the modern cornerstones is the Eigenstate Thermalization Hypothesis (ETH) as pioneered by Srednicki and Deutsch \cite{Deutsch1991Feb, Srednicki1994Aug, Srednicki1996Feb, Srednicki1999Feb}, which provides a sufficient condition for thermalization. Whether ETH serves as a necessary condition or not remains an area of intense research \cite{D'Alessio2016May}. For this reason, any analytically solvable model (such as the SYK) that is non-integrable and maximally chaotic, plays a significant role in addressing the question of thermalization. Since the SYK model allows for full treatment in the thermodynamic limit, it is an attractive choice for the study of such effects. The thermalization in SYK model from the perspective of ETH has been studied in Refs. \cite{Garcia-Garcia2016Dec, Sonner2017, Altland2018May} where they find that $q/2$-body interacting SYK model with $q>2$ indeed is an ergodic model. Furthermore, both Majorana and complex SYK models in large-$q$ limit have been shown to present instantaneous thermalization with respect to the Green's functions, when quenching to a single-term Hamiltonian in the final state \cite{Eberlein2017Nov, louw2022}. Other (quench) dynamics inducing non-equilibrium behaviors and the consequent approach towards thermalization are studied in Refs. \cite{bhattacharya2017, haldar2020, Cheipesh2021Sep, Zanoci-coupled-to-thermal-baths}, including a variant of SYK model, namely the Yukawa-SYK model \cite{Hosseinabadi2023Sep}, where connections to holography have been attempted and studied in Refs. \cite{magan, Cotler2017May, Garcia-mbl-syk-holography, Almheiri2024Aug}.

Thermalization in closed quantum systems is heavily studied both in cold atomic gases \cite{cold-atom-1, cold-atom-2, cold-atom-3} as well as in the foundational issues of quantum statistical mechanics \cite{D'Alessio2016May}. Nonetheless, most quantum systems that can be studied in the thermodynamic limit belong to the class of quantum integrable models which do not thermalize \cite{Vidmar2016Jun}. The approaches for quantum ergodic systems are severely impeded due to lack of analytical tractability of models \cite{dmft-eckstein} thereby resorting to numerical techniques accompanied by extrapolation to the thermodynamic limit. A method that has been developed to deal with this issue, is that of non-equilibrium dynamical mean-field theory \cite{dmft-eckstein, eckstein-1, eckstein-2, eckstein-3, eckstein-4} where the problem of lattices are mapped onto a self-consistent impurity problem, avoiding finite-size effects. Although, this becomes exact only in infinite dimensions \cite{dmft-exact-1, dmft-exact-2}, it still provides a useful \textit{approximation} for three dimensional systems in the thermodynamic limit but fails for lower dimensional systems \cite{dmft-eckstein}. Other thermalization studies that are exact in finite-dimensional systems tend to have instantaneous thermalization in the thermodynamic limit \cite{Eberlein2017Nov, louw2022}. These limitations and difficulties highlight the importance of our work: we study the thermalization of a closed quantum system in the thermodynamic limit where analytical equations obtained are exact in nature that we proceed to solve numerically. We further quantify finite thermalization rates as well as final temperatures for the quench dynamics.

In this work, we combine two Majorana SYK terms and study its thermalization properties from quench-induced non-equilibrium dynamics. The first term is a large-$q$ SYK model (referred to as $\text{SYK}_q$) \cite{Maldacena2016Nov}, initially prepared in equilibrium, while the second one is a random hopping Hamiltonian (referred to as $\text{SYK}_2$) \cite{Magan2016Jan, magan}, which is switched on at $t=0$. As expected, the quenches leading to a single final term (either $\text{SYK}_q$ or $\text{SYK}_2$) in the Hamiltonian, instantaneously thermalizes with respect to the Green's function \cite{Eberlein2017Nov}. However, we find that combining both the terms leads to finite thermalization time and the corresponding rich thermalization dynamics. The Kadanoff-Baym equations derived for this system are highly nonlinear and non-Markovian integro-differential equations whose solutions cannot be obtained under general conditions. Yet, we present two limiting cases which can indeed be solved analytically and serve as useful benchmarks for numerical calculations. 

Given the significant computational difficulty of obtaining these numerical solutions \cite{dmft-eckstein}, we develop our own numerical integrator based on a predictor-corrector algorithm \cite{meirinhos2022} and an analysis scheme to estimate the final temperature and thermalization rate. To help with our analyses, we also track the time dependent energy over the entire duration of the dynamical relaxation. Using these tools, we find that the Kadanoff-Baym solutions for the Green's functions do indeed reach a stationary state which we establish to correspond to thermal equilibrium and there emerges a complicated dependence of the thermalization rate on the final temperature. We now proceed to present the main results of this work as well as the structure of the paper.

 \subsection{Summary and results}\label{subsec. results}

 The closed Majorana SYK model considered here, consists of a large-$q$ SYK term (the so-called ``interaction'' term) and a random hopping term (the so-called ``kinetic'' term). The interaction term is initialized in equilibrium and the kinetic term is switched on at time $t=0$, which induces the non-equilibrium dynamics. We refer to this procedure as the ``mixed quench'' throughout this work. 
 
 The model is presented in Section \ref{sec. model}. We analyze the non-equilibrium dynamics in real time using the Kadanoff-Baym equations. Accordingly, we review the mathematical formalism of Keldysh formulation and derive the associated equations in the thermodynamic limit in Section \ref{sec. analytical methods}. Appendices \ref{app. review of matsubara and keldysh formalisms} - \ref{app. Evaluating energy in the Keldysh contour} compliment the main text where we provide detailed steps for all calculations involved. Afterwards, having obtained the Kadanoff-Baym equations for the general non-equilibrium case, they are reduced to the mixed quench case. These equations are highly nonlinear and non-Markovian in nature, thereby limiting the possibility of obtaining analytic solutions. However, in Section \ref{subsec. limiting cases} we present two limiting cases of the mixed quench whose solutions can indeed be obtained analytically. These serve as useful benchmarks for the numerical computation to follow (see section \ref{sec. benchmarking}). Then we proceed to the general case of mixed quench where we resort to numerically solving the non-equilibrium Kadanoff-Baym equations. The system is specified by the initial temperature and the quench parameter (which is the strength of the kinetic term that is switched on at $t=0$). First, we present an intuitive picture by providing visualizations of the non-equilibrium dynamics in Section \ref{sec. mixed SYK}. Then we quantify these observations in Section \ref{sec. thermalization}, where we show that the solution to the Kadanoff-Baym equations indeed reaches a stationary state corresponding to thermal equilibrium. Implying that we observe thermalization in real time in the thermodynamic limit for our (quenched) closed SYK system. 

 We study the dependence of these final equilibrium states on the initial temperatures and the quench parameters, by calculating their final temperatures and thermalization rates. Our results show that, as expected, stronger quenches lead to higher final temperatures, eventually reaching the infinite temperature regime. On the other hand,  we find a complicated dependence of the thermalization rates on the final temperature. For strong quenches, we find that thermalization is achieved in fairly short timescales. However, for weaker quenches, the thermalization rates quickly become several orders of magnitude smaller. This agrees with the expectation that lower final temperatures/weaker quenches thermalizes slower than higher final temperatures/stronger quenches. This behavior is more pronounced if we start from a higher initial temperature but the effect persists even in systems initiated close to the ground state. 
 
 Appendix \ref{app. additional figures} further supports the claims presented in the main text for other values of quench conditions. We conclude and discuss the outlook in Section \ref{sec. conclusion}. 

 \section{Model}
 \label{sec. model}

 We consider the time-dependent mixed SYK model whose Hamiltonian is given by
\begin{equation}
\begin{aligned}
    \Hh(t) = & \i^{q/2}\sum_{\{i_q\}_{\leq} } j_{q;\{ i_q \}} f_q(t) \psi_{i_1}...\psi_{i_q} \\
    &\hspace{4mm}+ \i   \sum_{\{l_2\}_{\leq}} j_{2;\{l_2\}}f_2(t)\psi_{l_1}\psi_{l_2} \\
    =& \text{SYK}_q + \text{SYK}_2,
    \end{aligned}
    \label{eq:mixed_syk_neq_hamiltonian}
\end{equation}
where the first term is the interaction term while the second term is the (quadratic) kinetic term (also known as random hopping term) and $q$ is an even number. Here $f_q(t)$ and $f_2(t)$ are arbitrary functions of time $t$ and the Majorana fields $\psi$ are also time-dependent. We introduce the following notation that is also used above:
\begin{equation}
\begin{aligned}
\{i_q\}_{\leq} &\equiv 1 \leq i_1 < i_2<\ldots<i_q\leq N,  \\
\{l_2\}_{\leq} &\equiv 1 \leq l_1 < l_2\leq N
\end{aligned}
\label{i less def}
\end{equation}
where $N$ is the total number of sites and $q$ denotes the $q/2$-body interaction in the system. Here $j_{q, \{i_q\}}$ and $j_{2,\{l_2\}}$ denote the coupling strengths of the interaction term and the kinetic term, respectively. To clarify the notation, we have clubbed the indices $\{ i_q\} = \{i_1, i_2, \ldots, i_{q-1}, i_q \}$ and therefore the interacting strength and the kinetic strength read respectively as 
\be
\begin{aligned}
j_{q; \{i_q\}} &\equiv j_{q; i_1, i_2, \ldots, i_q} \\
j_{2;\{l_2\}} &\equiv j_{2; l_1, l_2}.
\end{aligned}
\ee
Here $q$ and $2$ are  before the semi-colon are labels referring to the interaction and kinetic terms, while the remaining symbols after the semi-colon ($;$) are indices which are summed over under the constraint $\{i_q\}_{\leq}$ and $\{l_2\}_{\leq}$ as defined in Eq. \eqref{i less def}.


As is usual for SYK-like models \cite{Maldacena2016Nov}, $j_{q} $ and  $j_{2}$ are independent random numbers that introduce disorder in the system and follow a Gaussian distribution with means
\begin{equation}
    \langle j_q \rangle = 0, \quad \langle j_2 \rangle = 0,
\label{means}
\end{equation}
and variances
\begin{equation}
    \begin{aligned}
        \langle j_q^2 \rangle &= \frac{J_q^2 (q-1)!}{N^{q-1}} = \frac{\cj_q^2 q!}{N^{q-1} 2^{1-q}q^2} \\
        \langle j_2^2 \rangle &= \frac{J_2^2}{N} = \frac{2 \cj_2^2 }{N q}
    \end{aligned}
\label{variances}
\end{equation}
where we have introduced new rescaled couplings
\begin{equation}
    \mathcal{J}^2_{q} \equiv 2^{1-q}q J_q^2, \quad \mathcal{J}_{2}^2 \equiv 2^{-1}qJ_{2}^2
    \label{rescaled coupling constants}
\end{equation}
in order to induce competition between the two terms in the Hamiltonian \cite{Jha2023Jun}. Therefore, $j_q$ and $j_2$ are drawn from the two \textit{normalized} Gaussian distributions given by
\begin{equation}
\begin{aligned}
    \Pp_q [ j_{q; \{i_q\}}] &= \sqrt{\frac{N^{q-1}q^2 2^{-q}}{\pi \Jj_q^2 q!}} \exp\left(-\frac{1}{2 \langle j_q^2\rangle} \sum_{\{i_q\}_{\leq} } j_{q; \{i_q\}}^2\right) \\
    \Pp_2 [ j_{2; \{l_2\}}] &= \sqrt{\frac{N q}{4\pi \Jj_2^2}} \exp\left(-\frac{1}{2 \langle j_2^2\rangle} \sum_{\{l_2\}_{\leq} } j_{2; \{l_q\}}^2\right).
\end{aligned}
\label{gaussian ensembles}
\end{equation} 

To simplify the equations, the time dependent functions $f_q(t)$ and $f_2(t)$ in the Hamiltonian in Eq. \eqref{eq:mixed_syk_neq_hamiltonian} are absorbed into the newly introduced coupling constants $\Jj_q(t)$ and $\Jj_2(t)$, thereby making them time-dependent. Their explicit time-dependence in this work is chosen as
\begin{equation}
    \Jj_q(t) = \Jj_q, \quad \Jj_2(t) = \Jj_2 \Theta(t)
    \label{quench couplings}
\end{equation}
where the interaction and kinetic strengths on the right-hand side ($\Jj_q$ and $\Jj_2$, respectively) are time-independent. This choice represents a system with an initial single large-$q$ Hamiltonian, that is quenched at $t=0$ by suddenly turning on a kinetic contribution and keeping both terms constant at subsequent times. Since the large-$q$ SYK model is known to instantaneously thermalize, both for Majorana and complex fermions \cite{Eberlein2017Nov, louw2022}, the Hamiltonian in Eq. \eqref{eq:mixed_syk_neq_hamiltonian} is in equilibrium for $t<0$ (due to the choice of the coupling strengths made in Eq. \eqref{quench couplings}) and quenching the kinetic term at $t=0$ leads to a non-equilibrium behavior. Similar setups in the large-$q$ limit have been studied in Refs. \cite{Eberlein2017Nov, louw2022} and have been shown to present instantaneous thermalization when quenching to a \textit{single term} in the post-quench Hamiltonian. By contrast, the quench to the \textit{two-term} Hamiltonian described above, referred to as the \textit{mixed quench}, presents rich relaxation dynamics which are the focus of the present work and the results are summarized in Section \ref{subsec. results}.

However before presenting these results, the following section gives an overview of the derivation of the Kadanoff-Baym (KB) equations in the thermodynamic limit for a more general situation with arbitrary time-dependence in the couplings. To this end, we present the Keldysh formulation and the derivation of the Schwinger-Dyson equations from disorder-averaged partition function. 

 \section{Analytical Methods}
 \label{sec. analytical methods}

 In this section, we apply the Keldysh formalism to derive the saddle-point equations in the thermodynamic limit (the so-called Schwinger-Dyson equations), which ultimately yield the Kadanoff-Baym equations. We will make use of two limits as is prevalent for SYK-like systems \cite{Maldacena2016Nov, Rosenhaus2019Jul, Eberlein2017Nov, louw2022}, namely $N \to \infty$ (ignoring all $\Oo(1/N$ corrections) followed by the large-$q$ limit (where we ignore all $\Oo(1/q^2$ corrections). We provide a self-contained review in Appendix \ref{app. analytical methods required in this work} of the methods and the analytical results used in the numerical implementation in this work. Detailed derivations are shifted to other appendices that have been cited at appropriate places. 

\begin{figure}
    \centering
\includegraphics[width=0.95\linewidth]{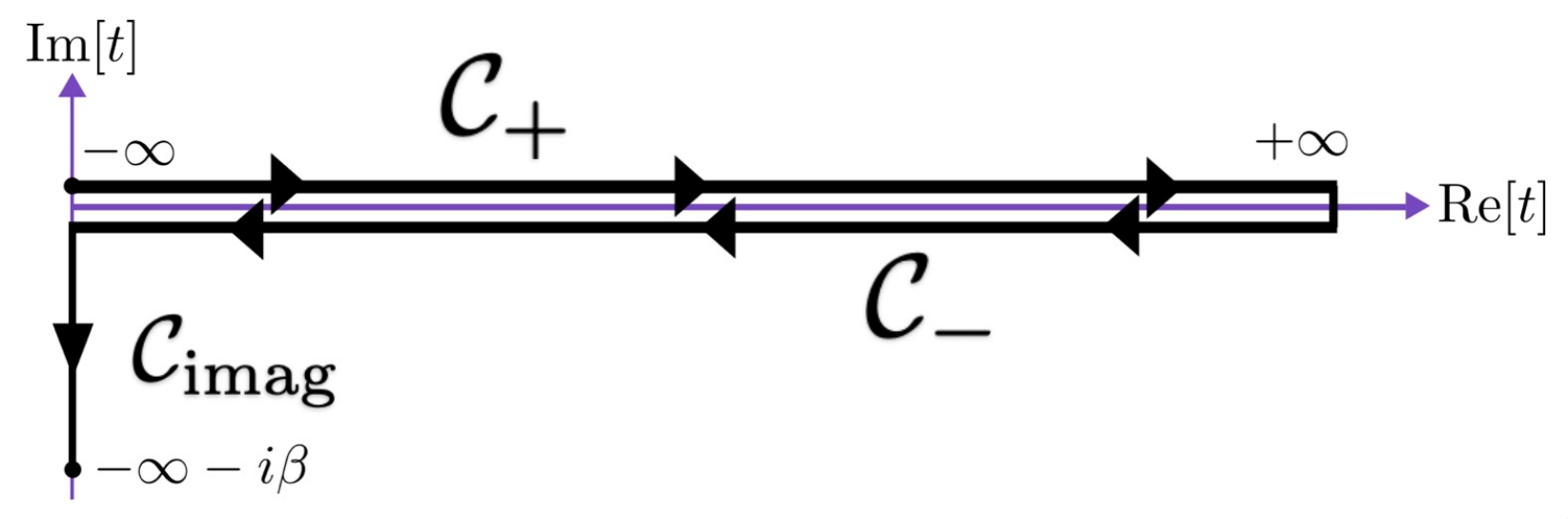}
    \caption{Sketch of the Schwinger-Keldysh contour $\Cc = \Cc_+ + \Cc_- + \Cc_{\text{imag}}$. The real time forwards and backwards paths go between $-\infty$ to $+\infty$ and are followed by the imaginary time branch which represents the equilibrium state. The branches running from $-\infty$ to $\infty$ are purely real and are shifted above and below the axis only for visualization purposes. The points on the vertical contour comes later than points lying on both the forward and the backward horizontal contours.}
    \label{fig:keldysh_contour}
\end{figure}

 The Keldysh plane is represented in Fig. \ref{fig:keldysh_contour} where the x- and y-axes are real and imaginary times axes respectively. The path shown through arrows is called the \textit{contour} $\Cc$ and it is divided into three parts: $\Cc = \Cc_+ + \Cc_- + \Cc_{\text{imag}}$. The imaginary part (vertical contour) of the contour represents the system that reaches equilibrium after the real time dynamics (horizontal contour). As mentioned in Appendix \ref{app. review of matsubara and keldysh formalisms}, if the system is prepared in equilibrium in the infinite past ($t_0 \to -\infty$) then, due to the Bogoliubov's principle of weakening of initial correlations, the imaginary part is deprecated. Having prepared the large-$q$ SYK Hamiltonian at equilibrium in the infinite past, this principle is employed throughout this work. Therefore the contour becomes $\Cc = \Cc_+ + \Cc_-$.

The structure of this section and the corresponding Appendix \ref{app. analytical methods required in this work} are as follows. In Appendix \ref{subsec. Keldysh Path Integral Formalism}, we review the basis of our calculations, namely the Keldysh path integral formulation and the associated partition function/generating functional that handles the non-equilibrium setup in real time. Then we evaluate the disorder-averaged partition function in Appendix \ref{subsec. Disorder-averaged partition function} and calculate the saddle point equations (also known as the Schwinger-Dyson equations) in the large-$N$ limit. We introduce the Green's functions in Keldysh contour in Appendix \ref{subsec. Green's functions in real-time} and use them to derive the Kadanoff-Baym equations from the Schwinger-Dyson equations in Section \ref{subsec. kadanoff-baym equations} in the main text. Finally we make use of the large-$q$ limit and introduce the large-$q$ ansatz for the Green's function in Section \ref{subsec. large q green's function ansatz} and simplify the Kadanoff-Baym equations to leading order in $1/q$. We analyze the equilibrium condition in Section \ref{subsec. equilibrium situation} and calculate the energy in the Keldysh contour in Section \ref{subsec. Energy in the Keldysh contour}. Finally after developing the entire formalism, we test our equations for two limiting cases in Section \ref{subsec. limiting cases} whose solutions can be obtained analytically. These will later also serve as useful benchmarks for our numerical implementation. 

\subsection{Kadanoff-Baym equations}
\label{subsec. kadanoff-baym equations}

With the real time Green's functions, the Schwinger-Dyson equations in Eq. \eqref{sd equations in terms of capital G} can be analytically continued to real time using the \textit{Langreth rules}. This leads to a set of nonlinear, non-Markovian partial differential equations known as the \textit{Kadanoff-Baym equations} \cite{Baym1961Oct, Langreth1976, Kadanoff2019Jun}. We provide a derivation of the Kadanoff-Baym equations starting from the Schwinger-Dyson equations in Appendix \ref{app. deriving kadanoff-baym equations} where we employ the Langreth rules which are briefly introduced in Appendix \ref{app. A brief note on the Langreth rules}. 

The complete set of Kadanoff-Baym equations in the most general conditions is given by Eq. \eqref{kb equation full}. These are a set of coupled differential equations of motion for the set of Green's functions $\Gg^>(t_1, t_2)$ and $\Gg^<(t_1, t_2)$. The definitions of Green's functions are provided in Appendix \ref{subsec. Green's functions in real-time}. The general conjugate relation in Eq. \eqref{general conjugate relation} already simplifies the computational complexity \cite{haldar2020, meirinhos2022} significantly, as only one Kadanoff-Baym equation is needed for $\Gg^>(t_1, t_2)$ and $\Gg^<(t_1, t_2)$ each, the other one is obtained by this conjugate relation. In addition, the Majorana fermions introduce further simplifications via the conjugate property in Eq. \eqref{Majorana conjugate relation}. Thus, in the Majorana case only a single Kadanoff-Baym equation is sufficient to find both $\Gg^>(t_1, t_2)$ and $\Gg^<(t_1, t_2)$. Furthermore, the Majorana self-energy $\Sigma(t_1, t_2)$ satisfies the same decomposition and the corresponding relations \cite{Babadi2015Oct}. As mentioned in Appendix \ref{app. deriving kadanoff-baym equations}, we choose to solve for $\Gg^>(t_1, t_2)$ from the Kadanoff-Baym equation
\begin{equation}
\begin{aligned}
    \i \partial_{t_1} \Gg^> (t_1,t_2) = \int_{-\infty}^\infty dt_3 &\left( \Sigma^R(t_1,t_3)\Gg^>(t_3,t_2) \right.\\
    &\left.+ \Sigma^>(t_1,t_3)\Gg^A(t_3,t_2)\right).
    \end{aligned}
    \label{main kb equation}
\end{equation}
For our system, as derived in Appendix \ref{app. deriving kadanoff-baym equations}, it takes the more specific form 
\begin{widetext}
    \begin{equation}
\begin{aligned}
\i \partial_{t_1} \Gg^>\left(t_1, t_2\right) =& \int_{-\infty}^{t_1} d t_3\left\{-\frac{(-1)^{q/2}}{2^{1-q}q} \cj_q\left(t_1\right) \cj_q\left(t_3\right)\left[G^{>}\left(t_1, t_3\right)^{q-1}-G^{<}\left(t_1, t_3\right)^{q-1}\right] G^{>}\left(t_3, t_2\right)\right. \\
& \left.+\frac{2}{q} \cj_2\left(t_1\right) \cj_2\left(t_3\right)\left[G^{>}\left(t_1, t_3\right)-G^{<}\left(t_1, t_3\right)\right]G^{>}\left(t_3, t_2\right)\right\} \\
& +\int^{t_2}_{-\infty} d t_3\left\{\frac{(-1)^{q/2}}{2^{1-q}q}\cj_{q}(t_{1}) \cj_ { q }(t_{ 3 } ) G^>\left(t_1, t_3\right)^{q-1}\left[G^>\left(t_3, t_2\right)-G^<\left(t_3, t_2\right)\right]\right. \\
& \left.-\frac{ 2 }{q} \cj_2\left(t_1\right) \cj_2\left(t_3\right)G^>\left(t_1, t_3\right)\left[G^>\left(t_3, t_2\right)-G^{<}\left(t_3, t_2\right)\right]\right\}.
\end{aligned}
\label{final kb equation for this paper}
\end{equation}
\end{widetext}
It is important to highlight that, due to the integrals on the right-hand side, the solution of Eq. \eqref{final kb equation for this paper} at each bi-local time ($t_1,t_2$) depends on all of the past points in time. This type of integro-differential equation (IDE) is referred to as a Volterra IDE; all Kadanoff-Baym equations follow this general form. This non-Markovian nature makes obtaining numerical solutions specially challenging \cite{dmft-eckstein}, however, in the following we demonstrate how introducing the large-$q$ expansion further reduces the computational complexity.

 \subsection{Large-\texorpdfstring{$q$}{Lg} Green's function ansatz}
\label{subsec. large q green's function ansatz}

Having taken the large-$N$ limit while deriving the Schwinger-Dyson equations in Eq. \eqref{sd equations in terms of capital G}, it is now possible to take the large-$q$ limit starting with the exponential ansatz \cite{Maldacena2016Nov, Eberlein2017Nov}
\begin{equation}
    \Gg^>(t_1, t_2) = -\frac{\i}{2} e^{g(t_1, t_2)/q}
    \label{greater green's function large q ansatz}
\end{equation}
where $g = \Oo(q^0)$ and satisfies the boundary condition and conjugation property (Eq. \eqref{general conjugate relation})
\begin{equation}
    g(t,t) = 0, \quad g(t_1, t_2)^\star = g(t_2, t_1)
    \label{little g boundary and conjugate relation}
\end{equation}
 respectively (Majorana or otherwise, in or out-of-equilibrium). For Majorana fermions specifically, the conjugate relation (Eq. \eqref{Majorana conjugate relation}) yields
\begin{equation}
    \Gg^<(t_1, t_2) = \frac{\i}{2} e^{g(t_1, t_2)^\star/q}.
    \label{lesser green's function large q ansatz}
\end{equation}
The KMS condition for the Green's function is then given by
\be g(t) = g(-t - i\beta_f),
\label{kms relation}\ee 
where $\beta_f$ corresponds to the inverse temperature (see Eq. (C.1) in Ref. \cite{Sorokhaibam2020Jul} and use Eqs. \eqref{greater green's function large q ansatz}, \eqref{little g boundary and conjugate relation} and \eqref{lesser green's function large q ansatz} from above where $t \equiv t_1 - t_2 $). In the large-$q$ limit, from the expansion $e^{g/q} = 1 + g/q + \Oo(1/q^2)$ only the terms to first order in $1/q$ are kept. Substituting in Eq. \eqref{final kb equation for this paper} and simplifying to first order in $1/q$ results in
\begin{widetext}
\begin{equation}
\begin{aligned}
    \partial_{t_1}g(t_1,t_2) = & -\int_{-\infty}^{t_1} d t_3\left\{\cj_q\left(t_1\right) \cj_q\left(t_3\right)\left[e^{g\left(t_1, t_3\right)}+e^{g^*\left(t_1, t_3\right)}\right]
 + 2\mathcal{J}_2\left(t_1\right) \mathcal{J}_2\left(t_3\right)\right\}  \\
    & +\int_{-\infty}^{t_2} dt_3\left\{2 \mathcal{J}_q\left(t_1\right) \mathcal{J}_q\left(t_3\right) e^{g\left(t_1, t_3\right)} 
    +2\mathcal{J}_2\left(t_1\right) \mathcal{J}_2\left(t_3\right) \right\}.
\end{aligned}
\label{kb equation for small g}
\end{equation}
\end{widetext}
where we used the fact that $q$ is even in, e.g. $(-1)^{q/2}\i^q = (-1)^{q/2}\left(\i^2\right)^{q/2} = (-1)^q = 1$. Conveniently, as a result of the large-$q$ simplification, the only dependence on $t_2$ is found on the integral limits of the second term on the right-hand side. Furthermore, notice that Eq. \eqref{kb equation for small g} is valid only when $t_1\geq t_2$ due to the definition of the greater Green's function $\Gg^>(t_1, t_2)$. Thus, only Eq. \eqref{kb equation for small g} need be solved, in order to find the full Green's function. Additionally, this equation also produces the equilibrium states of the system as is shown next.

\subsection{Equilibrium}
\label{subsec. equilibrium situation}

It is known that in equilibrium, the Green's function becomes stationary in the sense that it becomes dependent only upon time differences. Imposing the equilibrium condition $g(t, t^\prime) = g(t-t^\prime)$ in Eq. \eqref{kb equation for small g}, assuming that the coupling constants $\Jj_q$ and $\Jj_2$ are time-independent ,the partial differential equation 
\begin{equation}
    \partial_{t_2} \partial_{t_1} g(t_1 , t_2) = 2  \mathcal{J}_q^2 e^{g\left(t_1,t_2\right)} 
    +2\mathcal{J}_2^2
\end{equation}
is obtained. Here, the second time derivative of Eq. \eqref{kb equation for small g} with respect to time $t_2$ is calculated using Leibniz integral rule. Substituting $(t_1, t_2) \to (t_1 - t_2)$ and $t_1 - t_2 = t$ gives the \textit{equilibrium ordinary differential equation}
\begin{equation}
    \frac{d^2 g(t)}{dt^2} = -2 \mathcal{J}_q^2  e^{g\left(t\right)} 
    -2\mathcal{J}_2^2 .
    \label{equilibrium ODE in real time}
\end{equation}
Unfortunately, no analytical solution is known for this differential equation, but its numerical solutions will later be useful in determining whether the solution obtained for the general non-equilibrium Kadanoff-Baym equation in Eq. \eqref{kb equation for small g} eventually reaches equilibrium.

This equation is formulated in real time $t$ but can be re-casted in terms of imaginary time (since this is an equilibrium situation) to make the temperature dependence more explicit. We use the analytic continuation \cite{Maldacena2016Nov}
\begin{equation}
    \frac{\tau}{\beta_f} \longrightarrow \frac{it}{\beta_f} + \frac{1}{2} = x + \frac{1}{2}
    \label{definition of x}
\end{equation}
where $\beta_f$ is the inverse of the final temperature (technically, the final temperature is $\Jj_q \beta_f $ but we will take $\Jj_q$ as constant throughout this work). The new Green's function in imaginary time is also obtained by analytic continuation via $\Tilde{g}(x) \equiv  g(t \to -i\beta_f x)$, resulting in a real-valued function and of order $\Oo(q^0)$ in the large-$q$ limit. We can use the KMS relation in Eq. \eqref{kms relation} and the definition $\Tilde{g}(x) = g(-i\beta_f x)$ to get: $\Tilde{g}(-x + 1) = \Tilde{g}(x)$. This implies that $\Tilde{g}(x)$ is axially symmetric around $x=1/2$. Then Eq. \eqref{equilibrium ODE in real time} takes the dimensionless form
\begin{equation}\label{equilibrium ODE in imaginary time}
    \frac{d^2 \Tilde{g}(x)}{dx^2} =  2(\beta_f \cj_q)^2 e^{\Tilde{g}(x)} + 2(\beta_f \cj_2)^2,
\end{equation}
where the Green's function $\Tilde{g}(x)$ also satisfies the following boundary conditions \footnote{Since $g(t,t) = 0$ in real time, therefore $\Tilde{g}(x=0) = 0$. Then using the periodicity of $\Tilde{g}$ as mentioned in the text, namely $\Tilde{g}(x) = \Tilde{g}(-x + 1)$, we get for $x=0$, the other boundary condition $\Tilde{g}(1) = 0$.}:
\begin{equation}
\Tilde{g}(0) = \Tilde{g}(1) = 0.
\end{equation}
Clearly $d^2 \Tilde{g}(x)/dx^2>0$ for non-vanishing values of $\Jj_2$, $\Jj_q$ and $\beta_f$, which implies that $ \Tilde{g}(x)$ is strictly concave. This together with the boundary conditions imply that $\Tilde{g}\leq 0$, with only one minima appearing at $x = 1/2$ due to the axial symmetry noted above. 

The numerical solution of Eq. \eqref{kb equation for small g} is the focus of this work, since other quantities can be obtained from it. In the following section, we derive an expression for the energy purely in terms of the Green's function, which is valid in the non-equilibrium setting of the mixed quench. This will allow us later to estimate the final thermalization temperature as well as the thermalization rate of the mixed quench. 
 
\subsection{Energy in the Keldysh contour}
\label{subsec. Energy in the Keldysh contour}

In the Keldysh formalism, the expectation values of an observable can be calculated from the functional derivatives of the partition function \cite{Kamenev2023Jan, haldar2020} by introducing the conjugate field $\eta (t)$ as explained in Appendix \ref{app. review of matsubara and keldysh formalisms} and mentioned above in Eq. \eqref{average expectation value formula}. The properly scaled energy operator is
\begin{equation}
   E(t_1) =q^2 \langle \hat{\Oo}(t_1) \rangle=  q^2 \langle H(t_1) \rangle 
   \label{energy definition}
\end{equation}
where $H(t)=\int dt \Hh(t)$ ($\Hh(t)$ is given in Eq. \eqref{eq:mixed_syk_neq_hamiltonian}). Without loss of generality, we chose time $t_1$ on the backward contour, namely $t_1 \in \Cc_-$ (see supplemental material of Ref. \cite{haldar2020}). As mentioned before, in the SYK, the observables are all obtained from the disordered averaged generating functional. This averaging procedure for $\Zz[\eta]$ is carried out in much the same way as in Appendix \ref{app. Deriving disorder-averaged partition function} for the partition function. In this section, we highlight the main steps, starting from the definition of the generating functional
\begin{equation}
     \Zz[\eta] = \int D\psi_i \exp{i S  + \i S_\eta }
     \label{generating functional definition}
\end{equation}
where $S$ is the bare action given in Eq. \eqref{eq:bareaction} and $S_\eta$ is the perturbed action given by $S_\eta = -\int dt \eta (t) \Hh(t)$. Then the averaged generating functional is given by 
\begin{equation}
    \overline{\Zz}[\eta] = \int \Dd j_q \Dd j_2 \Pp_q[j_q] \Pp_2[j_2] \Zz[\eta]  
\end{equation}
where $\Pp[j_q]$ and $ \Pp[j_2]$ are the Gaussian distributions in Eq. \eqref{gaussian ensembles} from which the coupling constants $j_q$ and $j_2$ are derived in the Hamiltonian, respectively. Re-doing the calculation and introducing the two bi-local fields, namely the Green's function (Eq. \eqref{green's function def}) and the self-energy via Eq. \eqref{self-energy def}, we get
\begin{equation}
     \overline{\Zz}[\eta] =  \exp\left[ \i N \overline{S}[\Gg, \Sigma] + \i N \overline{S}_\eta[\Gg, \Sigma,\eta] \right]
\end{equation}
where $\overline{S}[\Gg, \Sigma]$ is the same as in Eq. \eqref{final effective action} and $\overline{S}_\eta[\Gg, \Sigma,\eta]$ contains all dependencies on $\eta(t)$. Using the large-$q$ ansatz for the Green's function from Section \ref{subsec. large q green's function ansatz}, we get the following energy density at large-$q$:
\begin{equation}
\begin{aligned}
\label{eq:NEQ_energy}
     \cE (t_1) =\frac{E(t_1)}{N}  = \text{Im} &\Big\{\int_{-\infty}^{t_1}\, dt_{2} \Big ( \cj_q(t_1) \cj_q(t_2) e^{g(t_1,t_2)} +  \\ 
    &   \cj_2(t_1) \cj_2(t_2) g(t_1,t_2)  \Big) \Big\}+ \mathcal{O}(1/q)
\end{aligned}
\end{equation}
where we can separate the interacting and kinetic energy densities $\Vv(t_1)$ and $\Kk(t_1)$ respectively as
\begin{equation}
\begin{aligned}
 \Vv(t_1) = &   \text{Im} \int_{-\infty}^{t_1}\, dt_{2} \cj_q(t_1) \cj_q(t_2) e^{g(t_1,t_2)}+ \mathcal{O}\left(\frac{1}{q}\right)\\
 \Kk(t_1) = &  \text{Im} \int_{-\infty}^{t_1}\, dt_{2}  \cj_2(t_1) \cj_2(t_2) g(t_1,t_2) + \mathcal{O}\left(\frac{1}{q}\right).
 \end{aligned}
 \label{interacting and kinetic energy densities}
\end{equation}
The detailed calculations involved in obtaining these energies, are included in Appendix \ref{app. Evaluating energy in the Keldysh contour}.

The expectation values of the energies, together with the large-$q$ Kadanoff-Baym equation in Eq. \eqref{kb equation for small g} are the main objects of study for this research. Using these, we are able to obtain all the quantities relevant to the thermalization of our system. We emphasise that the entire formalism in this section has been developed without assuming any particular form for the coupling constants $\Jj_q(t)$ and $\Jj_2(t)$. We now proceed to present two analytically solvable cases, by choosing explicit forms for their time dependence: a single large-$q$ SYK term in equilibrium as well as a quench to a single random-hopping Hamiltonian. Both represent limiting cases of the mixed quench as presented in the following sections and will further serve us to benchmark its numerical solution. 

\subsection{Limiting Cases}
\label{subsec. limiting cases}

We consider two analytically solvable limiting cases here, namely (1) a single large-$q$ SYK model in equilibrium where $\Jj_q(t) = \Jj_q$ and $\Jj_2(t) = 0$, as well as (2) a quench to a single kinetic term where $\Jj_q(t) = \Theta(-t) \Jj_q$ and $\Jj_2(t) = \Theta(t) \Jj_2$, referred to as the \textit{kinetic quench}.

\subsubsection{Limiting case 1: \texorpdfstring{$\Jj_q(t) = \Jj_q$}{Lg} and \texorpdfstring{$\Jj_2(t) = 0$}{Lg}}

In this case, we have a single large-$q$ Majorana SYK Hamiltonian with time-independent coupling $\Jj_q$, corresponding to a mixed quench where $\cj_2\ll \cj_q$. A single large-$q$ SYK model has been studied in the literature and has been shown to thermalize intantly with respect to the Green's functions both for the Majorana as well as the complex cases \cite{Eberlein2017Nov, louw2022}. As stated above, we set up the initial state in equilibrium in the infinite past and remain as such throughout the time evolution to the infinite future. Therefore the Kadanoff-Baym equation in Eq. \eqref{kb equation for small g} reduces to
\begin{equation}
\begin{aligned}
     \partial_{t_1} g\left(t_1, t_2\right)= & -\int_{-\infty}^{t_1} d t_3 \, \cj_q^2\left[e^{g\left(t_1, t_3\right)}+e^{g^*\left(t_1, t_3\right)}\right] \\
     &+ \int_{-\infty}^{t_2} dt_3 2 \, \mathcal{J}_q^2 e^{g\left(t_1, t_3\right)} 
\end{aligned}
     \label{kb equation for single large q SYK}
\end{equation}
Then we impose the equilibrium condition as explained in Section \ref{subsec. equilibrium situation} to get the second-order ordinary differential equation
\begin{equation}
    \frac{d^2 g(t) }{dt^2} = -2\cj_q^2 e^{g(t)}.
\end{equation}
where $g(t)$ satisfies the initial condition $g(0) =0$. This can be analytically solved and the solution is given by
\begin{equation}
    g(t) = \ln\left\{ \frac{c_1}{4\cj_q^2}  \left(1-\tanh^2\left(\frac{1}{2}\sqrt{c_1(t+c_2)^2}\right)\right)\right\}
    \label{large q syk green's function solution 1}
\end{equation}
where $c_1$ and $c_2$ are integration constants that are determined by the initial condition. It is convenient to re-define them as
\begin{equation}
    \sigma \equiv \sqrt{c_1}/2, \quad \theta \equiv \frac{-\i c_2\sqrt{c_1}}{2}
\end{equation}
such that the initial condition $g(0)=0$ becomes
\begin{equation}
    \cos \theta = \frac{\sigma}{\Jj_q}
\end{equation}
which indicates that $\sigma$ is bounded by the coupling constants, $-\Jj_q \leq \sigma \leq +\Jj_q$. The solution in Eq. \eqref{large q syk green's function solution 1} takes the form
\begin{equation}
    g(t) = 2 \log\left\{ \frac{\sigma}{\cj_q} \sech(\i \theta + \sigma t )  \right\}.
    \label{large q syk green's function solution 2}
\end{equation}
It is clear that the initial state is completely determined by the ratio $\sigma/\Jj_q$. Moreover, the temperature of the system can be determined by applying the KMS condition for the Green's function given in Eq. \eqref{kms relation}. A quick check shows that if 
\begin{equation}\label{eq:initial temperature}
    \cj_q\beta_f = 2\frac{\theta}{\sigma / \Jj_q}
\end{equation}
then $g(t)$ fulfills KMS. Since $\theta = \cos^{-1}(\sigma/\Jj_q)$, we further restrict $\sigma$ as $0\leq \sigma \leq \Jj_q$ for the temperature to remain always positive where we use the identity $\cos^{-1}(-x) = \pi - \cos(x)$ for $x \in [-1,1]$.

The corresponding energy is obtained by plugging Eq. \eqref{large q syk green's function solution 2} into Eq. \eqref{eq:NEQ_energy}, changing bi-local $(t_1, t_2) \to (t_1 - t_2)$, substituting $t_1 - t_2 = t$ that changes the integration limit from $0$ to $+\infty$, resulting in
\begin{equation}\label{eq:E0}
    \cE = - \sigma \tan{\theta} = - \cj_q \sqrt{1- \left(\frac{\sigma}{\cj_q}\right)^2},
\end{equation}
where all time dependency has been integrated out. This equation indicates that the energy is bounded from above by zero, with the zero temperature ($\beta_f \to \infty$) being the ground state at energy density $\cE = -1$ ($\sigma/\cj_q=0$) and its maximum value being $\cE=0$ at infinite temperature ($\sigma/\cj_q=1$). 

The large-$q$ SYK in equilibrium is both a limiting case and the initial state of the system for the mixed and kinetic quench as will now be proceeding to.

\subsubsection{Limiting case 2: \texorpdfstring{$\Jj_q(t) = \Theta(-t) \Jj_q$}{Lg} and \texorpdfstring{$\Jj_2(t) = \Theta(t) \Jj_2$}{Lg}}

\begin{figure}
    \centering
\includegraphics[width=0.90\linewidth]{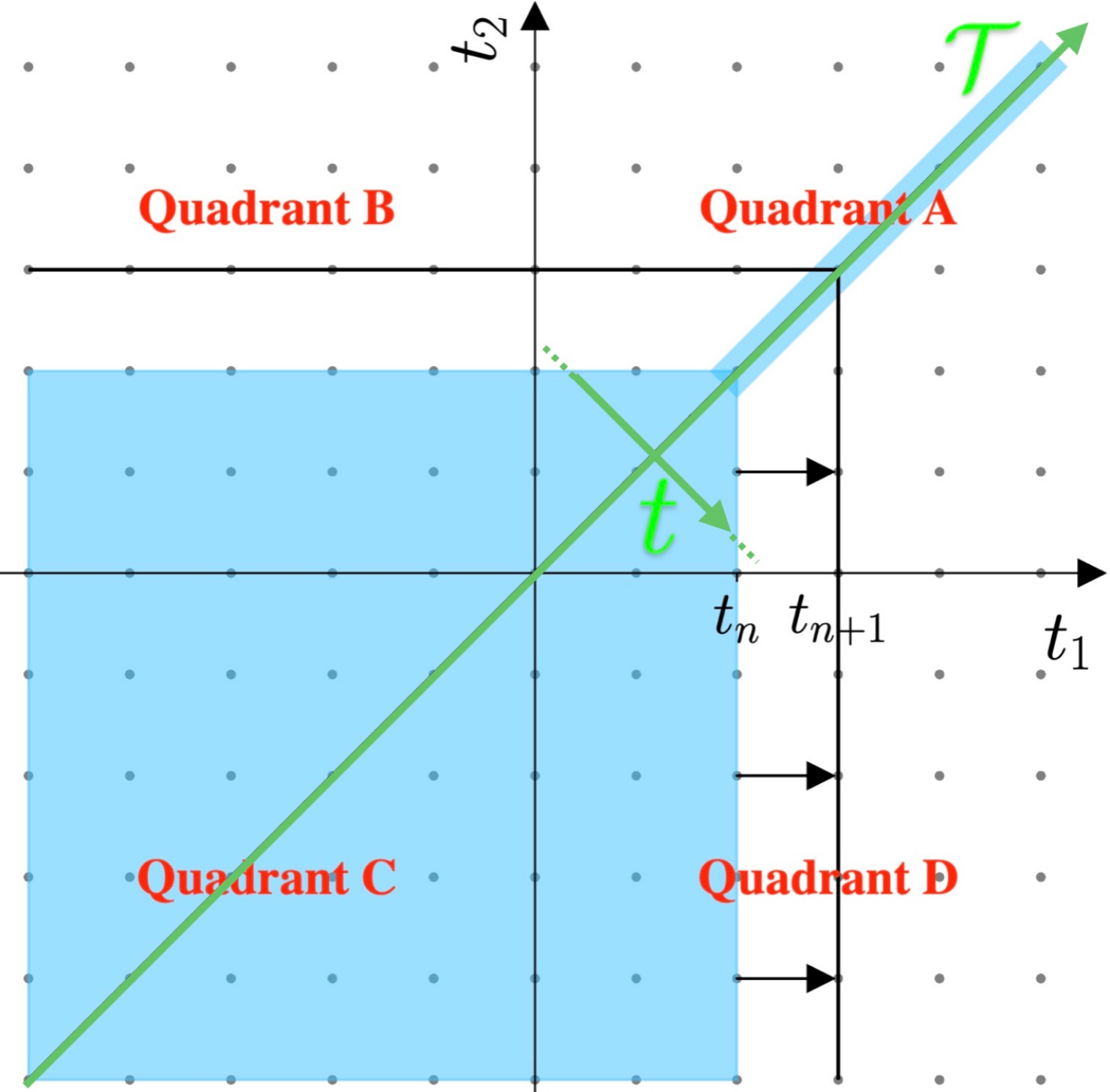}
    \caption{Sketch of the fan-like propagation in the two time plane induced by
\textit{causal stepping} as introduced in Section \ref{sect. numerical techniques}. All points inside the blue shaded areas are known. The time coordinates $(\mathcal{T}, t)$ are the Wigner coordinates defined in Eq. \eqref{eq:wigner_coords}.}
    \label{fig:causal_stepping}
\end{figure}

The so called kinetic quench consists of an initial large-$q$ SYK term in equilibrium, which is quenched at $t=0$ to a single random-hopping term. This case corresponds to the mixed quench where the final state is dominated by the kinetic term. Due to the quench, the Kadanoff-Baym equations have to be broken up into four quadrants in the two-time plane: Quadrant A,  where $t_1, t_2 \geq 0$; Quadrant B,  where $t_1 \leq 0, t_2 \geq 0$; Quadrant C,  where $t_1,t_2 \leq 0$ and Quadrant D; where $t_1 \geq 0, t_2 \leq 0$. Each quadrant has to be solved separately, and afterwards the integration constants are found by demanding consistency at the boundaries. A schematic can be found in Fig. \ref{fig:causal_stepping}.

Now we start to solve the system in all four quadrants. Quadrant C is the initial equilibrium large-$q$ SYK, so its solution $g_c(t_1,t_2)$ is just given by Eq. \eqref{large q syk green's function solution 2} where $t \to (t_1 - t_2)$, with initial temperature $\beta_0$ and energy $\cE_0$ given by Eqs. \eqref{eq:initial temperature} and \eqref{eq:E0}.

On the other hand, quadrants D and B are not in equilibrium, therefore the integro-differential equation (Eq. \eqref{kb equation for small g}) cannot be reduced to an ordinary differential equation. Fortunately, due to the conjugate property of $g(t_1, t_2)$ in Eq. \eqref{general conjugate relation}, solving in one quadrant is enough as the conjugation of that solution will solve the other quadrant. Let's consider quadrant D where exploiting the properties of the step-functions in the coupling constants, we can directly integrate Eq. \eqref{kb equation for small g} where $t_1 \geq 0$ and $t_2 \leq 0$
\begin{widetext}
    \begin{equation}
        \begin{aligned}
 \partial_{t_1} g\left(t_1, t_2\right) = & -\int_{-\infty}^{0} d t_3\left\{\cj_q\left(t_1\right) \cj_q\left(t_3\right)\left[e^{g\left(t_1, t_3\right)}+e^{g^*\left(t_1, t_3\right)}\right] + 2 \cj_2(t_1)\cj_2(t_3) \right\}  \\
 & -\int_{0}^{t_1} d t_3\left\{\cj_q\left(t_1\right) \cj_q\left(t_3\right)\left[e^{g\left(t_1, t_3\right)}+e^{g^*\left(t_1, t_3\right)}\right] + 2 \cj_2(t_1)\cj_2(t_3) \right\} \\
 & +2 \int_{-\infty}^{t_2} d t_3\left\{\cj_q\left(t_1\right)\cj_q\left(t_3\right)e^{g(t_1,t_2)} + \cj_2(t_1)\cj_2(t_3) \right\},
\end{aligned}
    \end{equation}
\end{widetext}
to finally get $ \partial_{t_1} g_D\left(t_1, t_2\right) = -2 \Jj_2^2 t_1$ that gives
\begin{align}
    g_D\left(t_1, t_2\right) = -\cj_2^2t_1^2 + D(t_2).
\end{align}
Here the integration constant is found from the boundary condition $g_D(0,t_2) = g_C(0,t_2)$, such that the solution becomes
\begin{equation}\label{eq:kinetic gD}
  g_D(t_1,t_2) = - \mathcal{J}_2^2 t_1^2 + 2\log{\Big[\frac{\sigma}{J\cosh{(i\theta - \sigma t_2)}}\Big]}.
\end{equation}
Notice how the Green's function cannot be expressed solely in terms of a time difference $(t_1 - t_2)$ as expected of an out-of-equilibrium state. This immediately gives the solution in quadrant B
\begin{equation}\label{eq:kinetic gB}
\begin{aligned}
    g_B(t_1,t_2)  &= g_D(t_2, t_1)^\star \\
    &= -\mathcal{J}_2^2 t_2^2 + 2\log{\Big[\frac{\sigma}{J\cosh{(i\theta + \sigma t_1)}}\Big]}.
    \end{aligned}
\end{equation}

Finally, the post-quench state in quadrant A is obtained, without assuming equilibrium, by directly integrating the corresponding Kadanoff-Baym equation and applying boundary conditions. We get $\partial_{t_1} g_A\left(t_1, t_2\right) = -2 \Jj_2^2 t_1 + 2 \Jj_2^2 t_2$ where $t_1, t_2 \geq 0$. This is integrated to get
\begin{equation}
    g_A(t_1, t_2) = - \Jj_2^2 t_1^2 + 2\Jj_2^2 t_2 + A(t_2)
\end{equation}
where the integration constant $A(t_2)$ is found via the boundary condition $g_A(t_2, t_2) = 0$. The final result is given by
\begin{equation}
  g_A(t_1,t_2) = - \mathcal{J}_2^2 (t_1-t_2)^2
  \label{eq:green's function for kinetic quench}
\end{equation}
which is always real and solely dependent upon time difference indicating that the final state instantly reaches thermal equilibrium with respect to the Green's function. Having found the Green's function in all four quadrants, we can now calculate the energy density pre- and post-quench using Eq. \eqref{eq:NEQ_energy}. The pre-quench initial energy density $\Ee_i(t_1 \leq 0)$ is given by the large-$q$ SYK energy density which is the same as in Eq. \eqref{eq:E0}. Given that $g_A(t_1, t_2)$ is purely real, the final energy density of the system is given by 
\begin{align}
     \cE_f(t_1 \geq 0) = \text{Im} \int_{-\infty}^{t_1}\, dt_{2}\,  \cj_2^2 g_A(t_1,t_2) = 0.
\end{align}
Therefore, the kinetic energy (in this case being the only component) \textit{instantaneously} adjusts to the new post-quench value (compare to the rich dynamics as studied below in Figs. \ref{fig:visualization_jtwo0.02} and \ref{fig:more_visualizations}). The energy density before the quench in quadrant C is decided by the initial inverse temperature $\beta_0$ (as we saw in the first limiting case above) and independent of the initial $\beta_0$, the final state after the quench is always at zero energy instantaneously corresponding to the infinite temperature state. Moreover, Eq.~\eqref{eq:green's function for kinetic quench} shows that the post-quench Green's function is not bi-local but a function of the time difference and therefore stationary. Furthermore, as shown by \cite{louw2022} the final state of the Green's function fulfills the KMS conditions, resembling a thermal state. Thus, having calculated both the Green's function and the energy density, we conclude that the kinetic quench is an example of instantaneous thermalization of large-$q$ SYK quenches to a single term in the final Hamiltonian where the final thermal state is always at infinite temperature, regardless of the pre-quench equilibrium conditions. The thermalization is with respect to the dynamics of the Green's function as further evidenced by the instantaneous saturation of post-quench kinetic energy density. This is in line with Refs.~\cite{Eberlein2017Nov, bhattacharya2019, haldar2020, louw2022} (also see Ref.~\cite{Magan2016Jan}). We emphasize that the instantaneous equilibration is with respect to the Green's function that satisfies the KMS relation and attains time-translational invariance immediately post-quench, however the final stationary state reached post-quench is not a thermal ensemble \cite{bhattacharya2019}.

Both of the limiting cases discussed in this section are relevant as they provide useful benchmarking for  the developed algorithm to solve for the general mixed quench ($\text{SYK}_q \to \text{SYK}_q + \text{SYK}_2 $). The analytical tractability of the two presented limiting cases sets the expectations for the numerical implementation of the mixed quench. For example, for weak quenches, the mixed SYK system is expected to have a final temperature close to the initial one. On the other hand, as the random-hopping term dominates over the interaction, the mixed quench should thermalize faster and to higher temperatures, approaching the behavior of the kinetic quench.

With these two analytic solutions in mind, we now turn to develop the numerical method used to solve and analyze the mixed quench.

\section{Numerical Techniques}
\label{sect. numerical techniques}

In a general quench setting, the Kadanoff-Baym equation are also split up in the four quadrants mentioned in the kinetic quench. The initially prepared state belongs to the quadrant C in Fig. \ref{fig:causal_stepping} where quench is introduced at the origin that leads to non-equilibrium behaviors in quadrants A, C and D. We are interested in studying thermalization which happens at late times in quadrant A. The casual structure of the two-time plane requires a special iterative procedure, often referred to as \textit{causal stepping} in order to avoid introducing retro-causal relations \cite{meirinhos2022} in the Green's function. A sketch of this iteration scheme can be seen in Fig. \ref{fig:causal_stepping} where all points inside the blue shaded area are known. As a consequence, most ready-made integrators cannot be used to solve the Kadanoff-Baym equations. While there are some dedicated packages \cite{meirinhos2022, nessi2020}, we developed our own algorithm for the Julia programming language and a methodology to analyze the numerical solution in Python. These are the subject of this section. First, an overview of a predictor-corrector integrator with causal-stepping is given. This is followed by presenting the results of the benchmarking with the limiting cases. Finally, we give an explanation of how the energy (Eq. \eqref{eq:NEQ_energy}) and its components can be used to calculate the final temperature of the system and the corresponding thermalization rate.

\subsection{Predictor-Corrector in the Two-Time plane}
\label{subsec predictor-corrector in the two time plane}

The initially prepared equilibrium solution for large-$q$ SYK model in quadrant C is already known from Eq. \eqref{large q syk green's function solution 2}. The initial temperature is decided by the ratio $\sigma/\Jj_q$ using Eq. \eqref{eq:initial temperature} where $\beta_f$ is the inverse of the initial temperature which we denote as $\beta_0$ from henceforth. This state is propagated after the quench into quadrants A, B and D, following a \textit{causal stepping} procedure whose sketch is provided in Fig. \ref{fig:causal_stepping}. As mentioned in the captions, all points inside the blue shaded region are known. The points in the $t_1=t_{n+1}$ line, where $t_2 = t_m\leq t_n$, are completely determined by the shaded region. We recall the boundary condition and the conjugate property of the Green's function from Eq. \eqref{little g boundary and conjugate relation} which tells us that the upper-triangular region is found from the complex conjugate of the lower-triangular region as per $g(t_n,t_m) = g^*(t_m,t_n)$. Therefore, only the horizontal step in the $t_1$ direction has to be solved. As a result the solution is propagated in a ``fan-like'' motion, where at each $t_n$ the known square region grows in both $(t_1, t_2)$ directions by one step. This way of calculating the solution guarantees that $g(t_1,t_2)$ does not depend on any future times, hence the name causal stepping.

Taking into account the causal stepping, we proceed to discretize the two-time plane as a symmetric grid, with the fixed-time step $\Delta t$ which is equal in both time directions. Given that the solution in quadrant C is already known while the solution in the upper triangular region (above $t_1 = t_2$ line in Fig. \ref{fig:causal_stepping}) is completely known using the conjugate property of $g(t_n, t_m)$, we only consider the region where $t_1 \geq 0$. Before discretizing the integrals, it is useful to combine the Kadanoff-Baym equations in quadrants A and D as ($t_1 \geq 0$)
\begin{align}
    \partial_{t_1} g\left(t_1=t_n, t_2=t_m \right) = & \int_{t_0}^{t_n} d t_3 F[g(t_n,t_3)] \\
    &+\int_{t_0}^{t_m} dt_3 K[g(t_n,t_3)],
\end{align}
where we have set a cut-off time $t_0$ in a distant past, long enough to minimize the error from the excluded terms and we define (recall $\Jj_2(t) = \Theta(t) \Jj_2$)
\begin{align}
   & F[g(t_n,t_3)] \equiv -\cj_q^2\left[e^{g\left(t_n, t_3\right)}+e^{g^*\left(t_n, t_3\right)}\right] - \cj_2(t_n)\cj_2(t_3) \\
    & K[g(t_n,t_3)] \equiv 2 \mathcal{J}_q^2  e^{g\left(t_n, t_3\right)} +2\mathcal{J}_2(t_n)\mathcal{J}_2(t_3).
\end{align}
To estimate the differential in the left-hand side in an interval $t_1 \in [t_n,t_{n+1}]$ along the $t_1$ axis in Fig. \ref{fig:causal_stepping}, we use the two-step predictor-corrector scheme described in Ref. \cite{haldar2020}: first, we calculate a prediction $g_P$ for the value at $(t_{n+1},t_m)$ by 
\begin{equation}
    g_P(t_{n+1},t_m) = g(t_n,t_m) + \Delta t P(t_n,t_m)
\end{equation}
where the slope $P(t_n,t_m)$ is given by
\begin{equation}
    P(t_n,t_m) = \int_{t_0}^{t_n} d t_3 F[g(t_n,t_3)] +\int_{t_0}^{t_m} dt_3 K[g(t_n,t_3)].
\end{equation}
Using this prediction, we get the corrected value
\begin{equation}
    g(t_{n+1},t_m) = g(t_n,t_m) + \frac{\Delta t}{2} \left( P(t_n,t_m) + P(t_{n+1},t_m) \right),
\end{equation}
where the slope at $t_{n+1}$ is obtained via
\begin{align}
    P(t_{n+1},t_m) =& \int_{t_0}^{t_{n+1}} d t_3 F[g_p(t_{n+1},t_3)] \\
    &+\int_{t_0}^{t_m} dt_3 K[g_p(t_{n+1},t_3)].
\end{align}
The integrals in these terms are calculated using the trapezoid rule for each $t_m<t_n$, given that the diagonal is always zero, namely $g(t_n, t_n) = 0$. After all the values in the vertical $t_1=t_{n+1}$ line (up to the diagonal) have been calculated, the horizontal $t_2 = t_{n+1}$ line is found using the conjugation property $g(t_m,t_{n+1}) = g^*(t_{n+1},t_m)$ and time is advanced by one step.

As mentioned in Ref. \cite{meirinhos2022}, it is useful to think of the two-time procedure as solving a system of vector-valued differential equations, by arranging the components along the vertical $t_1=t_n$ axis in vector form
\begin{equation}
    \Vec{g}(t_{n}) = \begin{pmatrix}
        g(t_n,t_0), & g(t_n,t_1), & \hdots, & g(t_n,t_{n-1}), & g(t_n,t_{n})
    \end{pmatrix}^T.
\end{equation}
where the last element $g(t_n,t_n) = 0$ due to the boundary condition of the Green's function (Eq. \eqref{little g boundary and conjugate relation}). Accordingly, the slope is written as
\begin{align}\label{eq:Ptn}
    \Vec{P}(t_n) &= 
    \begin{pmatrix}
        P(t_n,t_0) \\ P(t_n,t_1) \\ \vdots \\ P(t_n,t_n)
    \end{pmatrix} \\
    &=
    \begin{pmatrix}
           \int_{t_0}^{t_n} d t_3 F[g(t_n,t_3)] \\
           \int_{t_0}^{t_n} d t_3 F[g(t_n,t_3)] \\
         \vdots \\
          \int_{t_0}^{t_n} d t_3 F[g(t_n,t_3)]
    \end{pmatrix} 
    +
    \begin{pmatrix}
           \int_{t_0}^{t_0} d t_3 K [g(t_n,t_3)] \\
           \int_{t_0}^{t_1} d t_3 K [g(t_n,t_3)] \\
         \vdots \\
          \int_{t_0}^{t_n} d t_3 K [g(t_n,t_3)]
    \end{pmatrix}
\end{align}
that gives the predicted value
\begin{equation}
    \Vec{g}_P(t_{n+1}) = \Vec{g}(t_n) + \Delta t \Vec{P}(t_n)
\end{equation}
and finally this leads to the corrected value
\begin{align}
    \Vec{g}(t_{n+1}) = \Vec{g}(t_{n}) + \frac{\Delta t}{2} \left( \Vec{P}(t_n) + \Vec{P}(t_{n+1}) \right).
    \label{corrected g value in vector form}
\end{align}
where the slope at $t_{n+1}$ is given by
\begin{align}\label{eq:Ptn+1}
    \Vec{P}(t_{n+1})
    =& \begin{pmatrix}
        P(t_{n+1},t_0) \\ P(t_{n+1},t_1) \\ \vdots \\ P(t_{n+1},t_n)
    \end{pmatrix}\\
   =& \begin{pmatrix}
           \int_{t_0}^{t_n} d t_3 F[g_p(t_{n+1},t_3)] \\
           \int_{t_0}^{t_n} d t_3 F[g_p(t_{n+1},t_3)]  \\
         \vdots \\
        \int_{t_0}^{t_n} d t_3 F[g_p(t_{n+1},t_3)] 
    \end{pmatrix} \\
    &+
    \begin{pmatrix}
           \int_{t_0}^{t_0} d t_3 K [g_p(t_{n+1},t_3)] \\
           \int_{t_0}^{t_1} d t_3 K [g_p(t_{n+1},t_3)] \\
         \vdots \\
          \int_{t_0}^{t_n} d t_3 K [g_p(t_{n+1},t_3)]
    \end{pmatrix}.
\end{align}
In this vector-valued differential equation in Eq. \eqref{corrected g value in vector form}, one can readily notice one of the main difficulties involved in solving the Kadanoff-Baym equations numerically. Given that $t_0\leq t_m\leq t_n$, the involved vectors in the differential equation grow by one row at each time-step. Hence, the computational time increases for each subsequent iteration in $t_n$ as more integrals and longer vectors need to be handled.

Another problematic feature of the Kadanoff-Baym equations worth highlighting at this point is their non-Markovian nature: the integrals in Eqs. \eqref{eq:Ptn} and \eqref{eq:Ptn+1} at each $t_n$ involve all past values of $g(t_n,t_m)$. For numerical methods, this implies that the errors picked up at the start of the algorithm stack over time until the solution becomes unstable. As a consequence, a relatively small $\Delta t$ has to be chosen in order to account for the fast dynamics shortly after the quench, resulting in very large matrix representations of the Green's functions. To account for this issue, some adaptive solvers such as the one developed in Ref. \cite{meirinhos2022} have been proposed which allow for efficient and accurate estimations of non-equilibrium Green's functions. However, the Kadanoff-Baym equations for the mixed SYK as considered in this work are comparatively simple to be solved by a fixed step method.

To summarize, the causal stepping has been implemented as follows: first, the exact initial state given in Eq. \eqref{large q syk green's function solution 2} is calculated and is saved in elements of the matrix representation of $g(t_1,t_2)$ in quadrant C, up to a cut-off negative time $t_0$ set in the distant past. This initial state is then propagated into quadrant D in the $t_1$ direction by calculating $g(t_{1}+\Delta t, t_2)$ (using the predictor-corrector) for each of the components of $\Vec{g}(t_1)$, starting at $t_2=t_0$ up to $t_2=t_1-\Delta t$, just below the diagonal in quadrant A. The time step $\Delta t$ is taken as a constant along both $t_1$ and $t_2$ axes in the two-time plane. The process is repeated until a maximum time $t_1 = t_{\text{max}}$. The values above the diagonal are obtained with $g(t_2,t_1) = g^*(t_1,t_2)$, which includes quadrant B. Iteratively, we cover the entire two-time plane while keeping track of causality and non-Markovian nature of the Kadanoff-Baym equations. 

Having described the integration algorithm used to solve the Kadanoff-Baym equations, ones needs to measure the quality of the post-quench state. One useful metric is the conservation of energy (Eq. \eqref{eq:NEQ_energy}) at each time-step. This is expected since the Hamiltonian in Eq. \eqref{eq:mixed_syk_neq_hamiltonian} has no dissipative terms. In addition, since the disorder in the kinetic and the interaction terms are uncorrelated in the Hamiltonian, the expectation value of the energy with respect to the final Hamiltonian should be the same as the initial one. Therefore, we track the difference in the pre- and post-quench total energies as measure the stability of this algorithm. We require numerically that the change in energy has to be below $1\%$ at every point in time. We now proceed to provide the benchmarking results against the analytically solvable limits as presented in Section \ref{subsec. limiting cases}.

\onecolumngrid

\begin{figure}
	\centering
	\subfloat[]{\label{fig:energy_noquench}\includegraphics[width=.40\linewidth]{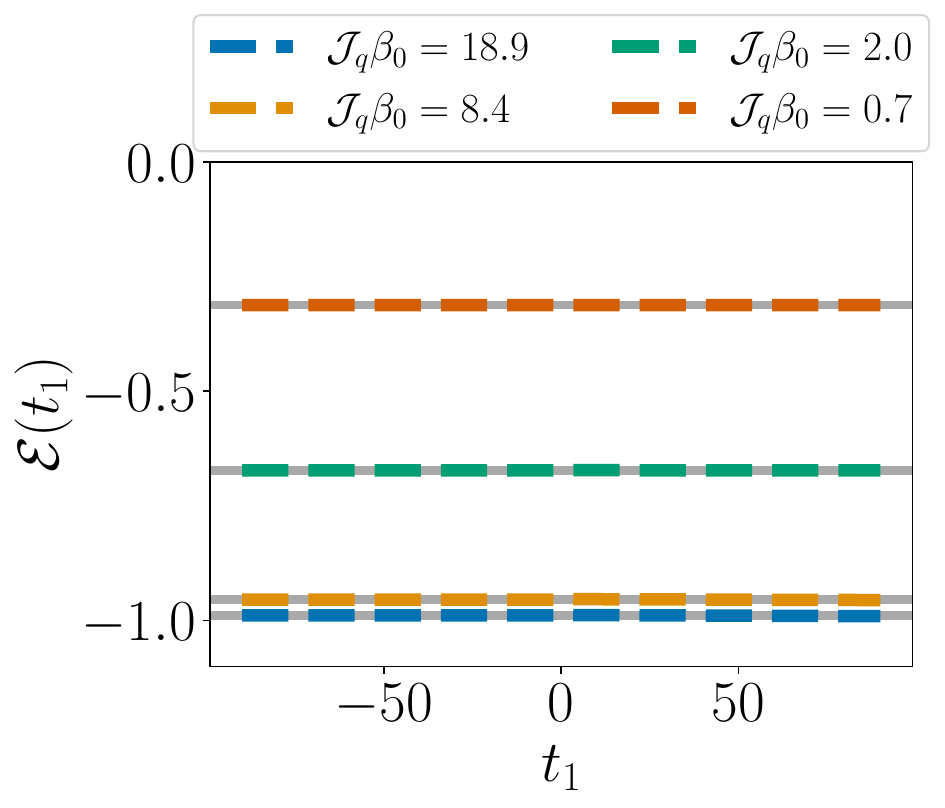}}
	\subfloat[]{\label{fig:errors_noquench}\includegraphics[width=.50\linewidth]{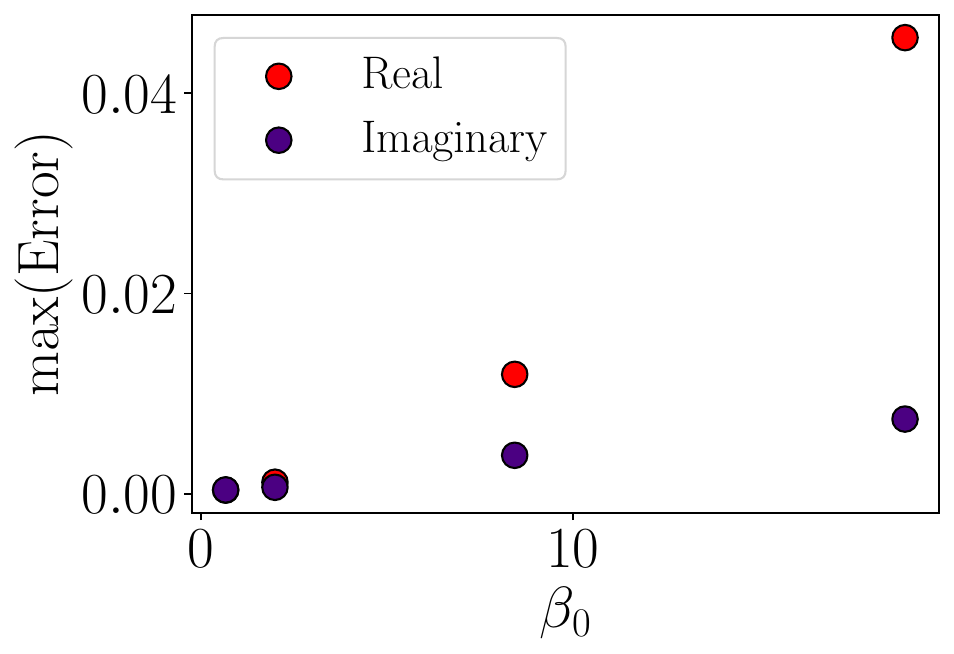}}\par
	\subfloat[]{\label{fig:energy_kineticquench}\includegraphics[width=.40\linewidth]{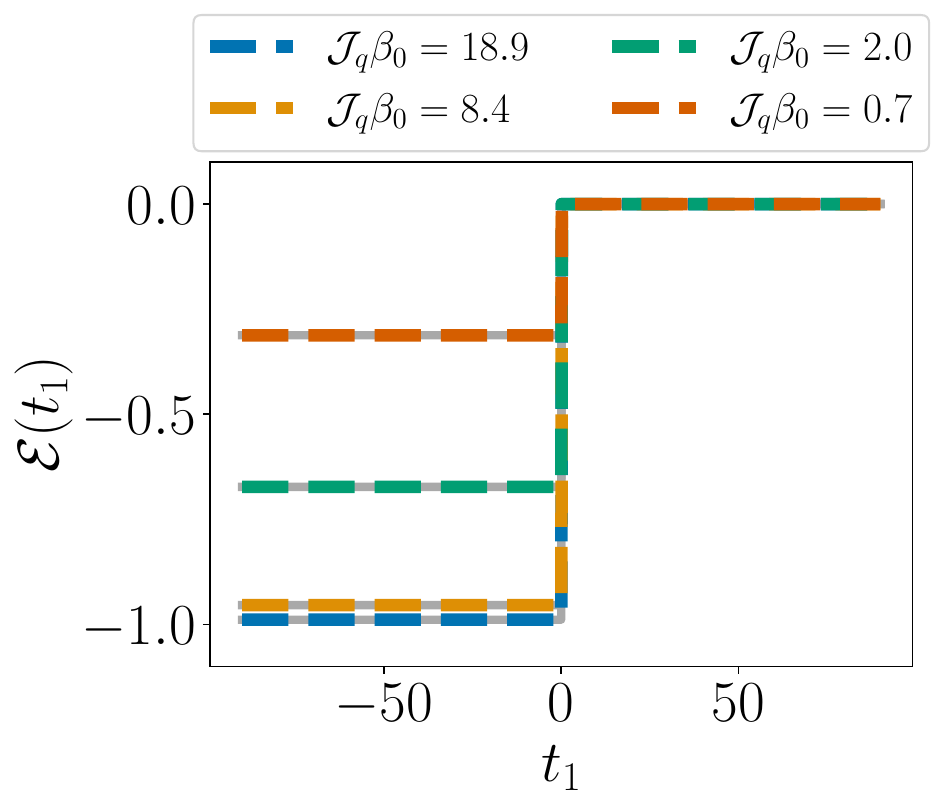}}
	\subfloat[]{\label{fig:errors_kineticquench}\includegraphics[width=.50\linewidth]{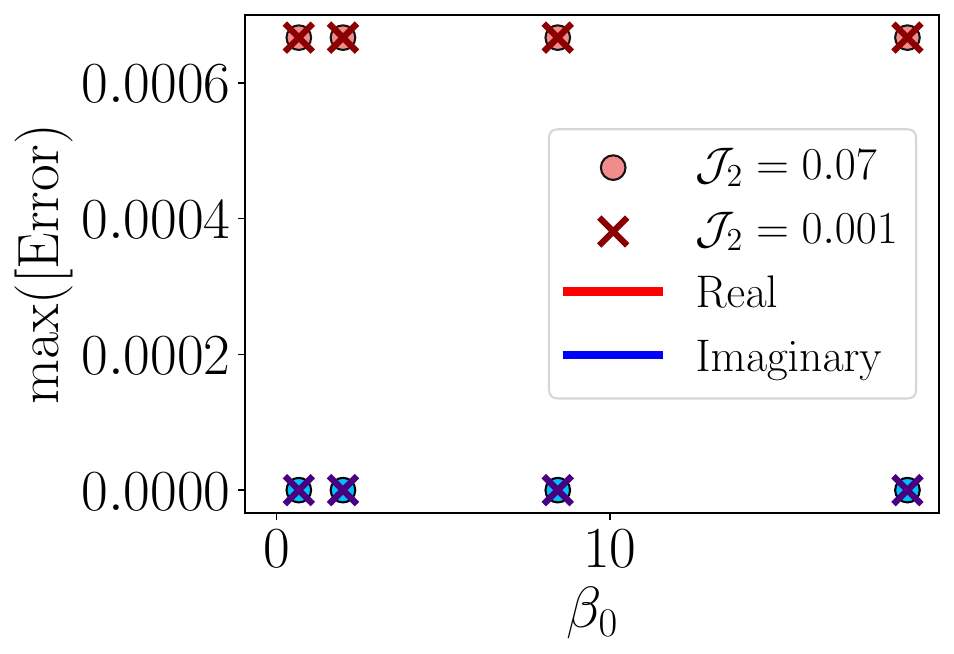}}
	\caption{Predictor-corrector performance in the limiting cases (as presented in Section \ref{subsec. limiting cases}) with $\Delta t = 0.06$. The dashed lines represent the numerical energy and the solid lines the corresponding analytical values. The red and blue dots represent the maximum relative error between the exact and numerical Green's function in the post-quench quadrants. Plots (a) and (b) depict the energies and errors for several temperatures, in the single large-$q$ SYK at equilibrium, where there is no quench ($\Jj_2 = 0$). Plots (c) and (d) show the same results for the kinetic quench. Plot (d) includes the errors for two different $\cj_2$ as round and x-shaped markers. Note that the x-markers ($\cj_2=0.001$) overlap with the circles ($\cj_2=0.07$) as the errors are significantly close to each other.}
	\label{fig:benchmark_results}
\end{figure}

\twocolumngrid

\begin{figure}
	\centering
	\subfloat[]{\label{fig:bench_Regn_kinquench}\includegraphics[width=.33\linewidth]{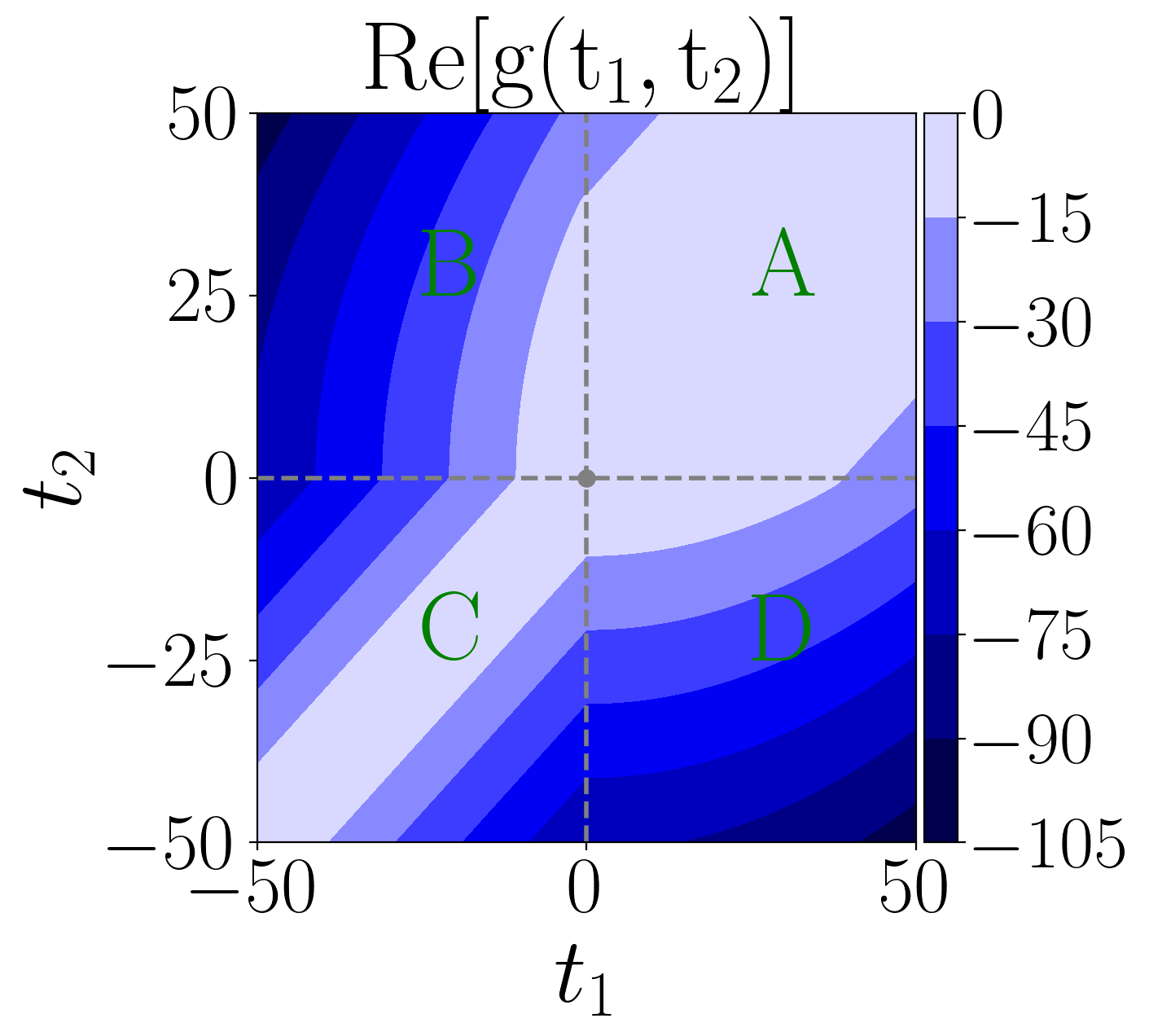}}
	\subfloat[]{\label{fig:bench_Imgn_kinquench}\includegraphics[width=.33\linewidth]{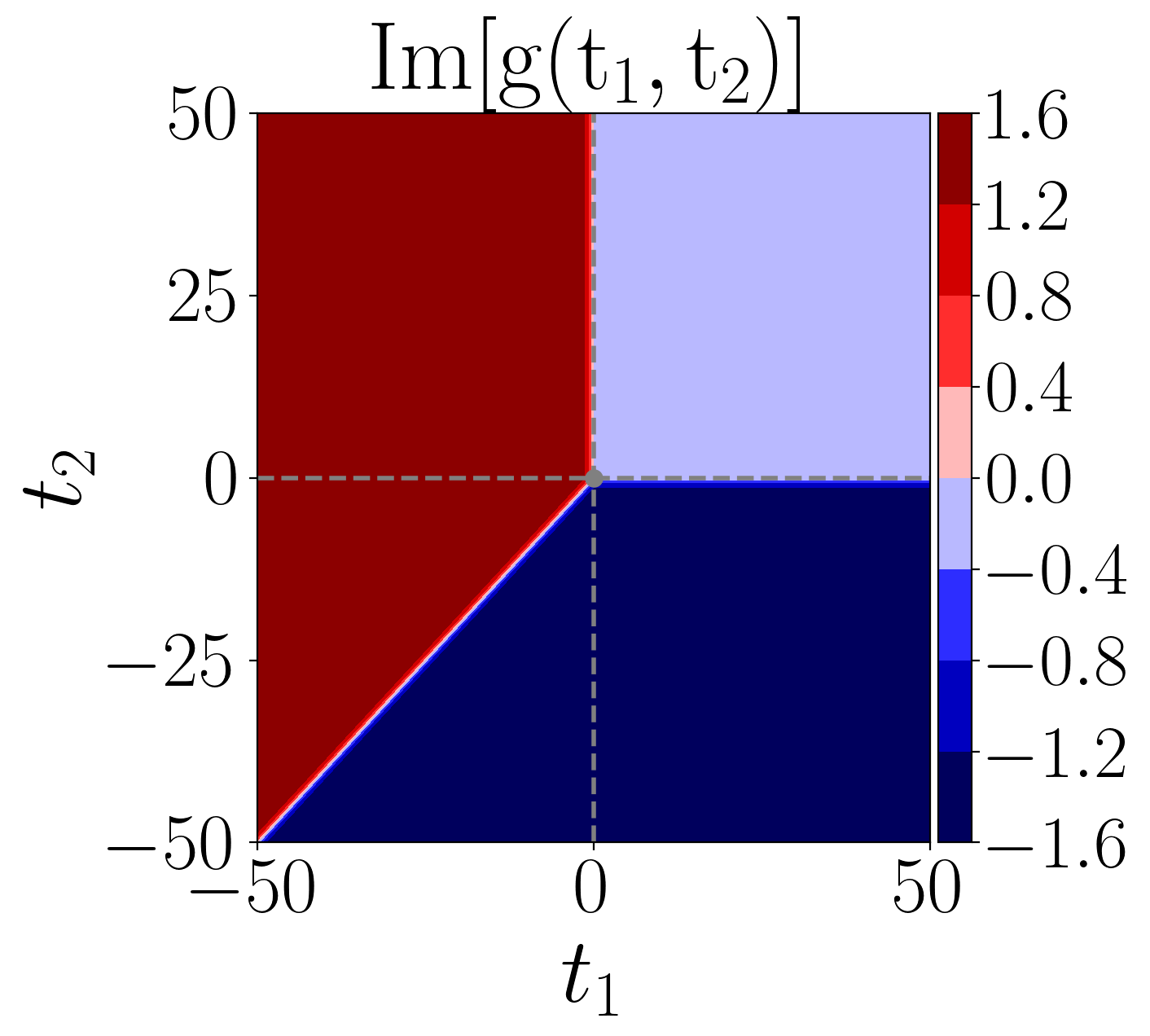}}
	\subfloat[]{\label{fig:bench_Errgn_kinquench}\includegraphics[width=.33\linewidth]{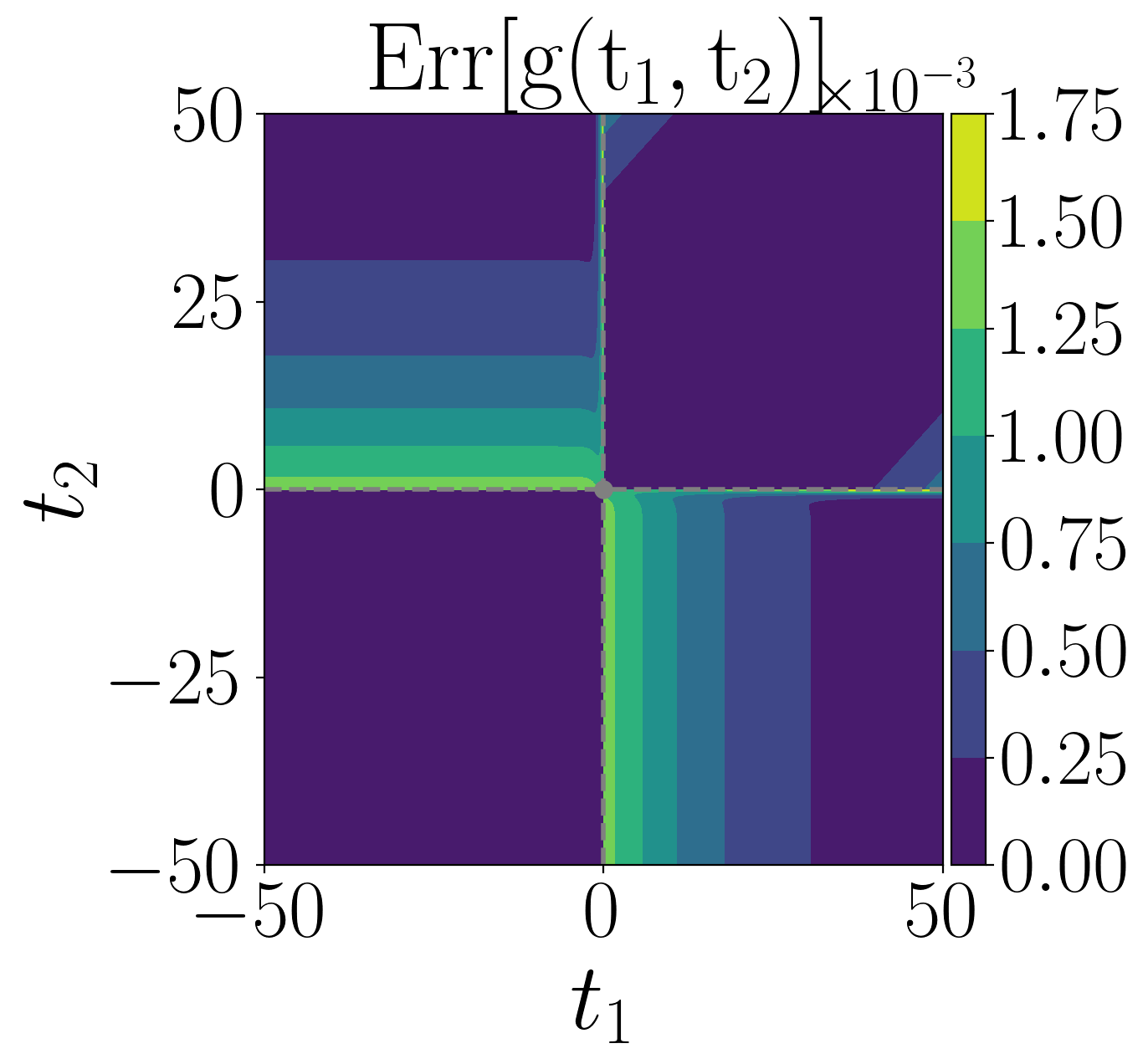}}\par
	\subfloat[]{\label{fig:bench_Regn_noquench}\includegraphics[width=.33\linewidth]{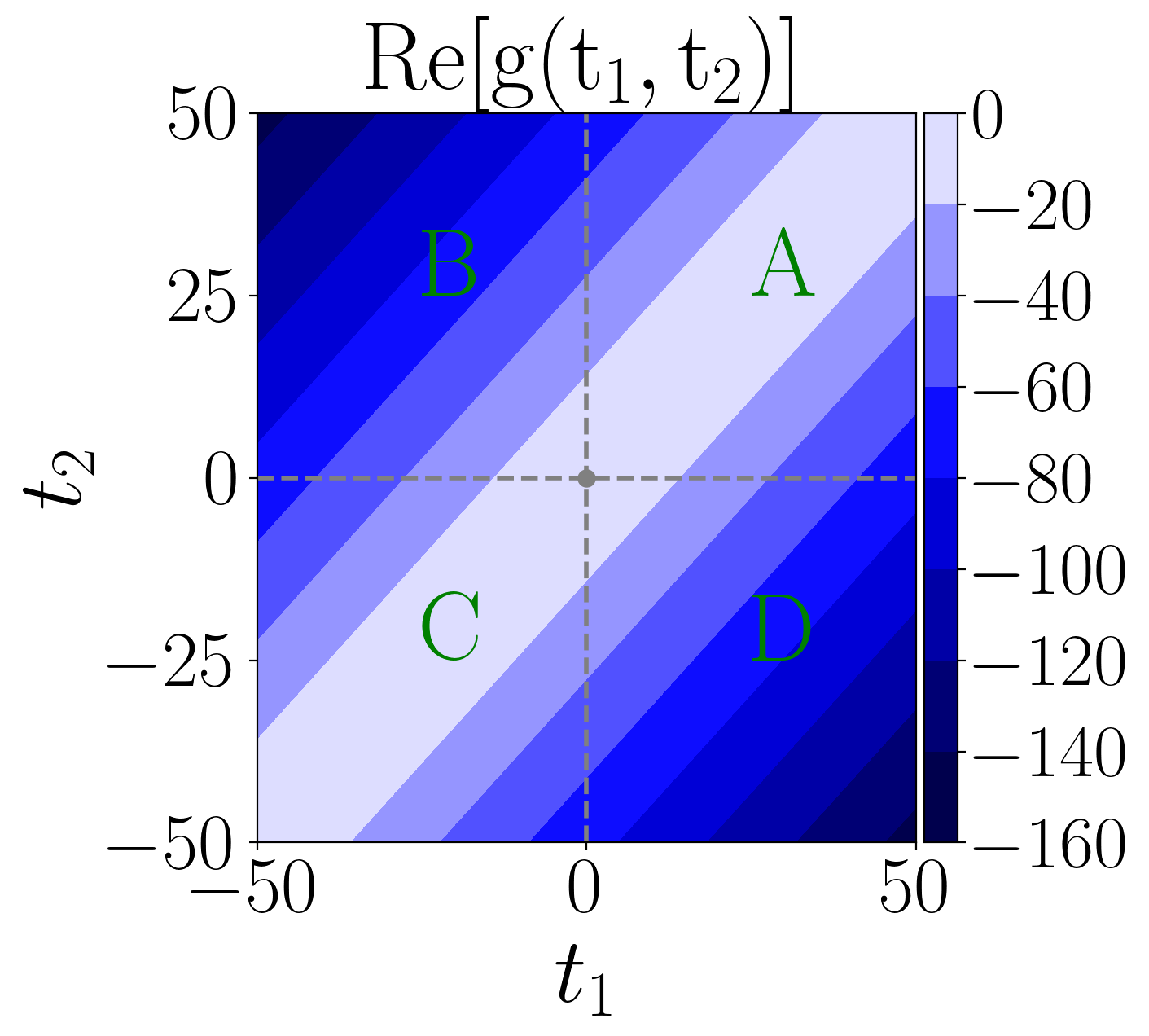}}
	\subfloat[]{\label{fig:bench_Imgn_noquench}\includegraphics[width=.33\linewidth]{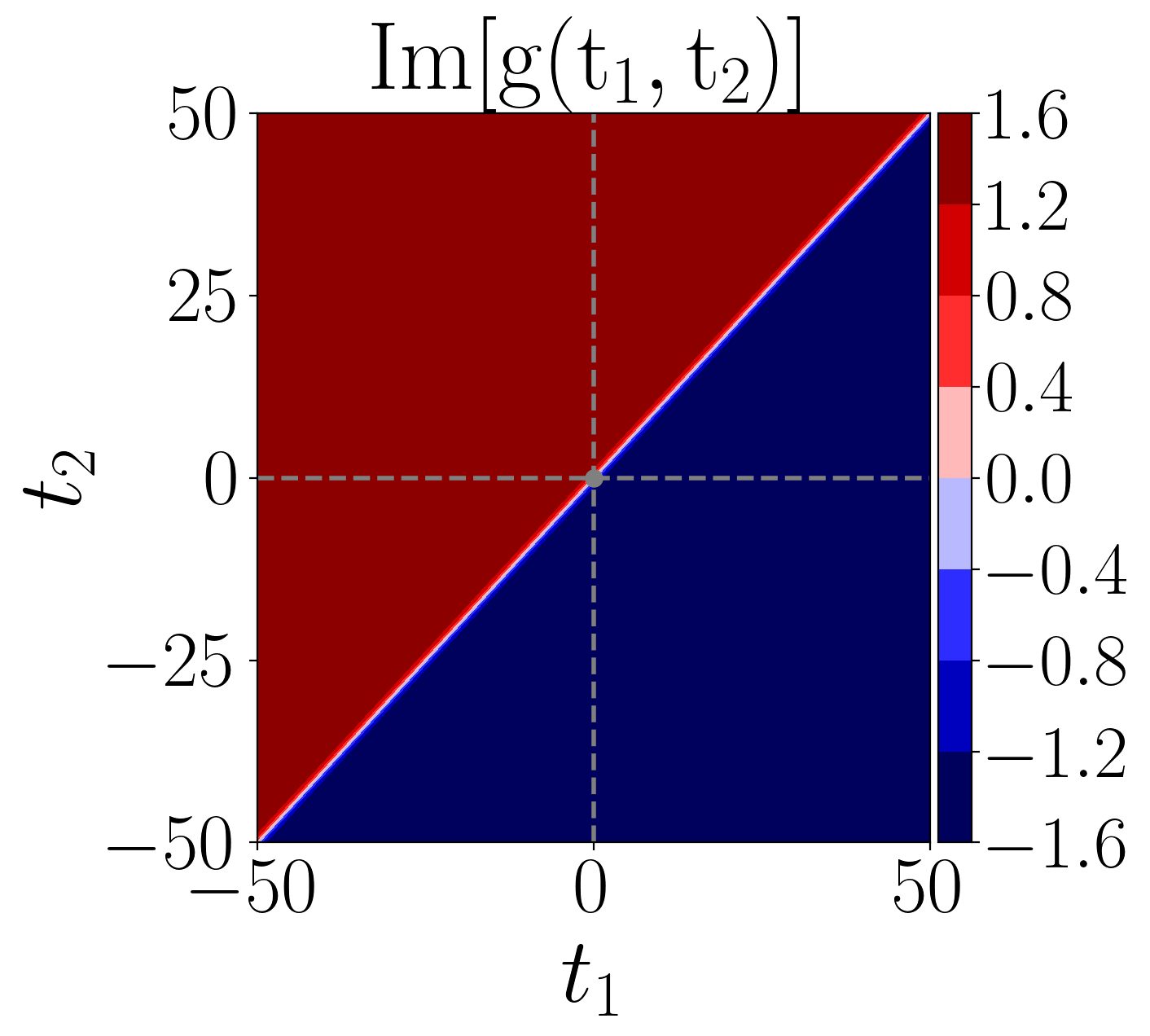}}
	\subfloat[]{\label{fig:bench_Errgn_noquench}\includegraphics[width=.33\linewidth]{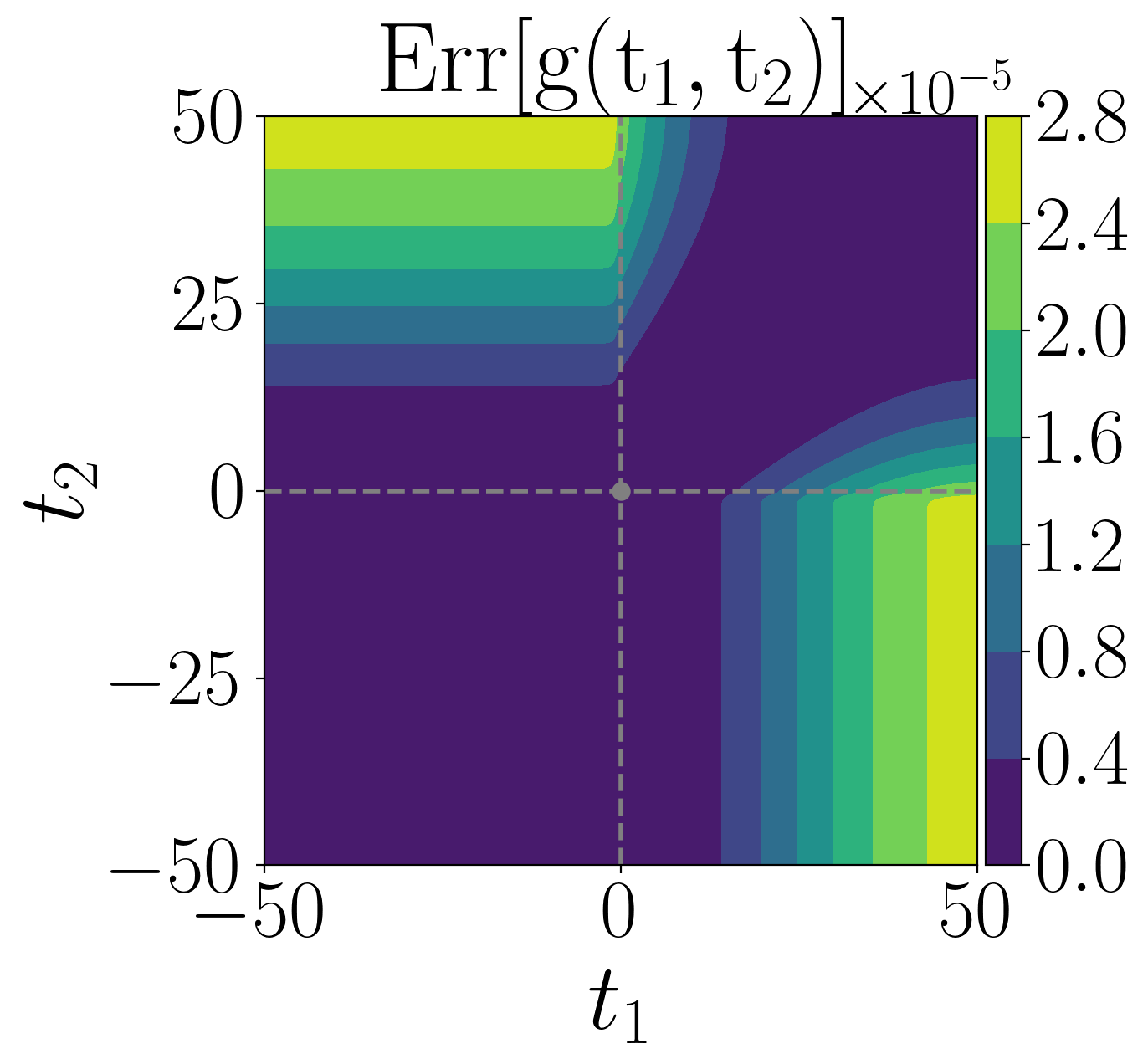}}
	\caption{Predictor-corrector solutions for the case of a kinetic quench as presented in Section \ref{subsec. limiting cases} with $\Delta t = 0.06$. (a)-(b) Real and imaginary components of $g(t_1,t_2)$ with $\cj_2 = 0.1$ and $\beta_0 = 2.0$. (c) Heat map of the absolute squared error between the numeric and analytic Green's function $g(t_1,t_2)$. (d)-(f) The same results but for the (non-quenched) limiting case of large-$q$ SYK model as presented in Section \ref{subsec. limiting cases} where nothing changes.}
	\label{fig:benchmark-green's functions}
\end{figure}

\subsection{Bench-marking and error-estimation}\label{sec. benchmarking}

The algorithm is benchmarked with respect to the analytically solvable cases described in Section \ref{subsec. limiting cases}. We find, as detailed in Fig. \ref{fig:benchmark_results}, that the predictor-corrector scheme is capable of maintaining a constant energy and obtaining a $g(t_1,t_2)$ well within $1\%$ of the analytic solution in both cases. Fig. \ref{fig:benchmark_results} shows the results at $\Delta t = 0.06$ for both limiting cases at several initial temperatures. For kinetic quench in particular, we also analyze for different values of $\Jj_2$ in addition to various initial temperatures. For both single large-$q$ SYK model in equilibrium as well as kinetic quench, the numerical $g(t_1,t_2)$ remains close to the analytic solutions. For large-$q$ SYK model without quench, the energy stays constant throughout. For the kinetic quench, the total energy immediately jumps to a constant value of zero after the quench signifying that the numerical solution for the Green's function $g(t_1,t_2)$ exhibits instantaneous thermalization to an infinite temperature state, regardless of the initial inverse temperature $\beta_0$ as well as the quench strength $\Jj_2$.

Furthermore, the numerical Green's function corresponding to both the limiting cases in Section \ref{subsec. limiting cases} are shown in Fig. \ref{fig:benchmark-green's functions} where we observe as expected an instantaneous thermalization for the kinetic quench.

For both the equilibrium and the kinetic quench, systems starting at lower initial temperatures exhibit larger errors. This behavior is especially relevant for the mixed-quench, as is discussed in the section \ref{sec. thermalization}. Overall, the benchmark with respect to the equilibrium $\text{SYK}_q$ and the kinetic quench indicates that the predictor-corrector algorithm achieves a good degree of accuracy and its solutions are capable of capturing the expected equilibrium behavior and instant thermalization.

We now provide the methodology used to determine the final temperature of the equilibrium state as well as the thermalization rate for the mixed quench. 

\subsection{Estimation of the final temperature}
\label{subsec Estimation of the final temperature}

There are different ways of describing the approach to equilibrium such as quantum scrambling \cite{paviglianiti2023}, the fluctuation-dissipation theorem \cite{larzul2022, Eberlein2017Nov, bhattacharya2019} and relaxation of observables \cite{Babadi2015Oct, dieplinger2023, haldar2020}.  It is this latter approach we will take for our functional definition of equilibrium: the post-quench system will be considered to be sufficiently close to equilibrium once its energy components have saturated to a constant value. 

The final temperature of the system can be estimated by comparing the non-equilibrium Green's function to one obtained from a system already prepared in equilibrium. The Green's function for the latter thermal system is given by Eq. \eqref{equilibrium ODE in imaginary time} where the imaginary time Green's function is related to real-time Green's function via $\Tilde{g}(x) = g(-i\beta_f x)$. Here $\beta_f$ is the final temperature at equilibrium (technically the final temperature is given by $\Jj_q \beta_f$ but we keep $\Jj_q = 1.0$ throughout this work). Unfortunately there is no known analytical solution to Eq. \eqref{equilibrium ODE in imaginary time}. However, numerical solutions for a given $\beta_f$ can be found using the power-series based solver developed in Ref. \cite{nick2020}. In this method, the Green's function is written as 
\begin{equation}\label{eq:power_series_thermal}
    \Tilde{g}(x) = \lim_{N\rightarrow \infty} \sum_{m=1}^N c_n \Big(x-\frac{1}{2}\Big)^{2n}.
\end{equation}
For given values of $\cj_q$, $\cj_2$ and $\beta_f$, the coefficients $c_n$ are found recursively from $c_0$, which is in itself determined (analogous to the shooting method) by choosing the value that best fulfills the boundary conditions $ \Tilde{g}(0) =  \Tilde{g}(1) = 0$. Notice that here $\beta_f$ is a free parameter. Having recursively solved for the equilibrium Green's function $\Tilde{g}(x)$, the correct $\beta_f$ for the quenched system can be determined by fitting the equilibrium energy to the long time limit of non-equilibrium energy in Eq. \eqref{eq:NEQ_energy}. The equilibrium energy corresponding to Eq. \eqref{equilibrium ODE in imaginary time} can be derived from Eq. \eqref{energy intemediate step 0 in appendix} by restricting to the imaginary part of the Keldysh contour $\Cc_{\text{imag}}$ in Fig. \ref{fig:keldysh_contour}. We derive this relation in Appendix \ref{app. Evaluating energy in the Keldysh contour} where the final result is given by
\begin{equation}
\begin{aligned}
\mathcal{E}_{\text{EQ}} (\beta_f) =& - \frac{\beta_f\cj_q^2}{2} \int_0^{1} dx \, e^{\tilde{g}(x)} \\
&- \frac{q \beta_f \cj_2^2}{4} - \frac{\beta_f\cj_2^2}{2}\int_0^{1} dx \, \tilde{g}(x),
\end{aligned}
\label{eq:computational_energy_thermal}
\end{equation}
We find the a kinetic contribution scaling as $q$ comes from the large-$q$ ansatz for the Green's function as explicitly shown in Appendix \ref{app. Evaluating energy in the Keldysh contour}. 
We note that there is no such term in the non-equilibrium kinetic energy density in Eq. \eqref{interacting and kinetic energy densities}. The reason is that the equilibrium energy density is evaluated on the imaginary contour in the Keldysh plane (see Fig. \ref{fig:keldysh_contour}), which we ignore in our non-equilibrium calculation due to the Bogoliubov's principle of weakening correlations. Had we considered the imaginary contour, there would be a leading order in $q$ contribution in the non-equilibrium kinetic energy density as well (see Appendix \ref{app. Evaluating energy in the Keldysh contour}). However this comes with a caveat that is explained in detail in Appendix \ref{app. Evaluating energy in the Keldysh contour} below Eq. \eqref{energy intemediate step 1 in appendix} (also see Fig. \ref{fig:keldysh_starting_zero}).
We note that the kinetic energy contribution that scales as $q$ does not contain any dynamical information and acts like a constant that can be scaled out. Accordingly, in line with Ref. \cite{meirinhos2022} and also considered in this work, we only deal with the interaction energy density given by
\begin{equation}\label{eq:iteraction_energy_thermal}
\cV_{\text{EQ}} (\beta_f) = - \frac{\beta_f\cj_q^2}{2} \int_0^{1} dx \, e^{\tilde{g}(x)}.
\end{equation}
Therefore we fit this $\cV_{\text{EQ}} $ to the non-equilibrium interaction energy density $\Vv(t_1)$ in Eq. \eqref{interacting and kinetic energy densities} that has been evolved for a long time for it to relax to a constant value. This fitting gives us the estimate for the final temperature $\beta_f$ of the quenched system if it thermalizes (implying that $\Vv(t_1)$ relaxes to a constant value).  

\subsection{Estimation of the thermalization rate}
\label{subsec Estimation of the thermalizing rate}

We use the kinetic energy density $\cK(t_1)$ in Eq. \eqref{interacting and kinetic energy densities} to estimate the thermalization rate $\gamma$ by fitting the exponential ansatz \cite{bhattacharya2019}
\begin{align}\label{eq:expo_ansatz}
    A(t_1) = B_{\cK} + C_{\cK}e^{-\gamma t_1},
\end{align}
with the protocol \textit{curve$\_$fit} provided by the Python package \textit{SciPy} \cite{SciPy}. The goodness of fit can be seen in Fig. \ref{fig:fit_accuracy} in Appendix \ref{app. additional figures} for several different mixed quenches. These indicate the validity of the ansatz. 

Having developed the predictor-corrector in the two-time plan, the energy error, as well as the means to calculate the inverse temperature $\beta_f$ and the thermalization rate $\gamma$ from the interaction and the kinetic energy densities respectively, we can now apply these concepts to the mixed quench. 


The results in the following sections for the case of mixed quench are all obtained with a discretization $\Delta t =0.05$, initial time in the distant past $t_0 = -100$, the maximum time of evolution to ensure relaxation of the system $t_{\text{max}}=1700$, the initial low temperature $\beta_0=18.9$ and interaction strength $\cj_q=1.0$, unless explicitly stated otherwise.

\begin{figure*}[htb]
    \centering 
 \begin{subfigure}{0.32\textwidth}\hfill
 \caption{}\label{fig:Re_two_time_Jtwo0.02_sig0.15}
  \includegraphics[width=\linewidth]{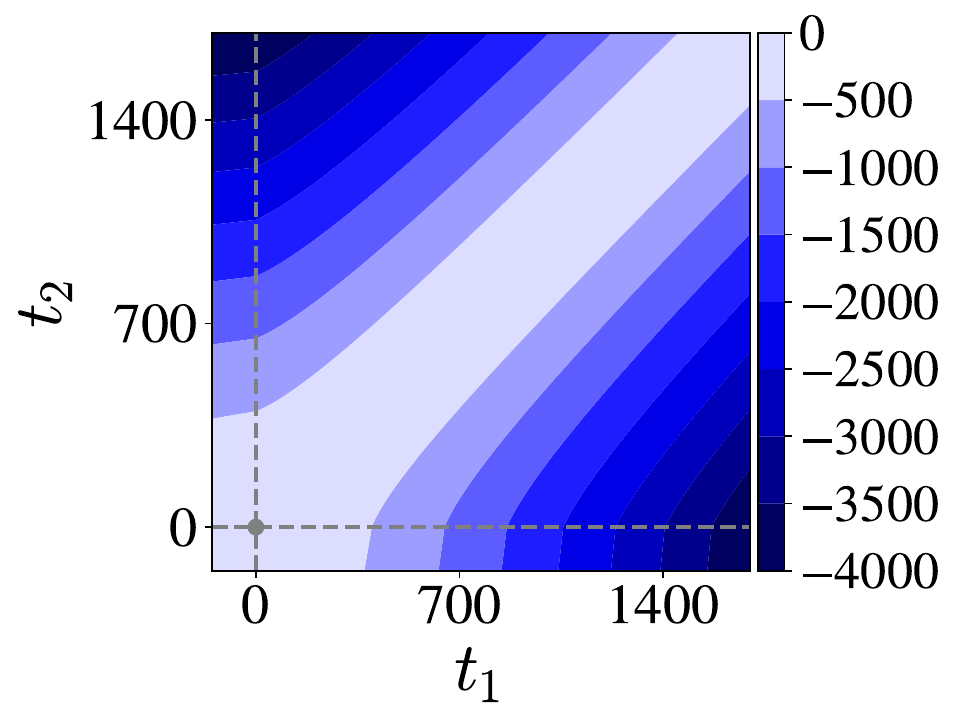}
\end{subfigure} %
\begin{subfigure}{0.31\textwidth}\hfill
  \caption{}\label{fig:Im_two_time_Jtwo0.02_sig0.15}
  \includegraphics[width=\linewidth]{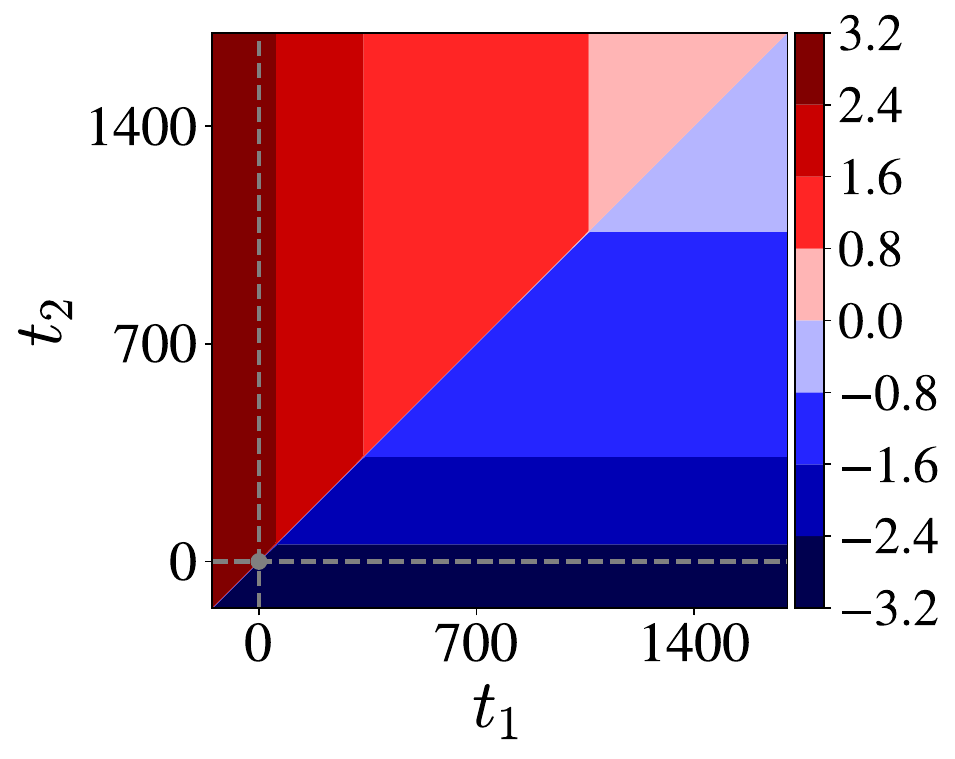}
\end{subfigure}%
\begin{subfigure}{0.36\textwidth}\hfill
  \caption{}\label{fig:E_mixed_Jtwo0.02_sig0.15}
  \includegraphics[width=\linewidth]{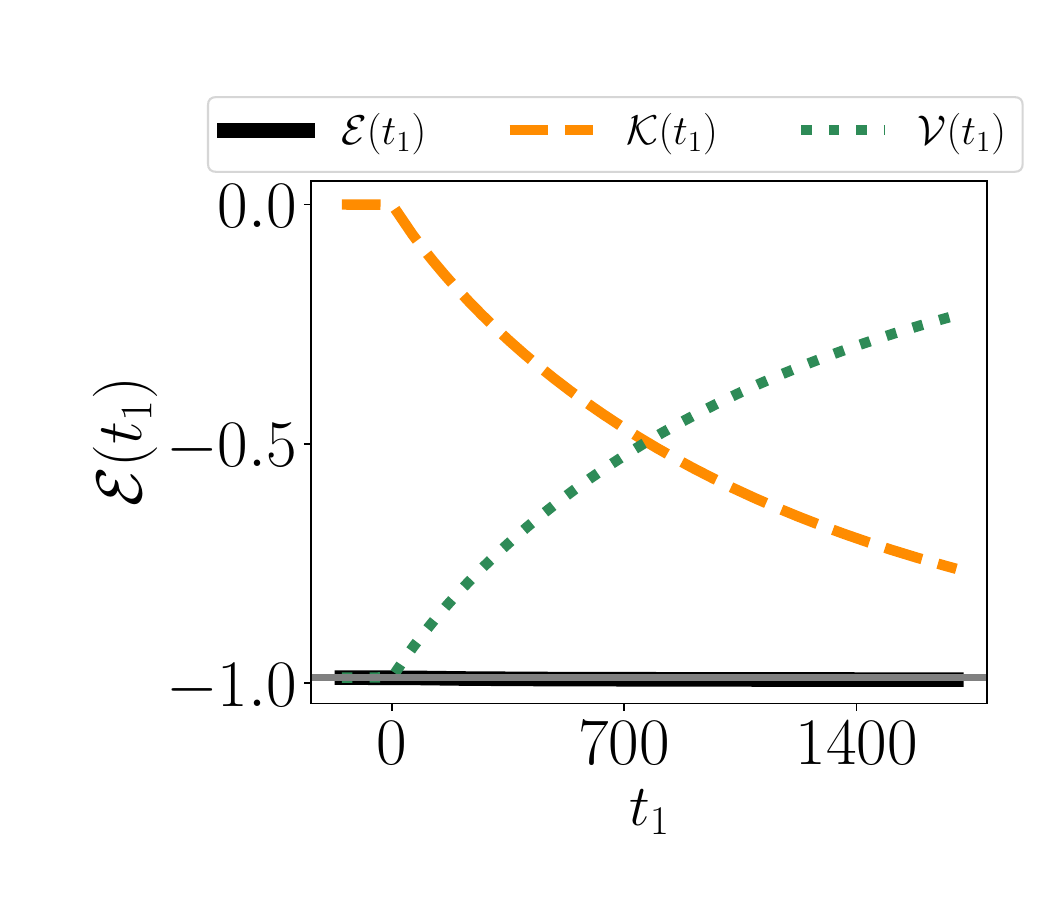}
\end{subfigure}
\caption{Visualization of the numerical solution for a mixed SYK quench with $\cj_2=0.02$ and $\beta_0 = 18.9$ as the initial temperature. Figures (a)-(b) depict the real and imaginary parts of $g(t_1,t_2)$. Figure (c) shows the time dependent energies obtained from the numerical solution. The dashed and dotted lines corresponds to kinetic and interaction contributions respectively and the solid black line corresponds to the total energy. Finally, the solid grey line indicates the value of the total energy in the initial equilibrium state which as clear from the graph overlaps with the solid black line. Here, the relative error between the initial and final total energies is $|(E_0 -E_f)/E_0|=0.0042$.}
\label{fig:visualization_jtwo0.02}
\end{figure*}

\section{Mixed SYK Quench Dynamics}\label{sec. mixed SYK}

An initial visual to the mixed SYK can be useful in order to get an intuitive understanding of the properties of its Green's function. The purpose of this section is to provide the visualization which is quantified in the next section. A plot of the in the two time plane already reveals important properties. Figs. \ref{fig:Re_two_time_Jtwo0.02_sig0.15} and \ref{fig:Im_two_time_Jtwo0.02_sig0.15} depict the real and imaginary parts of $g(t_1,t_2)$ for a quench with strength $\cj_2=0.02$. For starters, the diagonals in $\rm{Re}[{g(t_1,t_2)}]$ seem to become constant fairly quickly, while in $\text{Im}[{g(t_1,t_2)}]$ they continue to change even at longer times. The presence of long-time dynamics can also be inferred from the time dependent energies in Fig. \ref{fig:E_mixed_Jtwo0.02_sig0.15}. Here, $\cK(t_1)$ and $\cV(t_2)$ have yet to saturate to a given value, so further dynamics would be expected. From this figure, one can also conclude that the solution is accurate, as the total energy remains constant well below $1\%$ at each timestep, as is the relative error between its initial and final values.

\begin{figure*}[htb]
    \centering 
 \begin{subfigure}{0.32\textwidth}\hfill
 \caption{}
  \includegraphics[width=\linewidth]{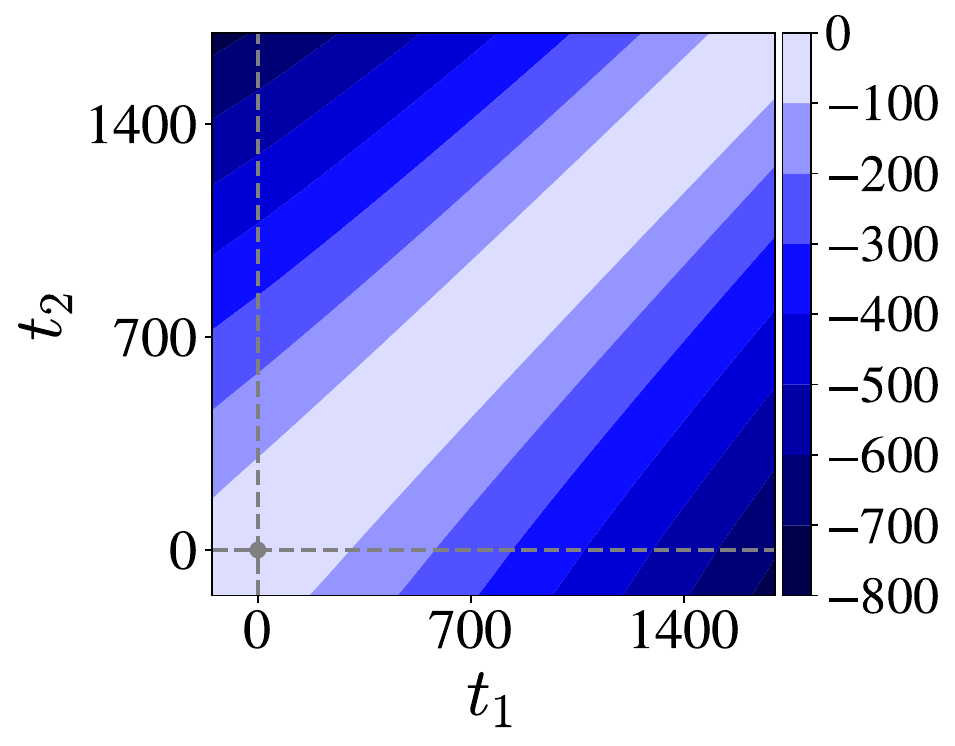}
\end{subfigure} 
\begin{subfigure}{0.31\textwidth}\hfill
  \caption{}
  \includegraphics[width=\linewidth]{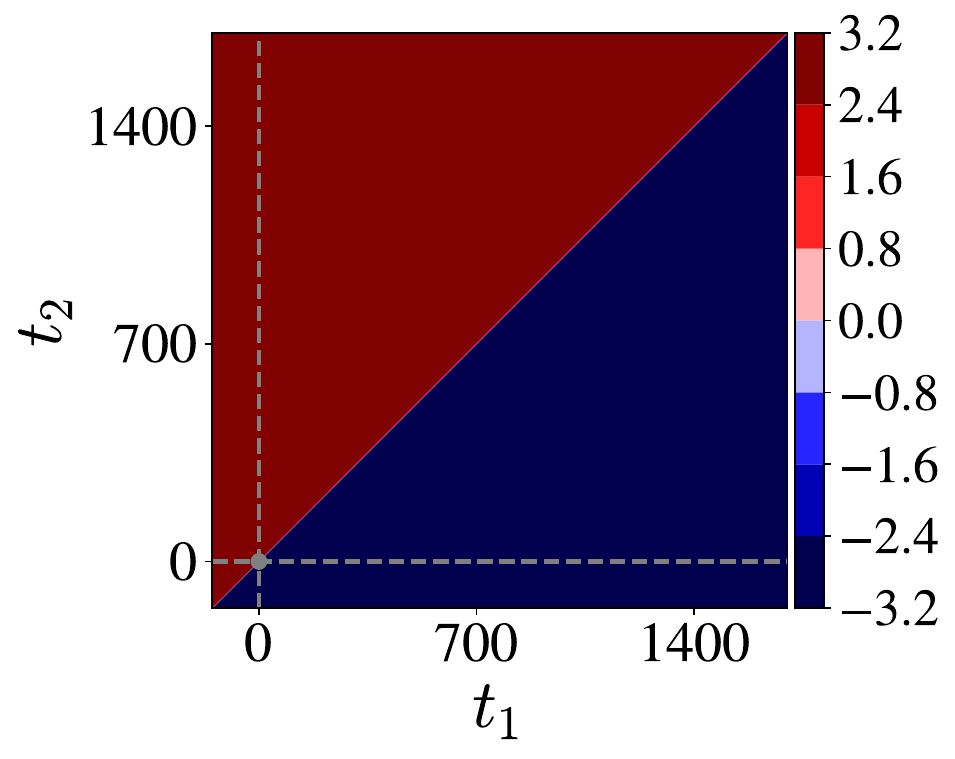}
\end{subfigure}
\begin{subfigure}{0.36\textwidth}\hfill
  \caption{}
  \includegraphics[width=\linewidth]{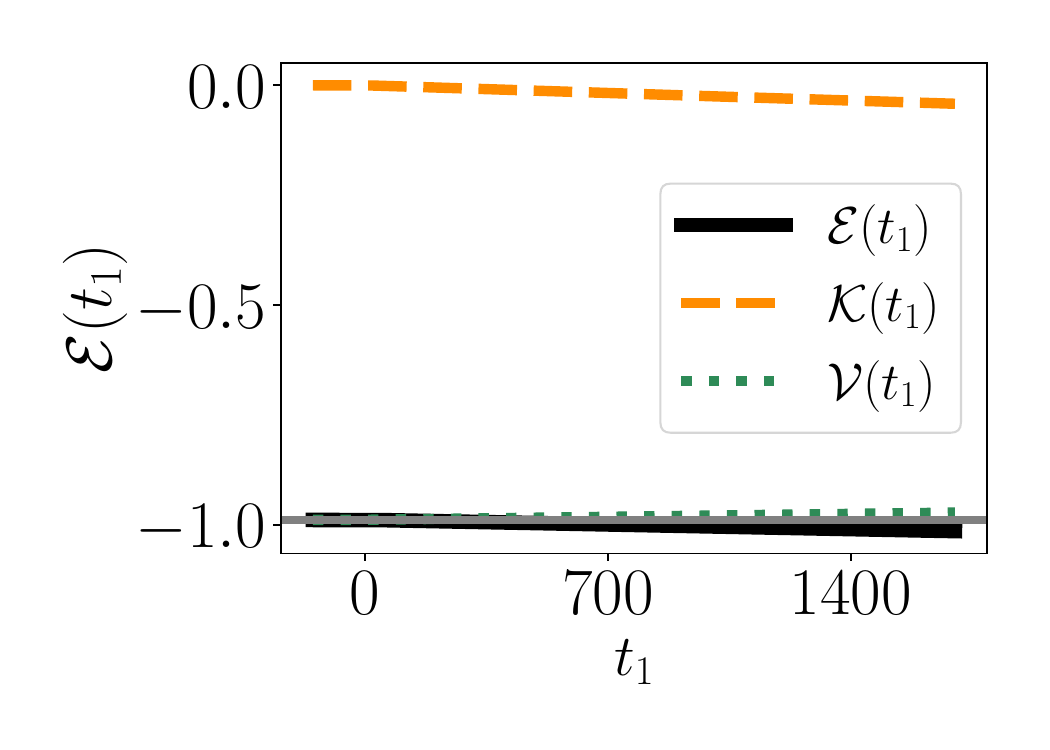}
\end{subfigure}
\smallskip
\centering 
 \begin{subfigure}{0.32\textwidth}\hfill
   \caption{}
  \includegraphics[width=\linewidth]{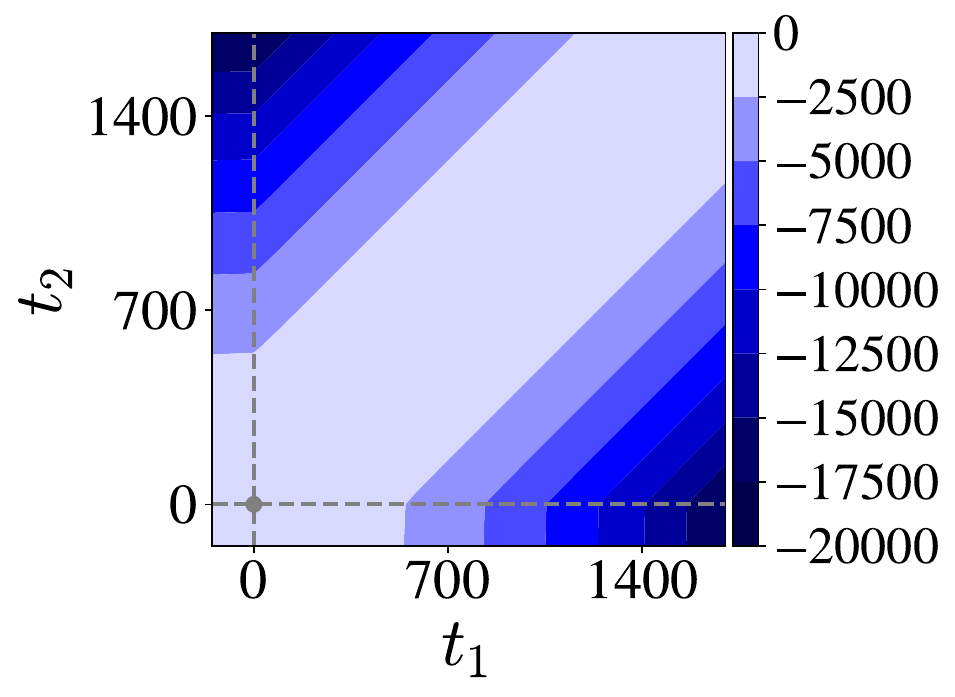}
\end{subfigure} 
\begin{subfigure}{0.31\textwidth}\hfill
  \caption{}
  \includegraphics[width=\linewidth]{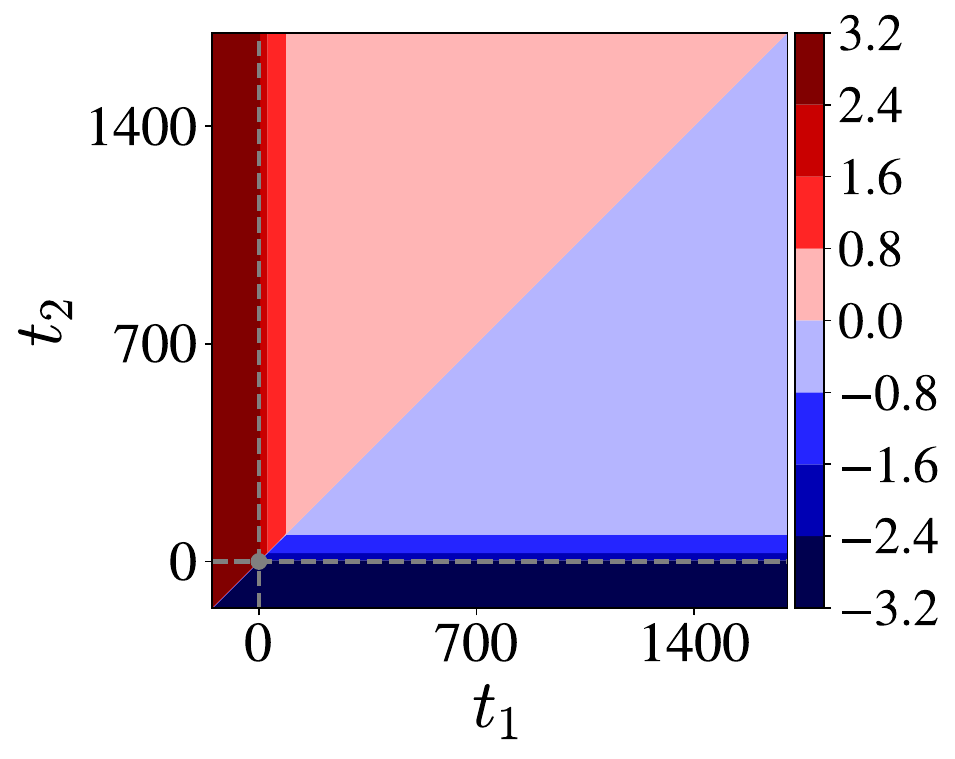}
\end{subfigure}
\begin{subfigure}{0.36\textwidth}\hfill
  \caption{}
  \includegraphics[width=\linewidth]{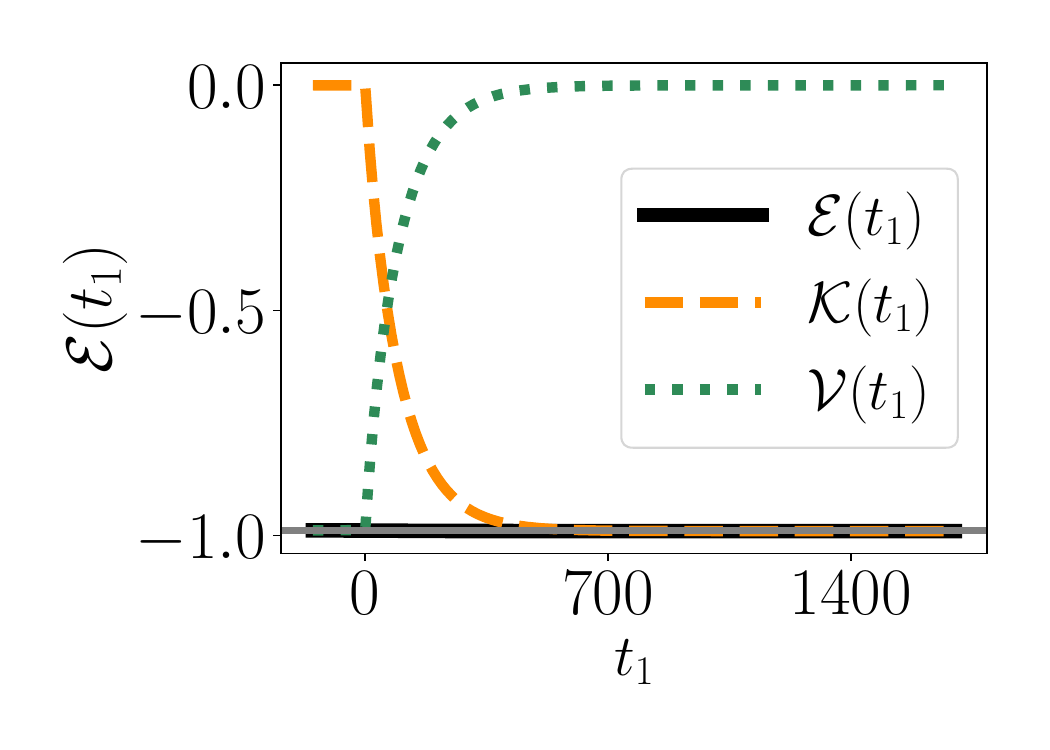}
\end{subfigure}
\caption{Additional visualizations of the numerical $g(t_1,t_2)$ of the mixed quench, for the initial temperature $\beta_0 = 18.9$ and two different $\mathcal{J}_2$. Plots (a)-(c) depict the real and imaginary parts of $g(t_1,t_2)$ and the corresponding non-equilibrium energies for $\cj_2=0.003$ with a relative error $|(E_0 -E_f)/E_0|=0.026$. Plots (d)-(f) show the same quantities for $\cj_2=0.07$, with relative error $|(E_0 -E_f)/E_0|=0.0016$. The color cording and line shapes follow the same convention as in figure \ref{fig:visualization_jtwo0.02}.}\label{fig:more_visualizations}
\end{figure*}

This same analysis is repeated for two other quenches in Fig. \ref{fig:more_visualizations} where the caption contains all the required details. In a weaker quench ($\cj_2 = 0.003$), it is clear from both the two-time plane and the energy that there are barely any dynamics for such small values. The relative error between the final and initial values of the energy is however somewhat larger in this case. On the other hand, the stronger quench ($\cj_2 = 0.07$) exhibits far quicker dynamics as one expects. Moreover, the fact that the diagonals become constant suggests that the the Green's function becomes dependent only upon time difference, and therefore, reaches a stationary state as is also corroborated by the saturation of the energy components. In contrast to the other cases, here the relative error in the energy is far smaller. 

From these three examples we can make three observations about the mixed quench: first, the Green's function reaches stationarity since, given enough time, it becomes dependent only upon time differences and its energy components become constant. Second, the time it takes the system to become stationary is strongly dependent on $\cj_2$ where the strong quenches result in faster dynamics than the weak quenches, as would be expected from the limiting cases. Finally, the error between the initial and the final total energies is larger for weaker quenches. As shown in Fig. \ref{fig:energy_errors} in appendix \ref{app. additional figures} (and as will be made clear in section \ref{sec. thermalization}), the error is dependent on the temperature of the system and is also present, albeit in smaller magnitude, in the limiting cases as depicted in Fig. \ref{fig:benchmark_results} in Section \ref{sec. benchmarking}.

We now turn to quantifying these observations and analyzing how the system responds to different quench strengths.

\section{Thermalization in the Thermodynamic Limit}\label{sec. thermalization}

This section provides an in depth analysis of the approach to equilibrium of the Green's function of the mixed quench. First we show that the Green's function indeed becomes stationary and that this state corresponds to thermodynamic equilibrium. Afterwards, we characterize the final equilibrium state by calculating the final temperature and thermalization rate, as a function of the initial temperature $\beta_0$ and the quench parameter $\Jj_2$.

\subsection{Stationary limit}

 In order to better determine the existence of stationary behavior of the Green's function, it is convenient to shift to Wigner coordinates \cite{larzul2022} defined as 
\begin{align}\label{eq:wigner_coords}
    \Tau \equiv \frac{1}{2}(t_1+t_2), & \quad t \equiv t_1-t_2,
\end{align}
which is equivalent to a $\pi/4$ rotation of the two-time plane (as also shown in Fig. \ref{fig:causal_stepping}). Here $\Tau$ is generally referred to as the \textit{average time}, moving along the diagonals, and $t$ as \textit{relative time}, leading away from them. Expressed in this form, the Green's function fulfills 
\begin{equation}\label{eq:equilibrium_condition}
    g(\Tau, t) \xrightarrow{ \Tau \to \infty } g_{\beta_f}(t)
\end{equation}
if it becomes stationary where $g_{\beta_f}(t)$ satisfies the (stationary) ordinary differential equation as given in Eq. \eqref{equilibrium ODE in real time}. The subscript $\beta_f$ denotes that we are referring to the final achieved stationary solution (if it ever gets achieved) corresponding to the final inverse temperature $\beta_f$. It is presupposed that $\beta_f$ has already been found by fitting the thermal ($\cV_{\text{EQ}} (\beta_f)$) and the non-equilibrium ($\cV_{\text{NEQ}} (t_1) $) energies as explained in Section \ref{subsec Estimation of the final temperature}. 

\begin{figure}[!ht]
\centering
    \begin{subfigure}{0.49\linewidth}
        \caption{}
        \includegraphics[width=\linewidth]{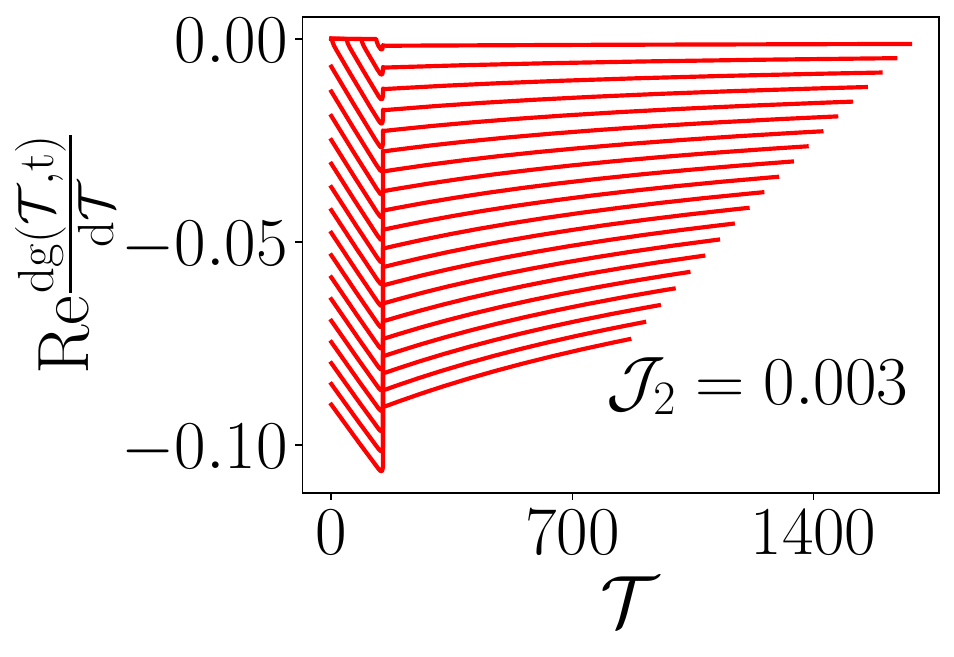}
    \end{subfigure}
\hfil
    \begin{subfigure}{0.49\linewidth}
        \caption{}
        \includegraphics[width=\linewidth]{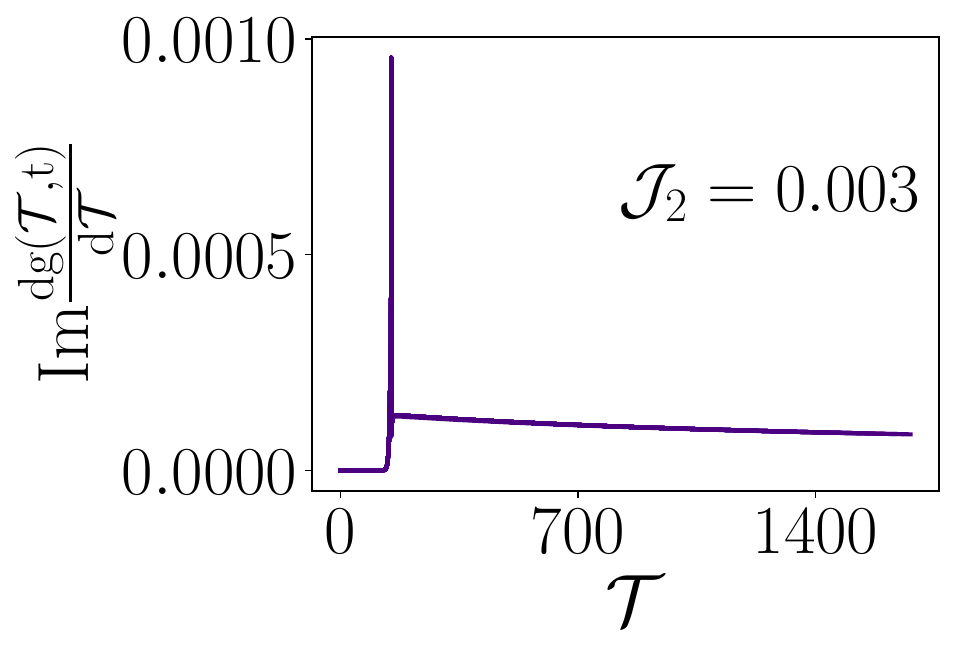}
    \end{subfigure}

    \begin{subfigure}{0.49\linewidth}
        \caption{}
        \includegraphics[width=\linewidth]{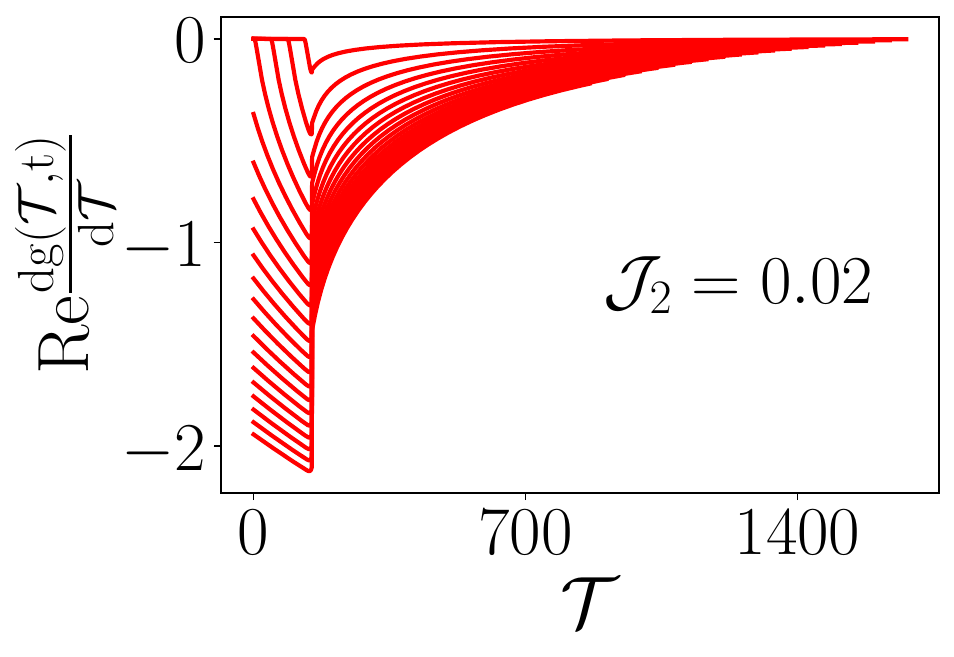}
    \end{subfigure}
\hfil
    \begin{subfigure}{0.49\linewidth}
        \caption{}
        \includegraphics[width=\linewidth]{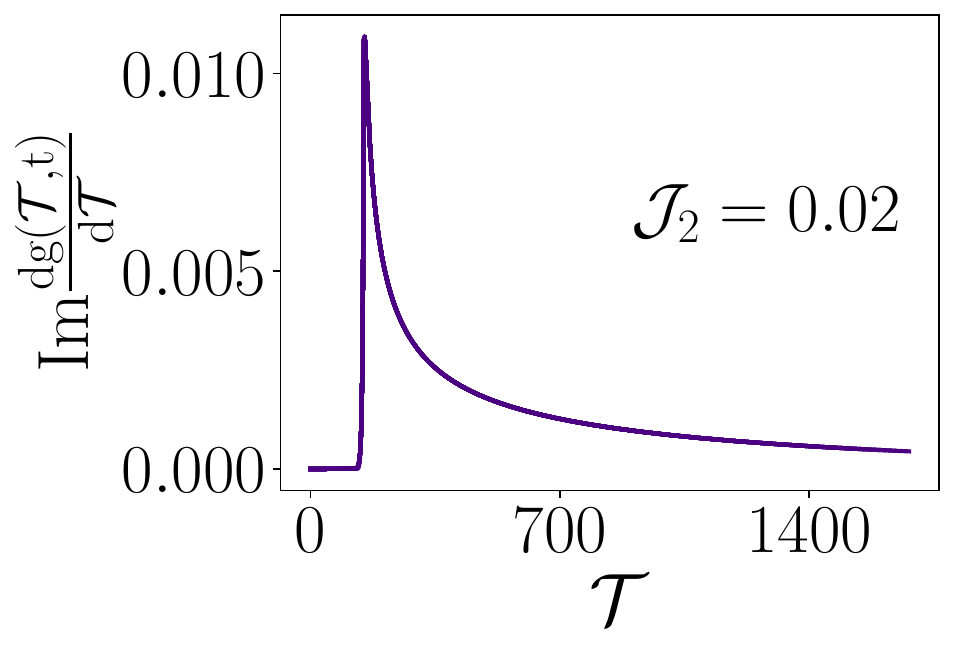}
    \end{subfigure}

    \begin{subfigure}{0.49\linewidth}
        \caption{}
        \includegraphics[width=\linewidth]{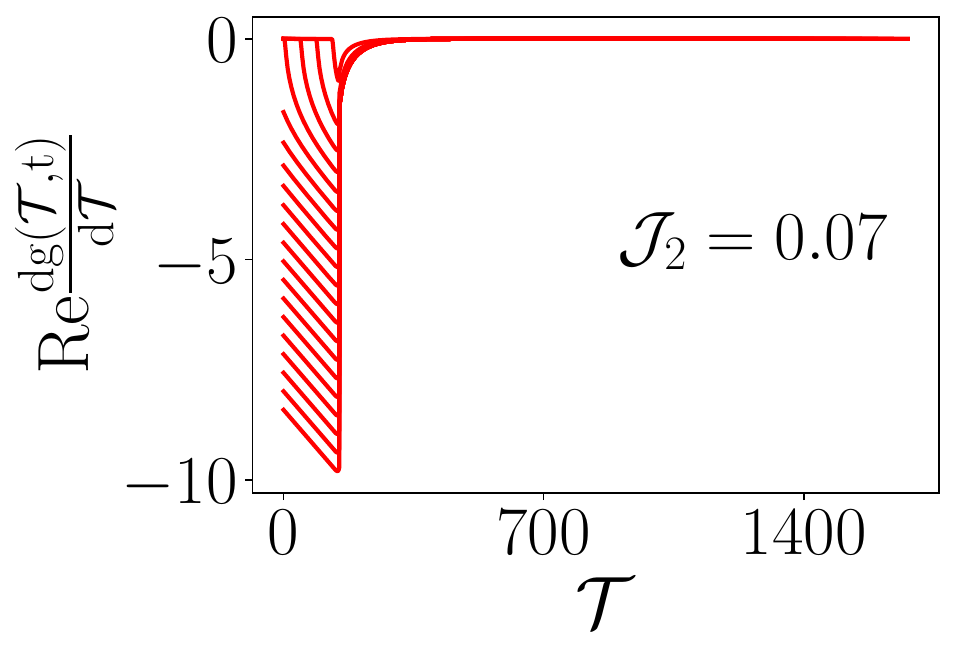}
    \end{subfigure}
\hfil
    \begin{subfigure}{0.49\linewidth}
        \caption{}
        \includegraphics[width=\linewidth]{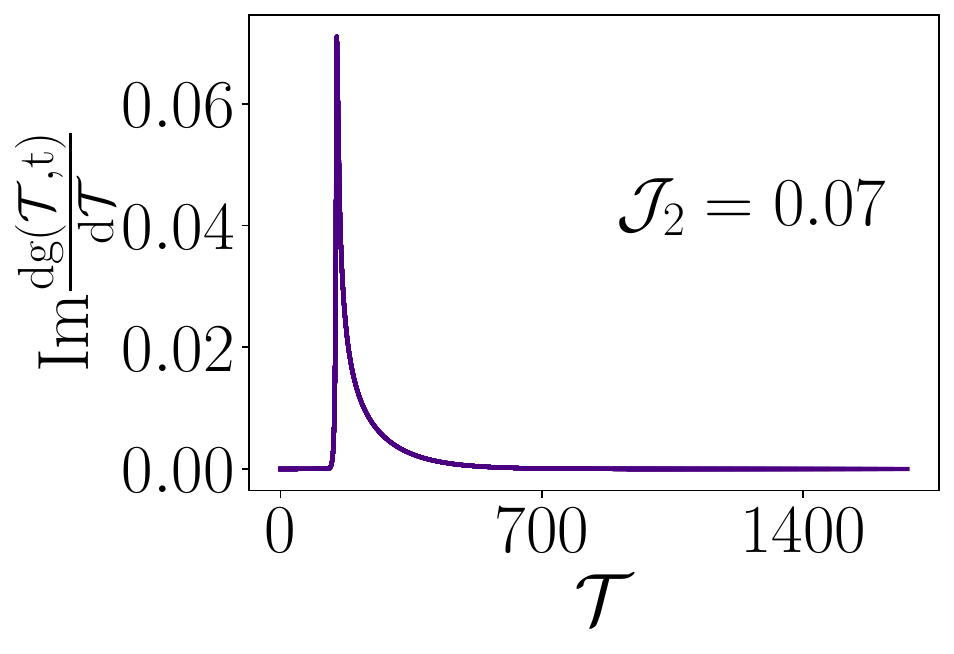}
    \end{subfigure}
\caption{Derivatives along the diagonals $\Tau$ of the real and imaginary components of $g(\Tau,t)$ (namely $ \frac{d g(\Tau, t)}{d\Tau} \bigr |_{t}$), corresponding to the cases shown in Figs. \ref{fig:visualization_jtwo0.02} and \ref{fig:more_visualizations} $\beta_0 = 18.9$. The different curves with same color in each graph correspond to the derivatives at different constant $t$.}
    \label{fig:diagonals}
    \end{figure}

We have to check Eq. \eqref{eq:equilibrium_condition} which we do by calculating the derivatives of $g(\Tau, t)$ along $\Tau$. If $\frac{d g(\Tau, t)}{d\Tau} \bigr |_{t} = 0$, then we confirm that the stationarity has been reached and the dependence of $\Tau$ has been eliminated in Eq. \eqref{eq:equilibrium_condition}. As presented in Fig. \ref{fig:diagonals} for the three examples discussed previously in Figs. \ref{fig:visualization_jtwo0.02} and \ref{fig:more_visualizations}, the diagonals along $\Tau$ do indeed approach zero thereby confirming the $\Tau$ independence of $g(\Tau, t)$ in Eq. \eqref{eq:equilibrium_condition}. Therefore, the Green's function indeed becomes stationary. In addition, we also conclude from Fig. \ref{fig:diagonals} that the real part of the derivative reaches stationary state faster than the imaginary part, just like the real part of the Green's function itself as mentioned above. Finally, we also find that the time it takes to reach stationary state is dependent on the strength of the quench parameter $\Jj_2$.

 Furthermore, we can verify whether or not the stationary state $g_{\beta_f}(t)$ in Eq. \eqref{eq:equilibrium_condition} corresponds to thermal equilibrium. The way this is done is that we wait for a long time $\Tau$ such that the relaxation dynamics have happened and then we substitute the Green's function $g(\Tau \to \text{ large}, t)$ into the equlibrium ordinary differential equation for $g_{\beta_f}(t)$ given in Eq. \eqref{equilibrium ODE in real time}. So we verify whether or not the left-hand side and the right-hand side of Eq. \eqref{equilibrium ODE in real time} is satisfied for $g(\Tau \to \text{ large}, t)$. 
 \begin{equation}\label{eq:LHSvsRHS}
    \frac{d^2 g(\Tau \to \text{ large}, t)}{dt^2} \stackrel{?}{=} - 2\cj_q^2 e^{g(\Tau \to \text{ large}, t)} -2 \cj_2^2,
\end{equation}
 If it does satisfy then we conclude that the stationary state given by $g(\Tau \to \text{ large}, t)$ indeed has reached the thermal equilibrium described by $g_{\beta_f}(t)$. 
 
\begin{figure}[!ht]
\centering
    \begin{subfigure}{0.40\linewidth}
        \caption{}
        \includegraphics[width=\linewidth]{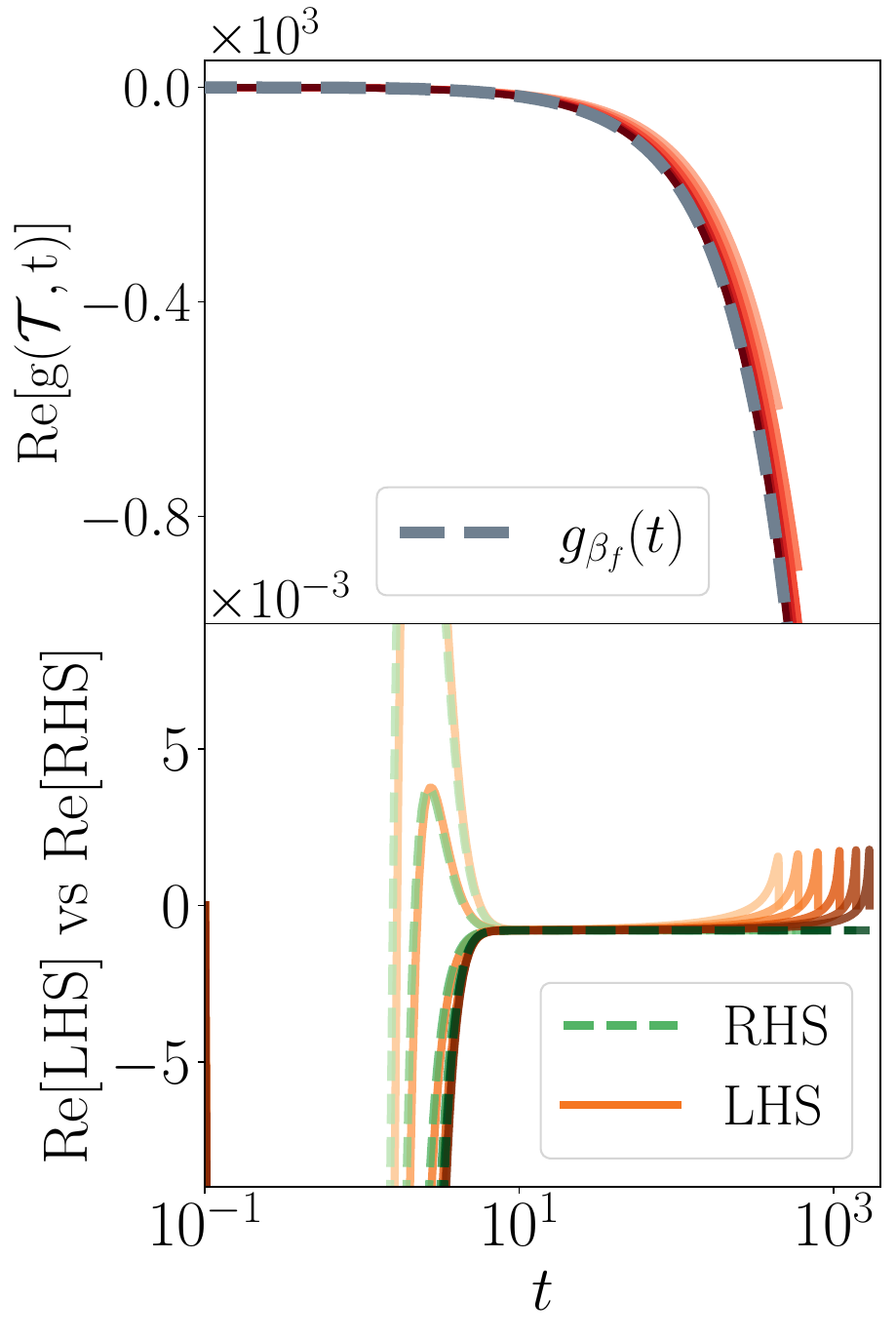}
    \end{subfigure}
\hfil
    \begin{subfigure}{0.445\linewidth}
        \caption{}
        \includegraphics[width=\linewidth]{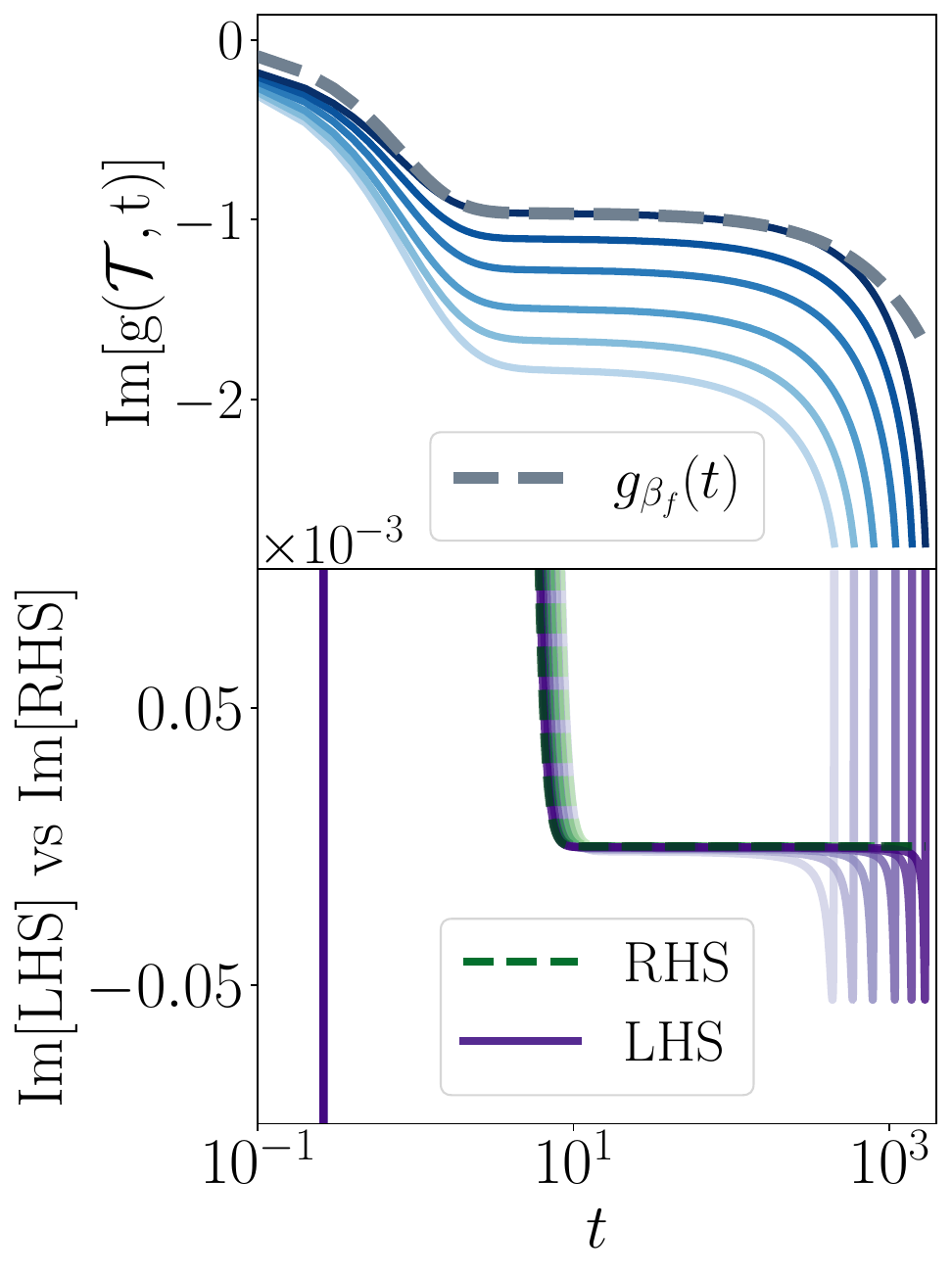}
    \end{subfigure}
\caption{Stationary limit of the Kadanoff-Baym solution $g(\Tau,t)$ compared to the corresponding thermal system with $\cj_2 = 0.02$ and $\beta_0 = 18.90$. The curves are time slices in Wigner coordinates corresponding to $\Tau\in\{ 225, 300, 400, 550, 700, 850 \}$. The figures in (a) depict the real components. The upper plot in (a) directly compares $g(\Tau, t)$ (solid lines each corresponding to slices at constant $\Tau$ which go from lighter to darker colors as $\Tau$ increases) to the thermal $g_{\beta_f}(t)$ (dashed line). The bottom plot in (a) depicts the left-hand side (solid line) and the right-hand side (dashed line) of Eq. \eqref{equilibrium ODE in real time} calculated using the Kadanoff-Baym solution. Again, just like the upper plot, the solid curves here correspond to different slices of $\Tau$ with darker colors corresponding to longer times. The imaginary components are shown in (b) following the same convention as (a). We find a remarkable match except at very short $t$ and at the very end, where the former is due to numerical error while the latter is a consequence of the finite simulation time. Both of these situations are explained in the text.}
    \label{fig:sationary_limit}
    \end{figure}

For this machinery to work, we need to solve the equilibrium ordinary differential equation Eq. \eqref{equilibrium ODE in real time}. We already know the initial condition $g_{\beta_f}(0)=0$ but we also need to knwo the first derivative of $g_{\beta_f}$ in order to uniquely solve the differential equation. In order to avoid guess-work, the initial condition of the derivative of $g(t)$ is calculated via
\begin{equation}\label{eq:derivative_init}
   \frac{d g_{\beta_f}(t)}{dt}\bigr|_{t=0}  = \frac{d g(\Tau_{\text{max}},t)}{dt}\bigr|_{t=0},
\end{equation}
where $\Tau_{\text{max}}= t_{\text{max}}/2$ (see the paragraph before Section \ref{sec. mixed SYK} where all parameter values are specified), which is then used to solve Eq. \eqref{equilibrium ODE in real time} numerically via a Runge-Kutta-4 method. Fig. \ref{fig:sationary_limit} shows that $g(\Tau,t)$ converges towards the thermal $g_{\beta_f}(t)$ as $\Tau$ increases. Also shown in the same figure, the equality in Eq. \eqref{equilibrium ODE in real time} holds remarkably well for $g(\Tau,t)$ in the stationary regime, except at very short $t$ and at the very end. The former is due to a numerical error induced by the fourth order approximation of the boundary terms in the derivative. The latter is a consequence of the finite simulation time: as $t$ increases (away from the diagonal in Fig. \ref{fig:causal_stepping}), it eventually reaches the regions of the two-time plane that are out-of-equilibrium and the equality ceases to hold. In principle, if $\Tau \to +\infty$, then we should never be out-of-equilibrium for however large value of $t$.

\onecolumngrid

\twocolumngrid

To summarize, we solved the Kadanoff-Baym equations for $g(\Tau, t)$ starting from the initial thermal state of large-$q$ SYK model where we evolved until the relaxation of energy is completed as denoted in Figs. \ref{fig:visualization_jtwo0.02} and \ref{fig:more_visualizations}. Then we used the equilibrium condition in Eq. \eqref{eq:equilibrium_condition} to argue that the Kadanoff-Baym solution has reached the stationary state (independent of $\Tau$). We did so by analyzing the derivatives of $g(\Tau, t)$ with respect to $\Tau$ at various fixed $t$ in Fig. \ref{fig:diagonals} where we saw the derivatives vanish in line with the expectation of stationary state. Further we established in Fig. \ref{fig:sationary_limit} that the stationary state reached indeed satisfies the equilibrium ordinary differential equation \eqref{equilibrium ODE in real time} for $g_{\beta_f}(t)$ where $\beta_f$ is the final inverse temperature of the equilibrium state which is estimated using the methodology explained in Section \ref{subsec Estimation of the final temperature}. Eq. \eqref{eq:derivative_init} was used to provide for the initial condition along with $g_{\beta_f}(t=0)=0$ to solve for Eq. \eqref{equilibrium ODE in real time} for $g_{\beta_f}(t)$ which is also compared to the Kadanoff-Baym solution $g(\Tau \to \text{ large}, t)$ in Fig. \ref{fig:sationary_limit}. This further confirms that the Kadanoff-Baym stationary solution is also the equilibrium solution with final inverse temperature $\beta_f$. Finding the unique solution for the Kadanoff-Baym that flows towards the equilibrium equation in Eq. \eqref{equilibrium ODE in real time} is in accordance with the uniqueness theorem, in particular the Picard-Lindel\"of theorem for initial value problem of ordinary differential equations \cite{Nandakumaran2017May}. Further examples for other quenches are shown in Figs. \ref{fig:sationary_limit_Jtwo0.003} and \ref{fig:sationary_limit_Jtwo0.07} in Appendix \ref{app. additional figures} that lead to the the same conclusion.

We finally proceed to calculating the thermalization rate corresponding to the non-equilibrium dynamics induced in the system due to the quenching of the random hopping term in the Hamiltonian at $t=0$. The following section characterizes the final equilibrium state in terms of its final temperature and the thermalization rate.

 \begin{figure*}[ht]
     \centering
     \begin{subfigure}{0.5\linewidth}
         \centering
         \caption{}
         \includegraphics[width=\linewidth]{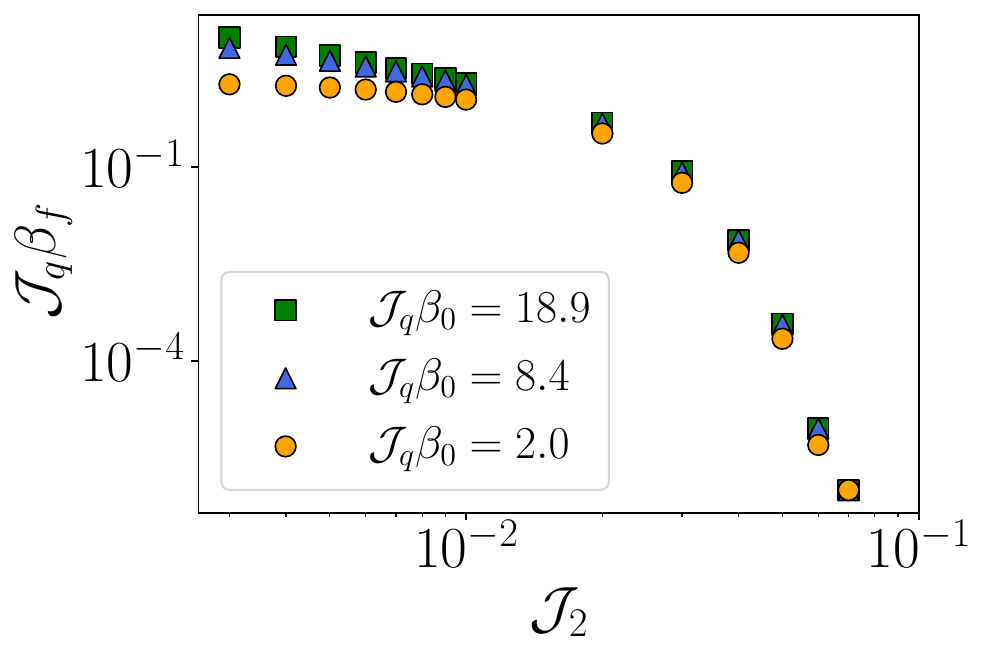}
         \label{fig:final_betas_J2}
     \end{subfigure}
     \begin{subfigure}{0.47\linewidth}
         \centering
         \caption{}
         \includegraphics[width=\linewidth]{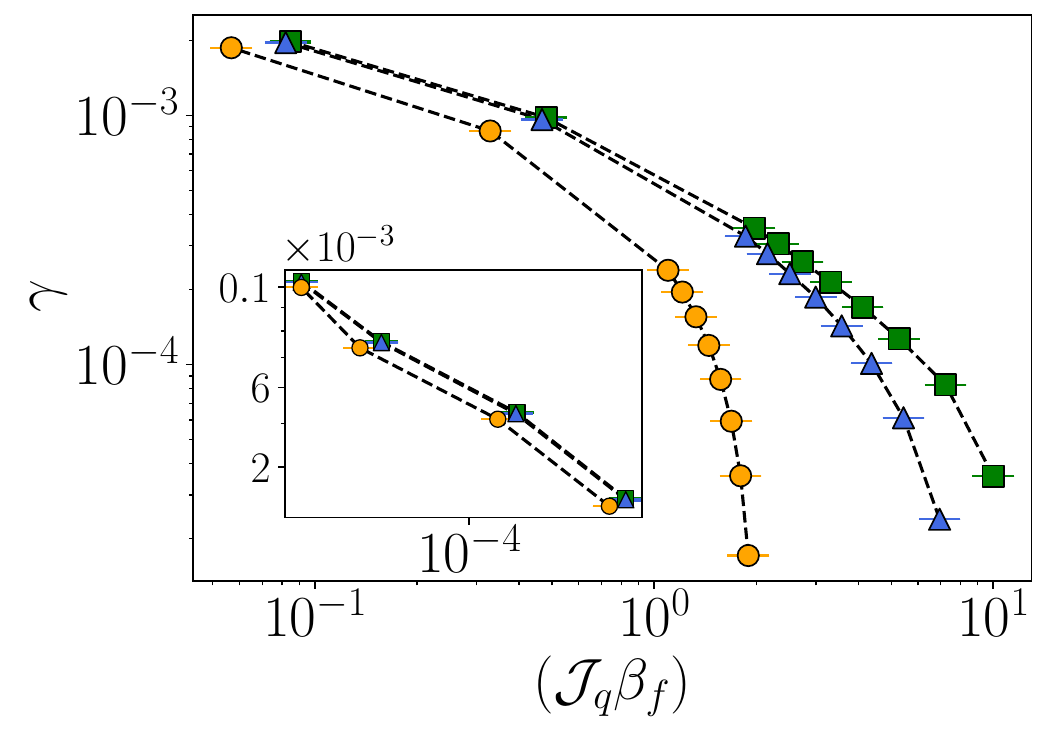}
         \label{fig:final_gammas_beta}
     \end{subfigure}
        \caption{Thermalization and approach to the final state of the mixed quench as a function of $\cj_2$ for different initial temperatures. The data for $\beta_0 = 2.0$ was obtained with $\Delta t = 0.06$ as a discretization while others have been obtained for the discretization corresponding to $\Delta t = 0.05$ (see the paragraph before Section \ref{sec. mixed SYK}). (a) Final inverse temperature $\cj_q \beta_f$ as a function of the quench strength $\cj_2$. (b) Thermalization rate $\gamma$ as a function of $\cj_q \beta_f$, the error bars correspond to the mean squared error between the fitted curve and $\cK(t)$ (see Section \ref{subsec Estimation of the thermalizing rate}). Inset (also in $\log-\log$ scale) shows the thermalization rates corresponding to higher final temperatures (stronger quenches) where they all coincide as expected and the deviation start to happen at lower final temperatures (weaker quenches).}
     \label{fig:beta_gamma}
\end{figure*}

\subsection{Thermalization rate and final temperature}

Having established that the predictor-corrector algorithm is capable of solving the mixed quench accurately and that the Kadanoff-Baym solution $g(t_1,t_2)$ reaches thermal equilibrium in the post-quench state, it is logical to calculate its final temperature $\cj_q\beta_f$ and the $\gamma$ thermalization rates. We have already explained the methodology for estimating the final temperature and the thermalization rate in Sections \ref{subsec Estimation of the final temperature} and \ref{subsec Estimation of the thermalizing rate}, respectively.

The data presented in Fig. \ref{fig:beta_gamma} corresponds to the initial temperatures ranging from close to zero ($\beta_0 = 18.9$ and $\beta_0 = 8.4$, so as to remove thermal noise effects as much as possible) up to relatively high values ($\beta_0 = 2.0$). 
Furthermore, at $\beta_0=18.9$, the energy is calculated to be $\cE_0=-0.989$ (obtained from Eqs. \eqref{eq:initial temperature} and \eqref{eq:E0}), implying the initial state is close to the ground state. 

Regarding the quench strength $\cj_2\in \{ 0.003,0.07 \}$, this range was chosen since negligibly weaker quenches do not produce significant dynamics and stronger ones already reach the infinite temperature state. The latter can be inferred from Fig. \ref{fig:final_betas_J2} where we can observe that the final inverse temperatures $\beta_f$'s corresponding to stronger quenches are of order $10^{-4}$ and approach identical values regardless of initial temperature. This type of behavior is reminiscent of the kinetic-quench as seen in Section \ref{subsec. limiting cases} where the final state is dominated by the kinetic term and we reached the infinite temperature state post-quench. It is however remarkable here that this happens at such small kinetic couplings as well. This is a consequence of the scaling in Eq. \eqref{variances} introduced to induce competition between both terms in the Hamiltonian in Eq. \eqref{eq:mixed_syk_neq_hamiltonian}. From this figure it is clear that weaker quenches have a somewhat larger error (as previously noted) because they correspond to systems with smaller final temperatures.

Continuing with the thermalization rates $\gamma$, shown in Fig. \ref{fig:final_gammas_beta}, it is evident that there are a wide range of timescales in the system. The $\gamma$'s corresponding to large $\cj_q\beta_f$ (weak quenches) are two to three orders of magnitude smaller than the lower $\cj_q\beta_f$ (strong quenches) counterpart. Moreover, there is a marked change in behavior at around $\cj_q\beta_f \approx 10^{-1}$, at which point the thermalization rates quickly drop to far smaller values, leading to longer thermalization timescales. This suggests that at weaker quenches (lower final temperatures), the system's relaxation to equilibrium is slower as expected. This effect appears to be less dramatic for lower initial temperatures (higher $\beta_0$) where the slope of the $\gamma$'s for larger $\cj_q\beta_f$'s does not change as rapidly, but still the difference remains easily appreciable. Therefore we observe rich thermalization dynamics in the thermodynamic limit for the non-equilibrium dynamics (induced by quenching as in Eq. \eqref{quench couplings}) in the closed SYK system as considered in this work where we characterized the final state in terms of the final temperature of the thermal state at the end of the relaxation dynamics as well as the thermalization rate. 

\section{Conclusion and Outlook}
\label{sec. conclusion}

We studied the relaxation dynamics of a closed quantum system in non-equilibrium in the thermodynamic limit, consisting of large-$q$ Majorana SYK model at equilibrium, quenched with a random hopping term. Unlike the large-$q$ SYK model that shows instantaneous thermalization with respect to the Green's functions both for Majorana and complex cases \cite{Eberlein2017Nov, louw2022}, the mixed quench shows a finite thermalization rate. We found a competition between the interaction and the kinetic terms that allows for a rich non-equilibrium dynamics.

We analytically derived the Kadanoff-Baym equations and the expectation values of the non-equilibrium energy in the Keldysh plane from first principles, which we employed to study the mixed quench. We made use of the large-$q$ Ansatz for the Green's function and the Kadanoff-Baym equations so obtained turned out to be nonlinear and non-Markovian in nature, making it impossible to solve the equations analytically in its full generality. We still considered two limiting cases of the mixed quench, namely (1) the interaction term of large-$q$ SYK model where there is no quench and the system remains in equilibrium from infinite past to infinite future, and (2) the interaction term prepared in equilibrium for $t<0$ and at $t=0$, the interaction term is switched off while the kinetic term is switched on (the so-called kinetic quench). Both limiting cases can fortunately be analytically solved for the Green's function in a closed form and this provided us with the benchmarking cases for the consequent numerical solution of the full Kadanoff-Baym equations corresponding to the mixed quench. 

We developed and implemented a numerical technique based on a predictor-corrector algorithm to calculate and analyze the Green's function in the thermodynamic limit. We showed that our predictor-corrector algorithm is capable of solving the Kadanoff-Baym equations both in the analytically solvable limiting cases, as well as the full mixed quench, with a high degree of accuracy. Furthermore, we developed an scheme to obtain the final temperature $1/\beta_f$ and thermalization rate $\gamma$ from the numerical Green's function that allowed us to characterize the final state of the system. Our numerical scheme has the advantage that the initial equilibrium state is obtained analytically for a finite temperature and that the estimation of the final temperature does not require assuming that fluctuation dissipation holds. Our results indicate that, given enough time to relax, the Green's function does indeed reach a final thermal state which is dependent on the initial temperature and the quench parameter. We found that stronger quenches lead to faster relaxation dynamics and higher final temperature (higher $\gamma$ and lower $\beta_f$), eventually reaching the infinite temperature regime at around $\cj_2=0.07$ ($\Jj_q$ has been fixed throughout at the value of $1.0$). This observation is consistent across all finite (from very low to high) initial temperatures. All parameter values used in this work have been mentioned in the captions of the figure as well as the paragraph before Section \ref{sect. numerical techniques}. 

In addition, we identified a complicated dependence of the thermalization rate on the final temperature, which suggest the existence of richer thermalization dynamics than the Fermi liquid to non-Fermi liquid transition of the $\text{SYK}_4+\text{SYK}_2$ quench \cite{larzul2022, bhattacharya2019} (discussed below in further details). For higher final temperatures (lower $\beta_f$, implying stronger quenches), we find that thermalization is achieved in fairly short timescales. However, as the final temperature decreases (higher $\beta_f$, implying weaker quenches), the thermalization rates quickly become several orders of magnitude smaller, in line with the expectation that lower final temperatures/weaker quenches relaxes slower than higher final temperatures/stronger quenches. This effect is more pronounced if we start from higher initial temperatures (lower $\beta_0$), but is also present in systems initialized close to the ground state. 

\begin{figure}
	\centering
	\includegraphics[width=0.90\linewidth]{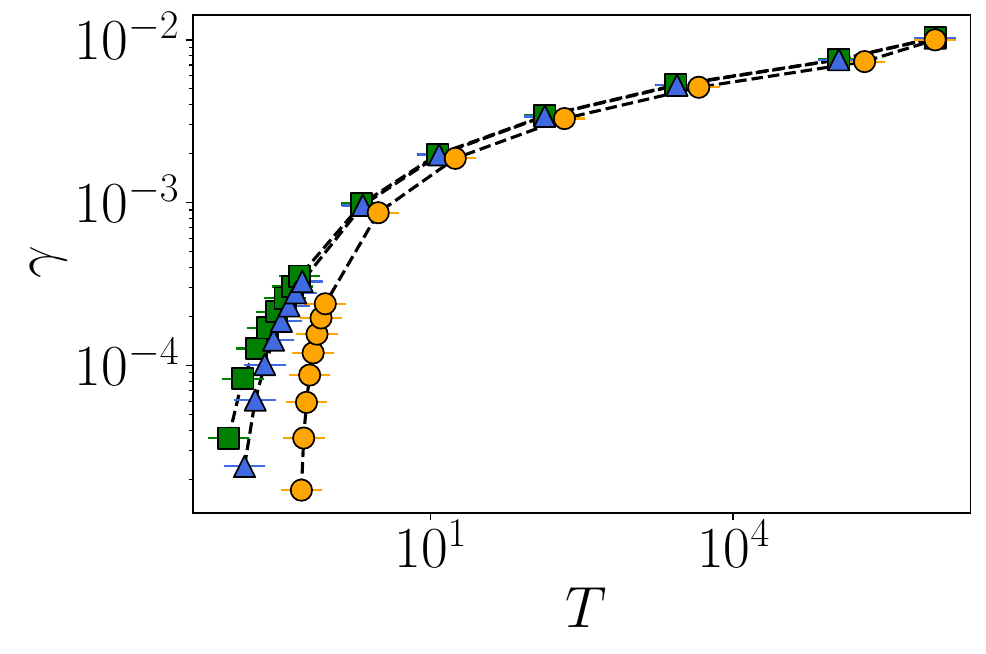}
	\caption{Thermalization rate plotted against temperature to complement the picture in Fig.~\ref{fig:final_gammas_beta} where inverse temperature is used. Color coding is the same as in Fig.~\ref{fig:beta_gamma}. We find a rich and complex thermalization dynamics, including but not limited  to the system being initiated close to the ground state. Moreover, there is a sudden drop in thermalization rate for the lower final temperatures.}
	\label{fig:T_gamma}
\end{figure}

We would like to point out the thermalization dynamics studied in Ref.~\cite{larzul2022} where the system was initialized with a kinetic SYK term and the quenches involved a SYK$_4$ term. Since no large-$q$ scaling is required in $\text{SYK}_4$, accordingly, Ref.~\cite{larzul2022} is able use the time-dependent spectral function to parametrize the non-equilibrium Green's function, such that they obtain the fluctuation-dissipation theorem (FDT) and through it, estimate the final effective temperature. However, as the authors themselves discuss, this procedure explicitly assumes that the system eventually thermalizes. We avoided this assumption as the goal was to deduce thermalization from our results. In addition, the FDT methodology is ineffective for our case due to a diverging delta function that appears at the leading order in $\Oo(1/q)$, indicating that the relation is dominated by the free term in the power series expansion of the Green's function. Instead, we used the kinetic and the interaction energy densities to estimate the thermalization rate and the final temperature, respectively. Moreover, Ref.~\cite{larzul2022} finds a quench-induced non-Fermi liquid (NFL) to Fermi liquid (FL) crossover, while we find a much richer and complex thermalization dynamics, including but not limited to the system being initiated close to the ground state. As an illustration, we provide the plot in Fig.~\ref{fig:T_gamma} of thermalization rate with temperature (in contrast to the inverse temperature in Fig.~\ref{fig:final_gammas_beta}). We can see that the thermalization rate drops really fast for lower final temperatures. Whether this drop-off in the $\gamma's$ is due to some crossover in physical behavior, or some more complex effect, is left as an outlook. 

\section*{Data and Code Availability}

All data and simulation codes used to generate the figures and the associated quantities presented in this study is available at Zenodo upon reasonable request \cite{code}. This includes all processed data for the examples shown in the main text as well as appendices, along with a .csv file containing the table of numerical values for the thermalization rates, temperatures and errors.

\section*{Acknowledgments}

RJ and SK would like to thank Deutsche Forschungsgemeinschaft (DFG, German Research Foundation) Grant No. 217133147 via SFB 1073 (Project No. B03). All authors are grateful for support from Deutsche Forschungsgemeinschaft (DFG, German Research Foundation) Grants No. 436382789, and No. 493420525, via large equipment grants (GOEGrid). 
The data that support the findings of this article are openly available \cite{code}.

\bibliography{thermalization.bib}


\appendix

\section{Analytical methods used in this work}
\label{app. analytical methods required in this work}

 \subsection{Keldysh path integral formalism}
 \label{subsec. Keldysh Path Integral Formalism}
 
Equilibrium interacting systems are connected to their non-interacting counter parts in an adiabatic manner in the infinite past as well as the infinite future. The formalism generally employed to deal with equilibrium systems is the \textit{imaginary-time formalism} (also known as the \textit{Matsubara formalism}) \cite{Stefanucci2013Mar}. For more general non-equilibrium dynamics, this adiabatic time evolution breaks down and the \textit{Keldysh formalism} (also known as the \textit{real-time formalism}) is employed \cite{Kamenev2023Jan}. A brief motivation for the use of the Keldysh formalism is provided in Appendix \ref{app. review of matsubara and keldysh formalisms}; here we state only the most important aspects required for our calculations.

We work in the Keldysh plane as shown in Fig. \ref{fig:keldysh_contour} and explained in the text in Section \ref{sec. analytical methods}. The central object of Keldysh formalism is the partition function, written in terms of quantum fields as (see Appendix \ref{app. review of matsubara and keldysh formalisms})
\begin{equation}
    \Zz =\int  \, D[\psi ,\overline{\psi}]e^{\i S[\psi, \overline{\psi}]},
    \label{eq:partitionfunc}
\end{equation}
where $\overline{\psi}$ denotes the conjugate field and $S[\psi ,\overline{\psi}]$ is the \textit{Keldysh action} given by
\begin{equation}
    S[\psi,\Bar{\psi}] = \int_{\cC} \, dt \left( \frac{\i}{2}\psi(t)\partial_{t}\bar{\psi}(t)-\Hh(t)\right).
    \label{eq:bareaction}
\end{equation}
We denote the unitary time evolution operator along the Keldysh contour as $\hat{U}_\Cc$. If the Hamiltonian is the same along $\Cc_+$ and $\Cc_-$ (the contours are \textit{unperturbed}), then $\hat{U}_\Cc = 1$ and the Keldysh partition function is normalized to unity: $\Zz = 1$. Despite this, the partition function still plays a significant role in extracting the operator expectation values, in an analogous way to equilibrium calculations. One can insert an operator along the backward (or the forward) contour and modify the Hamiltonian as $\Hh^- = \Hh - \hat{\Oo}\eta(t)$, where $\eta(t)$ is any conjugate field, while keeping it the unchanged along the backward contour $\Hh^+ = \Hh$. As a result, the partition function becomes the generating functional $\Zz [\eta]$ from which the operator expectation values can be extracted as
\begin{equation}
    \langle \hat{\Oo}(t)\rangle = \i \left. \frac{\delta \Zz[\eta]}{\delta \eta (t)} \right|_{\eta=0}.
    \label{average expectation value formula}
\end{equation}
Note that unlike the equilibrium case, one does not need to take the logarithm of the generating functional to get the expectation values. This simplifies the calculations significantly, especially for disordered systems where the disorder-average needs to be taken.
A detailed overview of this contour deformation by modifying the Hamiltonian (Eq. \eqref{modified hamiltonian}) in SYK-like systems is given in Section IV of Ref. \cite{jha2024currentcorrelationsconductivitysyklike}, where it has been used to analytically calculate current-current correlations and conductivity in three SYK-like systems. 

\subsection{Disorder-averaged partition function and Schwinger-Dyson equations}
\label{subsec. Disorder-averaged partition function}

The disorder-averaged partition function is calculated as
\begin{equation}
    \overline{\Zz} = \int \Dd j_q \int \Dd j_2 \Pp_q[j_{q;\{i_q\}}] \Pp_2[j_{2;\{l_2\}}] \Zz
    \label{disorder averaging defined}
\end{equation}
where $\Pp_q[j_{q;\{i_q\}}]$ and $\Pp_2[j_{2;\{l_2\}}]$ are given in Eq. \eqref{gaussian ensembles}. This procedure is justified because the SYK model has quenched disorder and, for many quantities, calculating for fixed values of $j_q$ and $j_2$ randomly drawn from their respective ensembles gives the same result as when averaged over different realizations of the coupling constants \cite{Rosenhaus2019Jul}. 

The disorder-average leads to the partition function being expressed in terms of two bi-local fields $\overline{\Zz}[\Gg, \Sigma]$, later identified as the Green's function $\Gg(t,t^\prime)$ and the self-energy $\Sigma(t, t^\prime)$. We note that the SYK model (both the Majorana and the complex versions) is a self-averaging model with no stable spin glass phase, which allows us to restrict ourselves to the replica diagonal subspace \cite{Sachdev2015Nov, Fu2016Jul, Gur-Ari2018Nov, Guo2019Jul}. Accordingly, the bi-local function later to be identified as the Green's function is defined as (following the convention of Ref. \cite{Eberlein2017Nov}, Eq. (2.6) as well as Ref. \cite{bhattacharya2019}, Eq. (2.11)) 
\begin{equation}
\Gg(t_1, t_2) \equiv  -\frac{\i}{N}\sum_{i=1}^N \psi_{i}(t_1) \psi_i (t_2), 
\label{green's function def}
\end{equation}
introduced with the self-energy as a Lagrange multiplier via the following identity:
\begin{widetext}
\begin{equation}
    \int D\Gg D \Sigma \exp{ \frac{-N}{2} \int_{\cC} dt_1 dt_2 \Sigma(t_1,t_2)\Big ( \Gg(t_1,t_2) + \frac{\i}{N} \sum_j \psi_j(t_1) \psi_j (t_2) \Big )} =1
    \label{self-energy def}
\end{equation}
\end{widetext}
as detailed in Appendix \ref{app. Deriving disorder-averaged partition function}. We quote the final result as follows where everything is exact and we still have not made use of either the $N\to \infty$ or the large-$q$ limit:
\begin{equation}
    \overline{\Zz} = \int \Dd \Gg \Dd \Sigma e^{\i N S_{\text{eff}}[\Gg, \Sigma]}
\end{equation}
where $S_{\text{eff}}[\Gg, \Sigma]$ is the effective action given by
\begin{widetext}
\begin{equation}
 S_{\text{eff}}[\Gg, \Sigma] \equiv \frac{-\i}{2}\log \det (\partial_t + \i \Sigma) +\frac{  \i (-1)^{q/2}  }{4  q^2 2^{-q}} \int dt dt^\prime \Jj_q (t)\Jj_q (t^\prime) \Gg(t, t^\prime)^q -\frac{\i }{2 q} \int dt dt^\prime \Jj_2(t)\Jj_2(t^\prime)  \Gg(t,t^\prime)^2 + \frac{\i}{2} \int dt dt^\prime \Sigma(t,t^\prime) \Gg(t,t^\prime).
\end{equation}
\label{final effective action}
\end{widetext}
This form is also known as the $\Gg-\Sigma$ action of the model, since the action has been reduced from containing $N$ fields to depending only upon the Green's function $G$ $\Gg (t, t^\prime)$ and the self-energy $\Sigma(t, t^\prime)$. With this action in a closed form, the model is in principle solved. The theory becomes semi-classical in the large-$N$ limit which we specialize now to. In the large-$N$ limit, the saddle point equations are calculated via the variational principle 
\begin{equation}
 \left.   \frac{\delta  S_{\text{eff}}[\Gg, \Sigma]}{\delta \Sigma}\right|_{\Gg} = 0, \quad   \left.   \frac{\delta  S_{\text{eff}}[\Gg, \Sigma]}{\delta \Gg}\right|_{\Sigma} = 0
 \label{laplace variational principle}
\end{equation}
which gives the \textit{Schwinger-Dyson equations} which is why we identify $\Gg(t, t^\prime)$ as the Green's function
\begin{equation}
\begin{aligned}
   \Gg(t,t^\prime) = &   \left[ \Gg_0^{-1}(t,t^\prime) - \Sigma (t,t^\prime)\right]^{-1}\\
   \Sigma(t,t^\prime)= &-\frac{(-1)^{q/2}}{2^{1-q}q} \Jj_q(t) \Jj_q(t^\prime) \Gg^{q-1}(t, t^\prime)\\
   &+\frac{2}{q} \Jj_2(t) \Jj_2(t^\prime) \Gg(t, t^\prime) 
    \end{aligned}
    \label{sd equations in terms of capital G}
\end{equation}
where $\Gg_0^{-1}(t, t^\prime) = \i \delta(t- t^\prime) \partial_t$ is the free Majorana Green's function. 

\subsection{Green's functions in real-time}
\label{subsec. Green's functions in real-time}

We are now interested in re-formulating the equations of motion on the real time Keldysh contour $\Cc$ as depicted in Fig. \ref{fig:keldysh_contour}, where the imaginary part is ignored, due to the aforementioned Bogoliubov's principle of weakening correlations. In this contour, the two time arguments in the bi-local Green's function can reside either on the forward contour $\Cc_+$ or the backward contour $\Cc_-$. This results in four different Green's functions (also true for any other function on the Keldysh contour) contained in a single Keldysh-contour-ordered Green's function $\Gg(t, t^\prime)$, defined as
\begin{equation}
\Gg\left(t, t^{\prime}\right)=\left\{\begin{array}{ll}
\Gg^{>}\left(t, t^{\prime}\right), & t \in \mathrm{C}_-, t^{\prime} \in \mathrm{C}_+ \quad (\therefore t>t^\prime)\\
\Gg^{<}\left(t, t^{\prime}\right), & t \in \mathrm{C}_+, t^{\prime} \in \mathrm{C}_- \quad (\therefore t<t^\prime)\\
\Gg_{\mathrm{t}}\left(t, t^{\prime}\right), & t, t^{\prime} \in \mathrm{C}_+ \quad (\text{time-ordered})\\
\Gg_{\tilde{\mathrm{t}}}\left(t, t^{\prime}\right), & t, t^{\prime} \in \mathrm{C}_- \quad (\text{anti-time-ordered})
\end{array} .\right.
\label{four green's functions defined}
\end{equation}
where the time-ordered and anti-time-ordered Green's functions
\begin{equation}
\begin{aligned}
\Gg_{\mathrm{t}}\left(t, t^{\prime}\right)    &\equiv +\Theta\left(t-t^{\prime}\right) G^{>}\left(t, t^{\prime}\right)+\Theta\left(t^{\prime}-t\right) G^{<}\left(t, t^{\prime}\right) \\
\Gg_{\tilde{\mathrm{t}}}\left(t, t^{\prime}\right)&\equiv +\Theta\left(t^{\prime}-t\right) G^{>}\left(t, t^{\prime}\right)+\Theta\left(t-t^{\prime}\right) G^{<}\left(t, t^{\prime}\right)
\end{aligned}
\label{time and anti-time ordered Green's functions defined}
\end{equation}
where $\Theta(t)$ is the Heaviside function. These definitions result in the identity
\be
\Gg_{\mathrm{t}}\left(t, t^{\prime}\right)   + \Gg_{\tilde{\mathrm{t}}}\left(t, t^{\prime}\right) = \Gg^{>}\left(t, t^{\prime}\right) + \Gg^{<}\left(t, t^{\prime}\right),
\ee
implying that there are only three linearly independent Green's functions in the Keldysh (real-time) plane. There exists a variety of conventions in the literature but for our purposes, the most suitable choice to work with are the following Green's functions $\{\Gg^>, \Gg^<, \Gg^R, \Gg^A \}$ where $G^R$ and $G^A$ are the retarded and the advanced Green's functions defined as
\begin{equation}
\begin{aligned}
G^{R}\left(t, t^{\prime}\right) & \equiv +\theta\left(t-t^{\prime}\right)\left[G^{>}\left(t, t^{\prime}\right)-G^{<}\left(t, t^{\prime}\right)\right] \\
G^{A}\left(t, t^{\prime}\right) & \equiv +\theta\left(t^{\prime}-t\right)\left[G^{<}\left(t, t^{\prime}\right)-G^{>}\left(t, t^{\prime}\right)\right]
\end{aligned}
\label{retarded and advanced green's functions def}
\end{equation}
and consequently  
\begin{equation}
    G^R(t,t^\prime) - G^A(t,t^\prime) = G^>(t,t^\prime) - G^<(t,t^\prime).
\end{equation}
As mentioned in the Appendix \ref{app. deriving kadanoff-baym equations} in Eq. \eqref{general conjugate relation in appendix}, the Green's functions satisfy the following general conjugate relation (Majorana or otherwise, both in and out-of-equilibrium) 
\begin{equation}
    \left[\Gg^\gtrless(t_1, t_2)\right]^\star = -\Gg^\gtrless(t_2, t_1).
    \label{general conjugate relation}
\end{equation}

Now, there is an additional constraint for Majorana fermions
\begin{equation}
\Gg^{>}\left(t_1, t_2\right)=-\Gg^{<}\left(t_2, t_1\right) \quad (\text{Majorana condition}).
\label{Majorana conjugate relation}
\end{equation}
 that is satisfied both in and out-of equilibrium. Consequently, only one Green's function is needed, in conjunction with Eqs. \eqref{retarded and advanced green's functions def} and \eqref{Majorana conjugate relation}, to get all the others. 

Having introduced the Keldysh Green's functions, we proceed in Section \ref{subsec. kadanoff-baym equations} to use the Schwinger-Dyson equations in Eq. \eqref{sd equations in terms of capital G} to find the equation of motion for $\Gg^>(t,t^\prime)$ in the Keldysh plane, that serves as the basis for the study of the quenched non-equilibrium behavior in this work. 

\section{A short review of Matsubara and Keldysh formalism}
\label{app. review of matsubara and keldysh formalisms}

We first summarize the Matsubara formulation (imaginary-time formalism) and proceed to motivate its shortcomings in non-equilibrium setup to further motivate the Keldysh formulation (real-time formalism). Details can be found in Ref. \cite{Kamenev2023Jan}. The motivation for Matsubara formulation comes from the von Neumann equation of motion
\begin{equation}
    \partial_{t}\hat{\rho}(t)=-i\big[ \hat{H}(t), \hat{\rho}(t) \big]
    \label{von neumann equation}
\end{equation}
 which has the formal solution
\begin{equation}
    \hat{\rho}(t) = \hat{U}_{t,-\infty}\hat{\rho}_{-\infty}\left[ \hat{U}_{t,-\infty} \right]^\dagger=  \hat{U}_{t,-\infty}\hat{\rho}_{-\infty} \hat{U}_{-\infty,t},
    \label{formal solution for rho}
\end{equation}
where $\hat{U}_{t^\prime,t}$ is the unitary time evolution operator. Here $\hat{\rho}_{-\infty}$ is the density matrix of a system that is initially (infinte past) prepared in some equilibrium state. The unitary evolution operator follows the following equations:
\begin{equation}
    \partial_t^\prime  \hat{U}_{t^\prime,t} = - \i \Hh(t^\prime) \hat{U}_{t^\prime,t} , \quad \partial_t  \hat{U}_{t^\prime,t} = + \i \hat{U}_{t^\prime,t}  \Hh(t) 
\end{equation}
where $\Hh(t)$ is the Hamiltonian of the system evaluated at time $t$ and satisfies the unitary condition $\hat{U}_{t^\prime,t} \left[\hat{U}_{t^\prime,t} \right]^\dagger = 1$. This gives the identity $\left[\hat{U}_{t^\prime,t} \right]^\dagger = \left[\hat{U}_{t^\prime,t} \right]^{-1} = \hat{U}_{t,t^\prime}$ which has been used in the last equality above in Eq. \eqref{formal solution for rho}. Then the operator expectation value of any observable $\hat{\Oo}$ is calculated as
\begin{equation}
\langle\hat{\mathcal{O}}(t)\rangle \equiv \frac{\operatorname{Tr}[\hat{\mathcal{O}} \hat{\rho}(t)]}{\operatorname{Tr}[\hat{\rho}(t)]}=\frac{\operatorname{Tr}\left[\hat{U}_{-\infty, t} \hat{\mathcal{O}} \hat{U}_{t,-\infty} \hat{\rho}_{-\infty}\right]}{\operatorname{Tr}[\hat{\rho}_{-\infty}]} 
\label{expectation value defined}
\end{equation}
where we used in the second equality the following: (a) Eq. \eqref{formal solution for rho}, (b) the cyclic property of trace, and (c) the fact that the trace of the initial density matrix does not change under unitary time evolution as dictated by the von Neumann equation (Eq. \eqref{von neumann equation}). Now we can re-write the first term in the numerator of Eq. \eqref{expectation value defined}, namely $\hat{U}_{-\infty, t}$, using the properties of the unitary operator that $\hat{U}_{-\infty, t} = \hat{U}_{-\infty, +\infty} \hat{U}_{+\infty, t} $ and therefore the numerator becomes
\begin{equation}
 \text{Numerator} =   \operatorname{Tr}[ \overbrace{\hat{U}_{-\infty, +\infty}}^{\text{backward evolution}} \underbrace{\hat{U}_{+\infty, t} \hat{\mathcal{O}} \hat{U}_{t,-\infty} \hat{\rho}_{-\infty}}_{\text{forward evolution}}]
  \label{forward and backward labeled equation}
\end{equation}
Now comes the central dogma of equilibrium systems: \textit{adiabatic continuity}. It states that the interacting systems are adiabatically connected to its non-interacting counterpart for $t\to \pm \infty$. Let the non-interacting state be denoted by $| 0\rangle$, then the assumption of adiabatic time evolution dictates for the forward evolution 
\begin{equation}
 \hat{U}_{+\infty, -\infty}   | 0\rangle = e^{\i \phi} | 0\rangle
 \label{adiabatic continuity assumption}
\end{equation}
or equivalently for the backward evolution $\hat{U}_{-\infty, +\infty}   | 0\rangle = e^{-\i \phi} | 0\rangle$ where $\phi $ is some phase. This makes the backward evolution trivial in Eq. \eqref{forward and backward labeled equation} up to a phase factor and has been argued in detail in Ref. \cite{Kamenev2023Jan}, the operator expectation value for equilibrium systems is solely given by the future time evolution without any need for backward time evolution as follows:
\begin{equation}
\langle\hat{\mathcal{O}}(t)\rangle^{\text{equilibrium}} =   \frac{\operatorname{Tr}\left[  \hat{U}_{+\infty, t} \hat{\mathcal{O}} \hat{U}_{t,-\infty} \hat{\rho}_{-\infty}\right]}{\operatorname{Tr}[\hat{\rho}_{-\infty}]} 
\end{equation}
Explicitly, $\rho_{-\infty} = e^{\beta \Hh}$ and $\hat{U}_{t,t^\prime} = e^{-i(t-t^\prime)\Hh}$. The Matsubara form is obtained by replacing the time argument by an imaginary variable $t\rightarrow-i\tau$ such that $0\leq \tau \leq \beta$. That's why the Matsubara formulation is also known as the imaginary-time formalism. This has the advantage of treating $\hat{\rho}_{-\infty}$ and $\hat{U}_{t,t^\prime}$ together in diagrammatic expansion \cite{bruus-flensberg}. 

Having clarified the Matsubara formulation, now we can immediately see why this fails for non-equilibrium systems. The central pillar of Matsubara formulation that avoids explicit calculation of backward evolution is the assumption of \textit{adiabatic continuity}. In non-equilibrium systems, this no longer holds. This implies that if the interactions are switched off at some time in the future, the final infinite future state will not go back to the original non-interacting state we started with in infinite past. In other words, Eq. \eqref{adiabatic continuity assumption} does not hold for non-equilibrium systems. Therefore the backward evolution must be considered in Eq. \eqref{forward and backward labeled equation} and together the forward and backward time contours is known as \textit{closed time contour}. This is formulated in real time (that's why the name real-time formalism) by what is known as Martin-Schwinger-Kadanoff-Baym-Keldysh formalism, after the authors who independently developed this methodology \cite{NEGF_conference}. We will refer to this as Keldysh formalism (or real-time formalism) for brevity. 

The Keldysh contour is shown in Fig. \ref{fig:keldysh_contour} where we see a forward real time evolution from $-\infty$ to $+\infty$, a backward time evolution from $+\infty $ to $-\infty$ and an additional imaginary time axis from $t_0 $ to $t_0 - \i \beta$ where $t_0 \to -\infty$. The imaginary branch represents the equilibrium state reached after real-time dynamics have occurred. For generic non-equilibrium dynamics, when the state is prepared in infinite past ($t_0 \to -\infty$), the imaginary branch is usually deprecated due to the \textit{Bogoliubov's principle of weakening correlations} \cite{bhattacharya2019}.
In this work, we have assumed throughout that the system is prepared in equilibrium at $t_0 \to -\infty$, and the imaginary branch is disposed off in all of the quenched cases.

The central object in Keldysh formalism is the partition function
\begin{equation}
    \Zz \equiv \frac{\Tr[ \hat{U}_{\mathcal{C}} \hat{\rho}_{-\infty}]}{\Tr\hat{\rho}_{-\infty}} 
\end{equation}
where $\hat{U}_{\mathcal{C}} \equiv \hat{U}_{-\infty, +\infty} \hat{U}_{+\infty, -\infty}$ (subscript $\Cc$ denotes the closed Keldysh contour as in Fig. \ref{fig:keldysh_contour} without the imaginary branch which has been deprecated due to the aforementioned Bogoliubov's principle of weakening correlations). We can see from the construction of the Keldysh contour that if the Hamiltonian is the same on the forward and the backward time evolution branch, then $\hat{U}_\Cc = 1$ and therefore the Keldysh partition function is normalized to unity ($\Zz = 1$). Still, as we have seen in this work, the partition function plays a significant role to compute physical quantities and the normalization condition serves as a useful check for the consistency of the calculations. 

We can express the partition function in terms of field variables $\psi$ as has been used throughout this work
\begin{equation}
    \Zz =\int  \, D[\psi ,\overline{\psi}]e^{\i S[\psi, \overline{\psi}]},
    \label{eq:partitionfunc in app.}
\end{equation}
where $\overline{\psi}$ denotes the conjugate field and $S[\psi ,\overline{\psi}]$ is the \textit{Keldysh action} given by
\begin{equation}
    S[\psi,\Bar{\psi}] = \int_{\cC} \, dt \left( \frac{\i}{2}\psi(t)\partial_{t}\bar{\psi}(t)-\Hh(t)\right).
    \label{eq:bareaction in app.}
\end{equation}
Here the integration takes over the complex Schwinger-Keldysh contour $\Cc$ as depicted in Fig. \ref{fig:keldysh_contour}. We can divide $\Cc = \Cc_+ + \Cc_-$ (ignoring the imaginary branch as argued above due to the Bogoliubov's principle of weakening correlations) where $\Cc_+$ and $\Cc_-$ denote the forward as well as the backward contours. 

In order to calculate the expectation values of any observable $\hat{\Oo}$, we insert the operator either in the forward or the backward contour by modifying the Hamiltonian $\Hh$ as follows \cite{Kamenev2023Jan}:
\begin{equation}
    \Hh^{\pm}_\eta = \Hh \pm \hat{\Oo} \eta(t)
    \label{modified hamiltonian}
\end{equation}
where $\eta(t)$ is the conjugate field. The plus sign denotes the forward contour and the minus sign denotes the backward contour and therefore, we have two different Hamiltonians for the forward and the backward contours. Accordingly $\hat{U}_\Cc \neq 1$ and therefore $\Zz[\eta]\neq 1$. The partition function becomes the \textit{generating functional} given by
\begin{equation}
\Zz[\eta] \equiv \frac{\operatorname{Tr}\left[\hat{U}_{\mathcal{C}}[\eta] \hat{\rho}_{-\infty}\right]}{\operatorname{Tr}[\hat{\rho}_{-\infty}]}
\end{equation}
which is nontrivial now ($\Zz[\eta] \neq 1$). Then the expectation value is calculated via the functional differentiation as (assuming the operator is inserted in the forward contour)
\begin{equation}
    \langle \hat{\Oo}(t)\rangle = \i \left. \frac{\delta \Zz[\eta]}{\delta \eta(t)} \right|_{\eta=0}
    \label{average expectation value formula in appendix}
\end{equation}

We end this appendix with a final note about the disordered systems such as the one considered in this work. Compared to the equilibrium systems where the expectation values of operators are calculated using the \textit{logarithm of the generating functional}, the Keldysh generating functional does not require to take the logarithm as is evident from Eq. \eqref{average expectation value formula in appendix} which significantly simplifies the calculations.

\onecolumngrid

\section{Deriving disorder-averaged partition function}
\label{app. Deriving disorder-averaged partition function}

We have the partition function from Eqs. \eqref{eq:partitionfunc} and \eqref{eq:bareaction} which we combine to write
\begin{equation}
    \Zz = \int \Dd \psi_i \exp\left[ - \frac{1}{2} \int dt \sum\limits_{i=1} \psi_i \partial_t \psi_i - \i \int dt \Hh \right]
\end{equation}
where we use the fact that Majorana fermions $ \psi_i(t)$ are their own conjugate and time-dependent. We avoid showing explicitly the time-dependence on the Majorana fermions unless absolutely required to avoid clutter. We consider the most general Hamiltonian as presented in Eq. \eqref{eq:mixed_syk_neq_hamiltonian}. Then the disorder-averaging is done over different realizations of coupling constants $j_q$ and $j_2$ as defined in Eq. \eqref{disorder averaging defined} where the Gaussian distributions $\Pp_q[j_{q;\{i_q\}}]$ and $ \Pp_2[j_{2;\{l_2\}}]$ are given in Eq. \eqref{variances}. Plugging $\Hh$ from Eq. \eqref{eq:mixed_syk_neq_hamiltonian} and Gaussian distributions from Eq. \eqref{gaussian ensembles} into the disorder-averaged partition function in Eq. \eqref{disorder averaging defined}, we get (showing integral sign only once for brevity)

    \begin{equation}
    \begin{aligned}
        \overline{\Zz} = \int \Dd j_q \Dd j_2 \Dd \psi_i \exp&\left[ - \frac{1}{2} \int dt \sum\limits_{i=1}^N \psi_i \partial_t \psi_i - \frac{N^{q-1}q^2 2^{-q} }{\Jj_q^2 q!} \sum\limits_{\{i_q\}_{\leq}} j_{q;\{i_q\}}^2  - \frac{N q}{4 \Jj_2^2} \sum\limits_{\{l_2\}_{\leq}} j_{2;\{l_2\}}^2 \right. \\ 
 &\left.- \i . \i^{q/2} \int  dt \sum\limits_{\{i_q\}_{\leq}}  j_{q;\{i_q\}} f_q(t) \psi_{i_1}\ldots \psi_{i_q} + \int dt \sum\limits_{\{l_2\}_{\leq}} j_{2;\{l_2\}} f_2(t) \psi_{l_1}\psi_{l_1}      \right]
    \end{aligned}
    \end{equation}
We can now integrate over $j_q$ and $j_2$ by separating the integral as
\begin{equation}
\begin{aligned}
    \overline{\Zz} = &\int \Dd \psi_i \exp\left[- \frac{1}{2} \int dt \sum\limits_{i=1}^N \psi_i \partial_t \psi_i \right] \\
    &\times \int \Dd j_q \exp\left[- \frac{N^{q-1}q^2 2^{-q} }{\Jj_q^2 q!} \sum\limits_{\{i_q\}_{\leq}} j_{q;\{i_q\}}^2 - \i . \i^{q/2} \int  dt  \sum\limits_{\{i_q\}_{\leq}} j_{q;\{i_q\}} f_q(t) \psi_{i_1}\ldots \psi_{i_q} \right] \\
    & \times \int \Dd j_2 \exp\left[ - \frac{N q}{4 \Jj_2^2} \sum\limits_{\{l_2\}_{\leq}} j_{2;\{l_2\}}^2  + \int dt \sum\limits_{\{l_2\}_{\leq}} j_{2;\{l_2\}} f_2(t) \psi_{l_1}\psi_{l_1}    \right] 
\end{aligned}
\end{equation}
where we use the standard Gaussian integral $\int dx e^{-a x^2 - b x - c} = \sqrt{\pi/a} e^{b^2/4a - c}$ where the coefficient $\sqrt{\pi/a}$ gets exactly canceled with the normalizing factor of the Gaussian ensembles in Eq. \eqref{gaussian ensembles}and we get
\begin{equation}
\begin{aligned}
\Rightarrow    \overline{\Zz} = \int \Dd \psi_i \exp\left[- \frac{1}{2} \int dt \sum\limits_{i=1}^N \psi_i \partial_t \psi_i \right] & \exp\left[ \frac{\Jj_q^2 q!}{4 N^{q-1} q^2 2^{-q}} \left(-\i . \i^{q/2} \int dt \sum\limits_{\{i_q\}_{\leq}} f_q(t) \psi_{i_1} \ldots \psi_{i_q} \right)^2 \right]  \\
& \times \exp\left[\frac{\Jj_2^2}{Nq} \left(\int dt \sum\limits_{\{l_2\}_{\leq}} f_2(t) \psi_{l_1} \psi_{l_2} \right)^2 \right]
\end{aligned}
\end{equation}
\begin{equation}
\Rightarrow    \overline{\Zz} = \int \Dd \psi_i \exp\left[- \frac{1}{2} \int dt \sum\limits_{i=1}^N \psi_i \partial_t \psi_i    - \frac{\i^q \Jj_q^2 q! }{4 N^{q-1} q^2 2^{-q}} \left( \int dt \sum\limits_{\{i_q\}_{\leq}} f_q(t) \psi_{i_1} \ldots \psi_{i_q} \right)^2 +\frac{\Jj_2^2}{Nq} \left(\int dt \sum\limits_{\{l_2\}_{\leq}} f_2(t) \psi_{l_1} \psi_{l_2} \right)^2 \right]
\label{intermediate step of deriving Z bar 1}
\end{equation}
Then we can write the squares of the fields as 
\begin{equation}
\left( \int dt \sum\limits_{\{i_q\}_{\leq}} f_q(t) \psi_{i_1} \ldots \psi_{i_q} \right)^2  = \int dt \int dt^\prime \sum\limits_{\{i_q\}_{\leq}} \left(f_q(t) \psi_{i_1}(t) \ldots \psi_{i_q}(t)\right) \times \left( f_q(t^\prime) \psi_{i_1}(t^\prime) \ldots \psi_{i_q}(t^\prime)\right)
\end{equation}
If we focus on the integrand and consider $q=4$ case for the sake of clarity, we get
\begin{equation}
\begin{aligned}
    \sum\limits_{1\leq i_1<i_2<i_3<i_4\leq N}^N \left( f_4\psi_{i_1} \psi_{i_2} \psi_{i_3}\psi_{i_4}\right)(t)\left( f_4\psi_{i_1} \psi_{i_2} \psi_{i_3}\psi_{i_4}\right)(t^\prime) &= \frac{1}{4!} \sum\limits_{i_1\neq i_2 \neq i_3 \neq i_4} \left( f_4\psi_{i_1} \psi_{i_2} \psi_{i_3}\psi_{i_4}\right)(t)\left( f_4\psi_{i_1} \psi_{i_2} \psi_{i_3}\psi_{i_4}\right)(t^\prime) \\
    &= \frac{1}{4!} f_4(t) f_4(t^\prime) \left( \sum\limits_{i=1}^N \psi_i(t) \psi_i(t^\prime)\right)^4
    \end{aligned}
\end{equation}
and for $q=2$, we get
\begin{equation}
\begin{aligned}
    \sum\limits_{1\leq i_1<i_2\leq N}^N \left( f_2\psi_{i_1} \psi_{i_2} \right)(t)\left( f_2\psi_{i_1} \psi_{i_2} \right)(t^\prime) &= \frac{1}{2!} \sum\limits_{i_1\neq i_2 } \left( f_2\psi_{i_1} \psi_{i_2} \right)(t)\left( f_2\psi_{i_1} \psi_{i_2} \right)(t^\prime) \\
    &= -\frac{1}{2!} f_2(t) f_2(t^\prime) \left( \sum\limits_{i=1}^N \psi_i(t) \psi_i(t^\prime)\right)^2
    \end{aligned}
\end{equation}
In both cases, the pattern is the the same: we matched the indices for the fermionic fields at times $t$ and $t^\prime$ and for every swapping of the fermionic fields, there is a minus sign due to their Grassmannian nature. This can be generalized to $q$ case where recall that $q$ is an even number
\begin{equation}
\begin{aligned}
    \sum_{\{i_q\}_{\leq}} \left( f_q\psi_{i_1} \dots \psi_{i_q}\right)(t) \left(f_q \psi_{i_1} \dots \psi_{i_q}\right)(t^\prime)  &= \frac{1}{q!} \sum_{\{i_q\}_{\neq}} \left( f_q\psi_{i_1} \dots \psi_{i_q}\right)(t) \left(f_q \psi_{i_1} \dots \psi_{i_q}\right)(t^\prime)\\
    &= \frac{(-1)^{q/2}}{q!} f_q(t) f_q(t^\prime) \left( \sum\limits_{i=1}^N \psi_i(t) \psi_i(t^\prime)\right)^q
    \end{aligned}
    \label{fermionic rearrangement}
\end{equation}
where we denote $ i_q =i_1 \neq i_2 \dots \neq i_q$ as $\{i\}_{\neq}$ for brevity. Next we use the definition of Green's function as given in the main text (Eq. \eqref{green's function def}) which we reproduce here for convenience
\begin{equation}
\Gg(t_1, t_2) \equiv  -\frac{\i}{N}\sum_{i=1}^N \psi_{i}(t_1) \psi_i (t_2), 
\end{equation}
which is used to replace the fermionic fields appearing in Eq. \eqref{fermionic rearrangement} as (recall $q$ is an even number)
\begin{equation}
   \frac{(-1)^{q/2}}{q!} f_q(t) f_q(t^\prime) \left( \sum\limits_{i=1}^N \psi_i(t) \psi_i(t^\prime)\right)^q = \frac{(-1)^{q/2}N^q}{q! (-\i)^q } \Gg(t,t^\prime)^q 
\end{equation}
Using these results and plugging them back in Eq. \eqref{intermediate step of deriving Z bar 1}, we get
\begin{equation}
\begin{aligned}
\Rightarrow    \overline{\Zz} = \int \Dd \psi_i \exp&\left[- \frac{1}{2} \int dt dt^\prime \sum\limits_{i=1}^N \psi_i (t) \delta(t - t^\prime )\partial_t^\prime \psi_i(t^\prime)    - \frac{\i^q (-1)^{q/2}\Jj_q^2  }{4 N^{q-1} q^2 2^{-q}} \int dt dt^\prime f_q(t) f_q(t^\prime) \left( \sum\limits_{i=1}^N \psi_i(t) \psi_i(t^\prime)\right)^q \right.\\
&\left.-\frac{\Jj_2^2}{2Nq} \int dt dt^\prime f_2(t) f_2(t^\prime) \left( \sum\limits_{j=1}^N \psi_j(t) \psi_j(t^\prime)\right)^2  \right].
\end{aligned}
\label{intermediate step of deriving Z bar 2}
\end{equation}
where we rewrote the first term on the right-hand side by inserting a delta-function to have two time dependencies. We now use the definition of the Green's function as mentioned above and introduce another bi-local field, the self-energy $\Sigma(t, t^\prime)$ via the identity mentioned in the main text (Eq. \eqref{self-energy def}) which we reproduce here for convenience: 
\begin{equation}
    \int \Dd\Gg \Dd \Sigma \exp{ \frac{-N}{2} \int dt_1 dt_2 \Sigma(t_1,t_2)\Big ( \Gg(t_1,t_2) + \frac{\i}{N} \sum_j \psi_j(t_1) \psi_j (t_2) \Big )} =1.
\end{equation}
By plugging this into Eq. \eqref{intermediate step of deriving Z bar 2}, we get
\begin{equation}
\begin{aligned}
\Rightarrow    \overline{\Zz} = \int \Dd \Gg \Dd \Sigma \Dd \psi_i  \exp&\left[- \frac{1}{2} \int dt dt^\prime \sum\limits_{i=1}^N \psi_i (t) \left( \delta(t - t^\prime ) \partial_t^\prime + \i \Sigma(t,t^\prime) \right)  \psi_i(t^\prime)    - \frac{ N (-1)^{q/2}\Jj_q^2   }{4  q^2 2^{-q}} \int dt dt^\prime f_q(t) f_q(t^\prime) \Gg(t, t^\prime)^q \right.\\
&\left.+\frac{N \Jj_2^2}{2 q} \int dt dt^\prime f_2(t) f_2(t^\prime) \Gg(t,t^\prime)^2 - \frac{N}{2} \int dt dt^\prime \Sigma(t,t^\prime) \Gg(t,t^\prime)\right].
\end{aligned}
\label{intermediate step of deriving Z bar 3}
\end{equation}
Now we integrate out the fermionic fields by using the following identity for the Grassmannian variables:
\begin{equation}
    \int \Dd \psi_i \exp\left[ - \frac{1}{2} \int dt dt^\prime \sum\limits_{i=1}^N \psi_i (t) \left( \delta(t - t^\prime ) \partial_t^\prime + \i \Sigma(t,t^\prime) \right)  \psi_i(t^\prime) \right] = \exp\left[ \frac{N}{2} \log \det (\partial_t + \i \Sigma)\right] = \left[ \det (\partial_t + \i \Sigma)\right]^{\frac{N}{2}}
\end{equation}
to finally get
\begin{equation}
\begin{aligned}
\Rightarrow    \overline{\Zz} = \int \Dd \Gg \Dd \Sigma \exp&\left[ \frac{N}{2}\log \det (\partial_t + \i \Sigma) - \frac{ N (-1)^{q/2}\Jj_q^2   }{4  q^2 2^{-q}} \int dt dt^\prime f_q(t) f_q(t^\prime) \Gg(t, t^\prime)^q \right.\\
&\left.+\frac{N \Jj_2^2}{2 q} \int dt dt^\prime f_2(t) f_2(t^\prime) \Gg(t,t^\prime)^2 - \frac{N}{2} \int dt dt^\prime \Sigma(t,t^\prime) \Gg(t,t^\prime)\right]
\label{final disorder-averaged partition function}
\end{aligned}
\end{equation}
We have derived this without assuming any limit for $N$ or $q$. We started evaluating the partition function as in Eq. \eqref{eq:partitionfunc} and the disorder-averaged partition function can be written as
\begin{equation}
    \overline{\Zz} = \int \Dd \Gg \Dd \Sigma e^{\i N S_{\text{eff}}[\Gg, \Sigma]}
\end{equation}
where $S_{\text{eff}}[\Gg, \Sigma]$ is the effective action can be read off by comparting against Eq. \eqref{final disorder-averaged partition function} 
\begin{equation}
    S_{\text{eff}}[\Gg, \Sigma] \equiv \frac{-\i}{2}\log \det (\partial_t + \i \Sigma) +\frac{  \i (-1)^{q/2}  }{4  q^2 2^{-q}} \int dt dt^\prime \Jj_q (t)\Jj_q (t^\prime) \Gg(t, t^\prime)^q -\frac{\i }{2 q} \int dt dt^\prime \Jj_2(t)\Jj_2(t^\prime)  \Gg(t,t^\prime)^2 + \frac{\i}{2} \int dt dt^\prime \Sigma(t,t^\prime) \Gg(t,t^\prime)
\end{equation}
where we have absorbed the time-dependent functions $f_q$ and $f_2$ into the coupling constants. This is what is used in the main text in Section \ref{sec. analytical methods} where the large-$N$ limit is taken to calculate the saddle point equations, also known as the Schwinger-Dyson equations. 

\twocolumngrid

\section{Deriving the Kadanoff-Baym Equations}
\label{app. deriving kadanoff-baym equations}

We start with the Schwinger-Dyson equations in Eq. \eqref{sd equations in terms of capital G} where we re-arrange the first equation (the Dyson equation) as
\begin{equation}
 \Gg_0^{-1}(t_1,t_3) =  \Gg^{-1}(t_1,t_3) + \Sigma (t_1,t_3).
\end{equation}
We take the convolution product from the right in the Keldysh contour with respect to $\Gg(t_3, t_2)$ to get
\begin{equation} 
\begin{aligned}
    \int_{\mathcal{C}} & dt_3 \,  G_0^{-1}(t_1,t_3)G(t_3,t_2)  \\
    =& \int_{\mathcal{C}} dt_3  \,  G^{-1}(t_1,t_3)G(t_3,t_2) + \int_{\mathcal{C}} dt_3  \,  \Sigma(t_1,t_3)G(t_3,t_2) \\
    =& \delta_{\mathcal{C}}(t_1,t_2)+ \int_{\mathcal{C}} dt_3  \,  \Sigma(t_1,t_3)G(t_3,t_2),
\end{aligned}
\label{eq:convolution_G}
\end{equation}
where $\delta_\Cc (t_1, t_2)$ is the delta-function on the Keldysh contour. Without loss of generality, we can take $t_1$ on the backward contour $\Cc_-$ and $t_2$ on the forward contour $\Cc_+$ in Fig. \ref{fig:keldysh_contour}. This implies that the delta-function vanishes. 

The left-hand side can be simplified further by using the real space representation for the free Majorana Green's function, namely $\Gg_0^{-1}(t, t^\prime) = \i \delta(t- t^\prime) \partial_t$. We get
\begin{equation}
    \text{Left-hand side} = \i \partial_{t_1} G(t_1^{\Cc_-},t_2^{\Cc_+}) = i\partial_{t_1} G^{>}(t_1,t_2)
\end{equation}
Next we simplify the right-hand side using the Langreth rule derived in Appendix \ref{app. A brief note on the Langreth rules} (Eq. \eqref{langreth rule 2}) to get the first Kadanoff-Baym equation in real time
\begin{equation}
\begin{aligned}
    \i \partial_{t_1} \Gg^> (t_1,t_2) = \int_{-\infty}^\infty dt_3 &\left( \Sigma^R(t_1,t_3)\Gg^>(t_3,t_2) \right.\\
    &\left.+ \Sigma^>(t_1,t_3)\Gg^A(t_3,t_2)\right).
    \end{aligned}
    \label{kb equation 1}
\end{equation}
where the retarded and the advanced functions are defined in Eq. \eqref{retarded and advanced green's functions def} (same definitions hold true for any other function on the Keldysh contour including the self-energy $\Sigma(t, t^\prime)$).

Similarly if we convolute the Dyson equation $ \Gg_0^{-1}(t_3,t_2) =  \Gg^{-1}(t_3,t_2) + \Sigma (t_3,t_2)^{-1}$ with $\Gg(t_1, t_3)$ from the left and take $t_1$ and $t_2$ on $\Cc_-$ and $\Cc_+$ respectively as before, we arrive at the second Kadanoff-Baym equation in real time
\begin{equation}
\begin{aligned}
   -\i\partial_{t_2} \Gg^> (t_1,t_2) = \int_{-\infty}^\infty dt_3 &\left( \Gg^R(t_1,t_3)\Sigma^>(t_3,t_2) \right.\\
   &\left.+ \Gg^>(t_1,t_3)\Sigma^A(t_3,t_2) \right)
    \end{aligned}
    \label{kb equation 2}
\end{equation}
Note that the Green's function satisfies the conjugacy relation in general (Majorana and otherwise, both in and out-of-equilibrium)
\begin{equation}
    \left[\Gg^\gtrless(t_1, t_2)\right]^\star = -\Gg^\gtrless(t_2, t_1)
    \label{general conjugate relation in appendix}
\end{equation}
which saves the computation effort of solving one of the two equations of motion for $\Gg^>(t_1, t_2)$, namely either Eq. \eqref{kb equation 1} or Eq. \eqref{kb equation 2}. We chose to solve Eq. \eqref{kb equation 1} in this work.

There are also a set of coupled equations of motion for $\Gg^<(t_1, t_2)$. We present the full Kadanoff-Baym equations
\begin{equation}
\begin{aligned}
    \i \partial_{t_1} \Gg^\gtrless(t_1,t_2) = \int_{-\infty}^\infty dt_3 &\left( \Sigma^R(t_1,t_3)\Gg^\gtrless(t_3,t_2) \right.\\
    &\left.+ \Sigma^\gtrless(t_1,t_3)\Gg^A(t_3,t_2)\right).
    \end{aligned}
\end{equation}
\begin{equation}
\begin{aligned}
   -\i\partial_{t_2} \Gg^\gtrless (t_1,t_2) = \int_{-\infty}^\infty dt_3 &\left( \Gg^R(t_1,t_3)\Sigma^\gtrless(t_3,t_2) \right.\\
   &\left.+ \Gg^\gtrless(t_1,t_3)\Sigma^A(t_3,t_2) \right)
    \end{aligned}
    \label{kb equation full}
\end{equation}
where the self-energy is given by the second Schwinger-Dyson equations in Eq. \eqref{sd equations in terms of capital G} and we need to use the Langreth rule (Eq. \eqref{langreth rule 3}) to analytically continue the self-energy to real time. 

Since we are dealing with Majorana fermions, we don't need the equations of motion for $\Gg^<(t, t^\prime)$ as this is completely determined by solving for $\Gg^>(t, t^\prime)$ and using the conjugacy relation 
\be 
\Gg^<(t, t^\prime) = - \Gg^>(t^\prime, t).
\label{Majorana conjugate relation in appendix}
\ee
This further simplifies the computation effort on top of Eq. \eqref{general conjugate relation in appendix} and therefore, solving only Eq. \eqref{kb equation 1} ensures that we have solved the system of Majorana fermions considered in this work.

Now we specialize to the model considered in this work. The self-energy is given by the second equation of the Schwinger-Dyson equations in Eq. \eqref{sd equations in terms of capital G} (using the Langreth rule in Eq. \eqref{langreth rule 3})
\begin{equation}
    \begin{aligned}
        \Sigma^\gtrless(t,t^\prime)= &-\frac{(-1)^{q/2}}{2^{1-q}q} \Jj_q(t) \Jj_q(t^\prime) \left( \Gg^\gtrless (t, t^\prime) \right)^{q-1}\\
   &+\frac{2}{q} \Jj_2(t) \Jj_2(t^\prime) \Gg^\gtrless(t, t^\prime).
    \end{aligned}
\end{equation}
We plug in the definition of the retarded and advanced functions from Eq. \eqref{retarded and advanced green's functions def} in Eq. \eqref{kb equation 1} and use the self-energy to get the following:
\begin{equation}
\begin{aligned}
    \i \partial_{t_1} &\Gg^>(t_1, t_2) \\
    =& \int\limits_{-\infty}^{+\infty} dt_3 \Theta(t_1 -t_3) \left[ \Sigma^>(t_1, t_3) - \Sigma^<(t_1, t_3)\right] \Gg^>(t_3, t_2) \\
    & + \int\limits_{-\infty}^{+\infty} dt_3 \Theta(t_2 -t_3) \Sigma^>(t_1, t_3) \left[\Gg^<(t_3, t_2) - \Gg^>(t_3, t_2) \right] .
    \end{aligned}
\end{equation}
We now plug in the expressions for the self-energies and simplify to get
\begin{widetext}
    \begin{equation}
\begin{aligned}
\i \partial_{t_1} \Gg^>\left(t_1, t_2\right) =& \int_{-\infty}^{t_1} d t_3\left\{-\frac{(-1)^{q/2}}{2^{1-q}q} \cj_q\left(t_1\right) \cj_q\left(t_3\right)\left[G^{>}\left(t_1, t_3\right)^{q-1}-G^{<}\left(t_1, t_3\right)^{q-1}\right] G^{>}\left(t_3, t_2\right)\right. \\
& \left.+\frac{2}{q} \cj_2\left(t_1\right) \cj_2\left(t_3\right)\left[G^{>}\left(t_1, t_3\right)-G^{<}\left(t_1, t_3\right)\right]G^{>}\left(t_3, t_2\right)\right\} \\
& +\int^{t_2}_{-\infty} d t_3\left\{\frac{(-1)^{q/2}}{2^{1-q}q}\cj_{q}(t_{1}) \cj_ { q }(t_{ 3 } ) G^>\left(t_1, t_3\right)^{q-1}\left[G^>\left(t_3, t_2\right)-G^<\left(t_3, t_2\right)\right]\right. \\
& \left.-\frac{ 2 }{q} \cj_2\left(t_1\right) \cj_2\left(t_3\right)G^>\left(t_1, t_3\right)\left[G^>\left(t_3, t_2\right)-G^{<}\left(t_3, t_2\right)\right]\right\}.
\end{aligned}
\label{final kb equation for this paper in appendix}
\end{equation}
\end{widetext}
This is the final Kadanoff-Baym equation that needs to be solved to completely solve the system in consideration. 

\section{A brief note on the Langreth rules}
\label{app. A brief note on the Langreth rules}

We derive the Langreth rule used in Appendix \ref{app. deriving kadanoff-baym equations} to derive the Kadanoff-Baym equations. We consider a general (out-of-equilibrium) Keldysh integral of the form
\begin{equation}
    Z(t_1, t_2) = \int_{\Cc} dt_3 X(t_1, t_3)  Y (t_3, t_2)
\end{equation}
where $Z, X$ and $Y$ are some arbitrary bi-local functions in time in the Keldysh plane and $\Cc $ is the Keldysh contour. 

We consider the first case where we take time $t_1$ on $\Cc_+$ and time $t_2$ on $\Cc_-$ in Fig. \ref{fig:keldysh_contour} (note that we have assumed throughout this work that the imaginary part of the contour has been deprecated due to the aforementioned reason of Bogoliubov's principle of weakening correlations). Then by construction, we get on the left-hand side $Z^<(t_1, t_2)$ where we used the definition of the lesser function from Eq. \eqref{four green's functions defined}. Accordingly we use the same definition on the right-hand side to get
\begin{equation}
\begin{aligned}
    \Rightarrow Z^<(t_1, t_2) = &\int_{\Cc_+} dt_3 X(t_1, t_3)  Y^< (t_3, t_2) \\
    &+ \int_{\Cc_-} dt_3 X^<(t_1, t_3)  Y (t_3, t_2)
    \end{aligned}
    \label{langreth intermediate step 1}
\end{equation}

\begin{figure}
    \centering
\includegraphics[width=\linewidth]{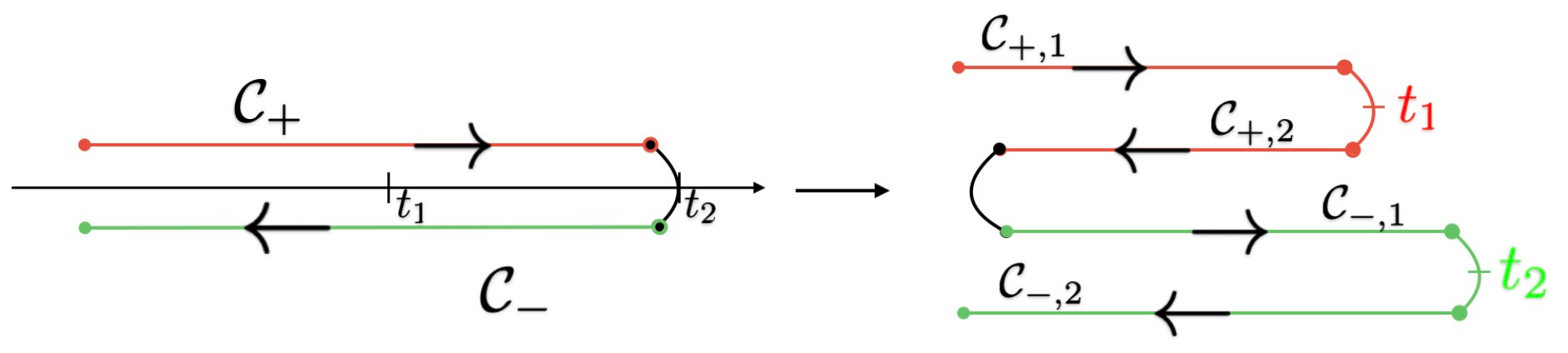}
    \caption{Sketch of the deformation of the Keldysh contour in Fig. \ref{fig:keldysh_contour}.}
    \label{fig:contour_de-deformation}
\end{figure}

Before we simplify further, we use the contour calculus to modify the Keldysh contour as in Fig. \ref{fig:keldysh_contour} to deform as in Fig. \ref{fig:contour_de-deformation}. This was originally devised in Ref. \cite{Danielewicz1990Jan} and we refer the readers to Section 4 of Ref. \cite{Hyrkas2019Apr} for a nice introduction. Therefore multiple forward-and-backward deformation loops can be added to the original Keldysh contour with each loop terminating at the maximum value of the real time that exists on that loop without changing the value of the integral. Accordingly using this contour deformation, we simplify the first term on the right-hand side of Eq. \eqref{langreth intermediate step 1} where $\Cc_+ = \Cc_{+,1} + \Cc_{+,2}$ as follows:
\begin{equation}
\begin{aligned}
    \int_{\Cc_+} & dt_3 X(t_1, t_3)  Y^< (t_3, t_2) \\
    =& \int_{\Cc_{+,1}} dt_3 X(t_1, t_3)  Y^< (t_3, t_2) \\
    &+  \int_{\Cc_{+,2}} dt_3 X(t_1, t_3)  Y^< (t_3, t_2) \\
    =& \int_{-\infty}^{t_1} dt_3 X^>(t_1, t_3)  Y^< (t_3, t_2) \\
    &+ \int_{t_1}^{-\infty} dt_3 X^<(t_1, t_3)  Y^< (t_3, t_2) \\
    =& \int_{-\infty}^{+\infty} dt_3 \Theta(t_1 - t_3)\left[X^>(t_1, t_3) - X^<(t_1, t_3) \right]  Y^< (t_3, t_2) \\
    =&\int_{-\infty}^{+\infty} dt_3 X^R(t_1, t_3)  Y^< (t_3, t_2) 
    \end{aligned}
\end{equation}
where we used the definition of the retarded function from Eq. \eqref{retarded and advanced green's functions def}. Similarly the second term on the right-hand side of Eq. \eqref{langreth intermediate step 1} simplifies to (where $\Cc_- = \Cc_{-,1} + \Cc_{-,2}$)
\begin{equation}
    \begin{aligned}
  \int_{\Cc_-} & dt_3 X^<(t_1, t_3)  Y (t_3, t_2)\\
  =& \int_{\Cc_{-,1}} dt_3 X^<(t_1, t_3)  Y (t_3, t_2) \\
  &+\int_{\Cc_{-,2}} dt_3 X^<(t_1, t_3)  Y (t_3, t_2) \\
  =&\int_{-\infty}^{t_2} dt_3 X^<(t_1, t_3)  Y^< (t_3, t_2) \\
  &+ \int_{t_2}^{-\infty} dt_3 X^<(t_1, t_3)  Y^> (t_3, t_2) \\
  =& \int_{-\infty}^\infty dt_3 \Theta(t_2-t_3) X^<(t_1, t_3)\left[Y^< (t_3, t_2) - Y^> (t_3, t_2)\right]\\
  =& \int_{-\infty}^\infty dt_3 X^<(t_1, t_3)Y^A (t_3, t_2)
    \end{aligned}
\end{equation}
where we used the definition of the advanced function from Eq. \eqref{retarded and advanced green's functions def}. Therefore, Eq. \eqref{langreth intermediate step 1} simplifies to give us the Langreth rule
\begin{equation}
\boxed{
\begin{aligned}
Z^<(t_1, t_2)  =  &   \int_{-\infty}^\infty dt_3 [X^R(t_1, t_3)  Y^< (t_3, t_2) \\
&+ X^<(t_1, t_3)Y^A (t_3, t_2)]
\end{aligned}}
\label{langreth rule 1}
\end{equation}
Similarly if we would have started with the choice that times $t_1$ lies on the backward contour $\Cc_-$ and $t_2$ lies on the forward contour $\Cc_+$, then we would similarly get the second Langreth rule
\begin{equation}
\boxed{
\begin{aligned}
Z^>(t_1, t_2)  =  &   \int_{-\infty}^\infty dt_3 [X^R(t_1, t_3)  Y^> (t_3, t_2) \\
&+ X^>(t_1, t_3)Y^A (t_3, t_2)]
\end{aligned}}
\label{langreth rule 2}
\end{equation}
These are the Langreth rules used in Appendix \ref{app. deriving kadanoff-baym equations} to derive the Kadanoff-Baym equations starting from the Schwinger-Dyson equations. 

Finally, we need another set of Langreth rules that appear in the structure of self-energy in the Schwinger-Dyson equations which we state here without proof. The structure on the Keldysh contour is of the type
\begin{equation}
   C (t,t^\prime) = A (t,t^\prime) B(t,t^\prime)
\end{equation}
which can be translated to the real time using the following Langreth rules:
\begin{equation}
\boxed{
\begin{aligned}
 C^\gtrless(t, t^\prime) =&    A^\gtrless(t, t^\prime)B ^\gtrless(t, t^\prime) \\
 C^R(t, t^\prime) =& A^{<}\left(t, t^{\prime}\right) B^{R}\left(t, t^{\prime}\right)+A^{R}\left(t, t^{\prime}\right) B^{<}\left(t, t^{\prime}\right) \\
 &+A^{R}\left(t, t^{\prime}\right) B^{R}\left(t, t^{\prime}\right)
\end{aligned}}
\label{langreth rule 3}
\end{equation}

\section{Evaluating energy in the Keldysh contour}
\label{app. Evaluating energy in the Keldysh contour}

We start with the definition of Eq. \eqref{generating functional definition}
\begin{equation*}
    \Zz[\eta] = \int D\psi_i \exp{i S  + \i S_\eta }
\end{equation*}
where $S_\eta = -\int dt \eta (t) \Hh(t)$. The disorder-averaged generating functional is given by 
\begin{equation}
    \overline{\Zz}[\eta] = \int \Dd j_q \Dd j_2 \Pp_q[j_q] \Pp_2[j_2] \Zz[\eta]  
\end{equation}
where we use Eq. \eqref{gaussian ensembles} and repeat the calculation as done in Appendix \ref{app. Deriving disorder-averaged partition function} to get
\begin{equation}
     \overline{\Zz}[\eta] =  \exp\left[ \i N \overline{S}[\Gg, \Sigma] + \i N \overline{S}_\eta[\Gg, \Sigma,\eta] \right]
\end{equation}
where $\overline{S}[\Gg, \Sigma]$ is given in Eq. \eqref{final effective action} and $\overline{S}_\eta[\Gg, \Sigma,\eta] $ contains the conjugate field dependency $\eta(t)$ given by
\begin{widetext}
    \begin{equation}
    \begin{aligned}
\overline{S}_\eta[\Gg, \Sigma,\eta] = \frac{\i (-1)^{q/2}}{4 q^2 2^{-q}}  & \int_\Cc dt dt^\prime \Jj_q(t) \Jj_q(t^\prime) \left[\eta(t) + \eta(t^\prime) + \eta(t)\eta (t^\prime)\right] \Gg(t, t^\prime)^q \\
& -\frac{\i }{2 q} \int_\Cc dt dt^\prime \Jj_2(t) \Jj_2(t^\prime) \left[\eta(t) + \eta(t^\prime) + \eta(t)\eta (t^\prime)\right] \Gg(t, t^\prime)^2
\end{aligned}
    \end{equation}
\end{widetext}
Therefore using Eqs. \eqref{expectation value defined} and \eqref{energy definition}, we get $E(t_1) = \i q^2 \lim_{\eta \to 0} \overline{\Zz}[\eta] (\i N) \frac{\delta \overline{S}_\eta[\Gg, \Sigma,\eta]}{\delta \eta(t_1)}$. Recall from Section \ref{subsec. Energy in the Keldysh contour} that without loss of generality we have assumed $t_1 \in \Cc_-$ (the backward contour). We know that the unperturbed Keldysh partition function is normalized to unity, i.e. $\Zz[\eta=0] = 1$ (see Appendix \ref{app. review of matsubara and keldysh formalisms}). Therefore we get
\begin{widetext}
    \begin{equation}
 E(t_1) = -\i N \Big[ \frac{(-1)^{q/2} }{ 2^{2-q}} \int_\Cc dt \Jj_q(t_1) \Jj_q(t) \left( \Gg(t_1, t)^q + \Gg(t, t_1)^q \right) - \frac{q}{2} \int_\Cc dt \Jj_2(t_1) \Jj_2(t) \left(\Gg(t_1, t)^2 + \Gg(t, t_1)^2 \right)  \Big]   
 \label{energy intemediate step 0 in appendix}
    \end{equation}
\end{widetext}
\onecolumngrid
We simplify this by expanding the Keldysh contour integral using $\Cc = \Cc_+ + \Cc_-$, namely $\int_\Cc dt (\ldots)= \int_{-\infty}^{+\infty} dt (\ldots) + \int_{+\infty}^{-\infty} dt(\ldots) $. To illustrate the point, we focus on the first term on the right-hand side (with $t_1\in \Cc_-$) which is also the interacting energy term $V(t_1)$ to get
\begin{equation}
    \begin{aligned}
        V(t_1) = -\i N \frac{(-1)^{q/2} }{ 2^{2-q}} & \left[ \int_{-\infty}^{+\infty} dt \Jj_q(t_1) \Jj_q(t)  \Gg^>(t_1, t)^q + \int_{+\infty}^{-\infty} dt \Jj_q(t_1) \Jj_q(t)  \Gg(t_1, t)^q \right.\\
        &\left. + \int_{-\infty}^{+\infty} dt \Jj_q(t_1) \Jj_q(t)  \Gg^<(t, t_1)^q + \int_{+\infty}^{-\infty} dt \Jj_q(t_1) \Jj_q(t)  \Gg(t, t_1)^q 
        \right]
    \end{aligned}
    \label{intermediate energy step 1 in appendix}
\end{equation}
We use the Majorana conjugate relation in Eq. \eqref{Majorana conjugate relation} to get $ \Gg^<(t, t_1)^q = \left[ - \Gg^>(t_1, t)\right]^q = \Gg^>(t_1, t)^q$ (since $q$ is even). So the first and the third terms in Eq. \eqref{intermediate energy step 1 in appendix} are the same. Then we identify $\Gg(t, t_1)^q$ appearing in the second and the fourth terms as the anti-time-ordered Green's function (since $t, t_1 \in \Cc_-$; see Eq. \eqref{four green's functions defined}) and use the definition of anti-time-ordered Green's function from Eq. \eqref{time and anti-time ordered Green's functions defined} as
\begin{equation}
    \Gg(t, t_1)^q = \left(\Theta(t_1 - t) \Gg^>(t, t_1) + \Theta(t-t_1) \Gg^<(t, t_1)\right)^q = \Theta(t_1 - t) \Gg^>(t, t_1)^q + \Theta(t-t_1) \Gg^<(t, t_1)^q
\end{equation}
where cross-terms don't contribute due to the properties of the Heaviside step function. Using the Majorana conjugate relation to bring all Green's functions in the form $\Gg^\gtrless(t_1, t)$, we get for Eq. \eqref{intermediate energy step 1 in appendix} the following interacting energy density
\begin{equation}
    \Vv(t_1) = \frac{V(t_1)}{N} = -\i  \frac{(-1)^{q/2} }{ 2^{2-q}} 2\left[ \int_{-\infty}^{t_1} dt \Jj_q(t_1) \Jj_q(t) \left( \Gg^>(t_1, t)^q - \Gg^<(t_1, t)^q  \right)\right]
\end{equation}
where the integration from $t_1$ to $+\infty$ gets cancelled out. Then substituting the large-$q$ ansatz for the Green's functions from Section \ref{subsec. large q green's function ansatz}, we get
\begin{equation}
    \Vv(t_1) = - \frac{\i}{2} \int_{-\infty}^{t_1} dt \Jj_q(t_1) \Jj_q(t) \left( e^{g(t_1, t)} - e^{g^\star(t_1, t)}  \right)+ \mathcal{O}\left(\frac{1}{q}\right) = \text{Im} \int_{-\infty}^{t_1}\, dt_{2} \cj_q(t_1) \cj_q(t_2) e^{g(t_1,t_2)}+ \mathcal{O}\left(\frac{1}{q}\right)
\end{equation}
which matches the expression in Eq. \eqref{interacting and kinetic energy densities}. Similarly we can evaluate the second term in Eq. \eqref{energy intemediate step 0 in appendix}, which is the kinetic energy contribution to get
\begin{equation}
\begin{aligned}
    \Kk(t_1) = \frac{K(t_1)}{N} = \i \frac{q}{2} \Big[ \int_{-\infty}^{+\infty} dt \Jj_2(t_1) \Jj_2(t) \Big( \Gg^>(t_1, t)^2 + \Gg^<(t, t_1)^2 &- [\Theta(t-t_1)  \Gg^>(t_1, t)^2 + \Theta(t_1 - t)  \Gg^<(t_1, t)^2 ] \\
    & - [\Theta(t_1-t)  \Gg^>(t, t_1)^2 + \Theta(t - t_1)  \Gg^<(t, t_1)^2]\Big) \Big].
\end{aligned}
\end{equation}
We again use $\Gg^>(t, t_1)^2 = (-\Gg^<(t_1, t))^2 = \Gg^<(t_1, t)^2$ and $\Gg^<(t, t_1)^2 = (-\Gg^>(t_1, t))^2 = \Gg^>(t_1, t)^2$ as well as the large-$q$ ansatz for the Green's function as introduced in Section \ref{subsec. large q green's function ansatz} to get
\begin{equation}
    \Kk(t_1) = \i \frac{q}{2} 2 \left(\frac{\i}{2}\right)^2 \Big [\int_{-\infty}^{t_1} dt \Jj_2(t_1) \Jj_2(t) \Big( e^{2g(t_1, t)/q} - e^{2g^\star(t_1, t)/q} \Big) \Big] = \Im \int_{-\infty}^{t_1}  dt \Jj_2 (t_1) \Jj_2(t) g(t_1, t) + \Oo\Big(\frac{1}{q}\Big).
\end{equation}
We kept to leading order in $1/q$ in the second equality by expanding the exponential which finally matches the expression for the kinetic energy density as in Eq. \eqref{interacting and kinetic energy densities}. 

\begin{figure}
    \centering
\includegraphics[width=0.95\linewidth]{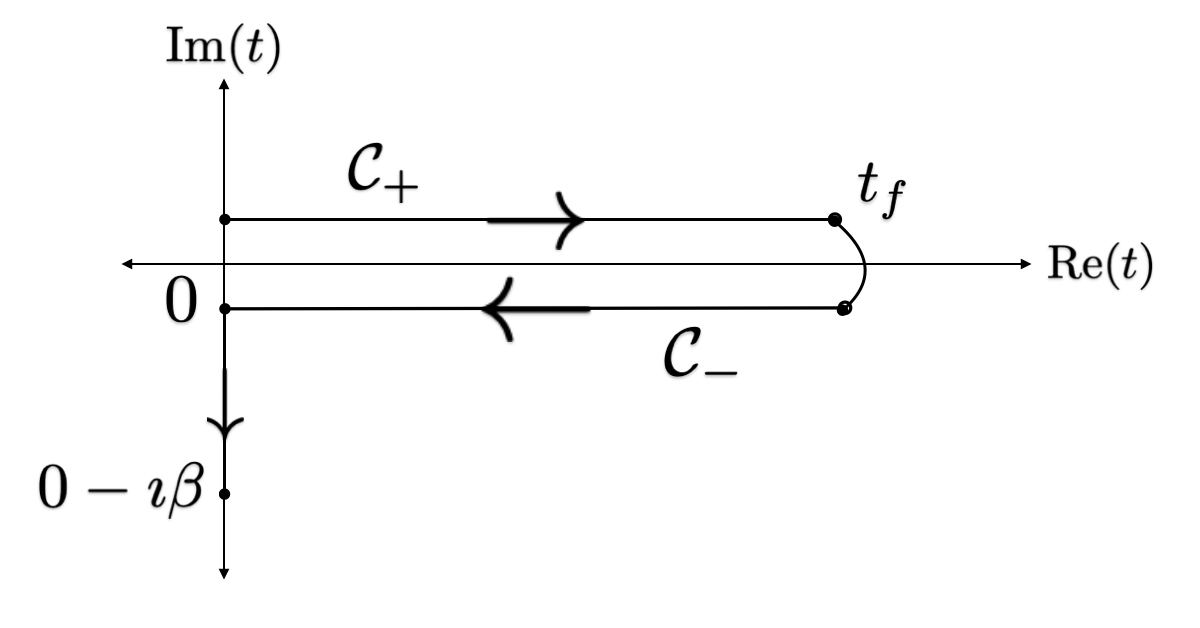}
    \caption{For our quenched case where $\text{SYK}_2$ term is switched on at $t=0$ starting from an $\text{SYK}_q$ equilibrium condition, accordingly the forward and the backward contour from time $-\infty$ to $0$ in Fig. \ref{fig:keldysh_contour} cancel each other and the vertical contour can accordingly be shifted to time $t=0$ which lies later than all points on the forward and the backward horizontal contours. This represents the equilibrium situation for our quench case and had we considered this vertical branch, we would have recovered the equilibrium energy density as in Eq. \eqref{eq:computational_energy_thermal} including the $q$ scaling of kinetic energy density contribution. Since our non-equilibrium energy density calculation explicitly considers real time dynamics and the full Keldysh contour as in Fig. \ref{fig:keldysh_contour} where the final vertical contour only has a $\text{SYK}_q$ term (since $\text{SYK}_2$ quench happens at $t=0$, so for any $t<0$, there is only $\text{SYK}_q$ term in equilibrium), this further justifies using interaction energy density for the estimation of final temperature as argued in the main text below Eq. \eqref{eq:iteraction_energy_thermal}.}
    \label{fig:keldysh_starting_zero}
\end{figure}

Having derived the general non-equilibrium expression, we also show now how to derive the equilibrium energy density as in Eq. \eqref{eq:computational_energy_thermal}. We start with Eq. \eqref{energy intemediate step 0 in appendix} where we restrict to the imaginary contour of $\Cc$ in Fig. \ref{fig:keldysh_contour} since we are in equilibrium. The imaginary contour runs from $t_0 $ to $t_0 - \i \beta$ in real time where $t_0 \to -\infty$. We impose the time translational invariance $(t, t_1) \to t-t_1 = t^\prime$ (say) where $t_1$ is just some constant. Then substitute $t = -\i x \beta_f$ where $\beta_f$ is the equilibrium inverse temperature. We take the definition of the Green's function from Section \ref{subsec. equilibrium situation} where $\Gg(-\i \beta_f x) = -\i \Tilde{\Gg}(x)$ where $\Tilde{\Gg}(x) = \frac{1 }{2} \text{sgn}(x) e^{\Tilde{g}(x)/q}$ is the imaginary time Green's function \cite{Maldacena2016Nov}. The anti-symmetric property in the imaginary time formalism is given by \cite{bruus-flensberg}
\begin{equation}
    \begin{aligned}
        \Tilde{\Gg}(x + 1 ) &= -  \Tilde{\Gg}(x), \quad x \in (-1,0) \\
         \Tilde{\Gg}(x - 1)&= - \Tilde{\Gg}(x), \quad x \in (0,1)
    \end{aligned}
\end{equation}
Accordingly the imaginary time Green's function $\Tilde{g}(x)$ (defined as $ g(-\i \beta_f x)$ in Section \ref{subsec. equilibrium situation} as a real-valued function) is periodic with unit periodicity. Therefore we use the relation $\int_{0}^p h(t) dt = \int_a^{a+p} h(t) dt$ for any real-valued continuous function $h$ satisfying $h(x+p) = h(x)$ $\forall$ $x$. Proof follows by considering $H(x) = \int_x^{x+p} h(t) dt$ and taking its derivative to get $dH/dx = h(x+p) - h(x) = 0$, so $H(x)$ is a constant and in particular, $H(a) = H(0)$. Therefore we shift the integral limits on $x$ from $0 \to 1$ to get
\begin{equation}
 \Ee_{\text{EQ}}(\beta_f) = - \beta_f \Big[ \frac{(-1)^{q/2} (-\i)^q}{ 2^{2-q}}\Jj_q^2 \int_0^1 dx  \left( \Tilde{\Gg}(-x)^q + \Tilde{\Gg}(x)^q \right) - \frac{q (-\i)^2}{2} \Jj_2^2 \int_0^1 dx  \left(\Tilde{\Gg}(-x)^2 + \Tilde{\Gg}(x)^2 \right)  \Big]   
 \label{energy intemediate step 1 in appendix}
    \end{equation}
where we plug in the full definition of $\Tilde{\Gg}(x)=\frac{1 }{2} \text{sgn}(x) e^{\Tilde{g}(x)/q}$ in the first integral and keep up to order $1/q$ in the second integral (recall that $\Tilde{g}(x) = \Oo(q^0)$). We also use above the Majorana conjugate relation $\Tilde{\Gg}(-x) = - \Tilde{\Gg}(x)$ that gives $\Tilde{g}(x) = \Tilde{g}(-x)$ to finally get Eq. 
\eqref{eq:computational_energy_thermal}. 
This calculation shows that in our non-equilibrium setup as considered in Eq. \eqref{energy intemediate step 0 in appendix}, had we considered the imaginary part of the contour (see Fig. \ref{fig:keldysh_contour}), we would have obtained a kinetic contribution that scales as $q$, similar to what we obtain in the equilibrium case in Eq. \eqref{eq:computational_energy_thermal}. Since we use the Bogoliubov principle of weakening correlations to ignore the imaginary contour in our non-equilibrium calculations, this is the reason why the non-equilibrium energy density does not have such a leading term in $q$. However there is a caveat. For the equilibrium calculation, both $\text{SYK}_q$ as well as $\text{SYK}_2$ terms exist from time $-\infty \to +\infty \to -\infty \to -\infty - \i \beta$ where the horizontal contours cancel each other (the system is always in equilibrium) leading to the only contribution coming from the vertical (thermal) contour. This is what we evaluated above in Eq. \eqref{energy intemediate step 1 in appendix}. However, the full Keldysh contour as shown in Fig. \ref{fig:keldysh_contour} for the non-equilibrium setup would have only involved the $\text{SYK}_q$ term on the vertical contour because the $\text{SYK}_2$ term is switched on at $t=0$. As such, to include the vertical contour, we need to make modifications as shown explicitly in Fig. \ref{fig:keldysh_starting_zero} where we use the fact that the equilibrium time evolution in the forward and the backward direction cancels each other (same reasoning that motivates contour deformation shown in Fig. \ref{fig:contour_de-deformation}) and accordingly we can remove this canceled section from the full Keldysh contour. What remains is shown in Fig. \ref{fig:keldysh_starting_zero} where the contour takes the path starting from time $t=0 \to \infty \to 0 \to -\i \beta$. However our non-equilibrium calculation is done keeping the full contour (Fig. \ref{fig:keldysh_contour}) which implies that only $\text{SYK}_q$ term exists on the imaginary contour; this further justifies the reasoning of using the interaction energy density to estimate the final temperature (see main text below Eq. \eqref{eq:iteraction_energy_thermal}). 
\\
    \twocolumngrid
    
\begin{figure}
    \centering
    \includegraphics[width=0.75\linewidth]{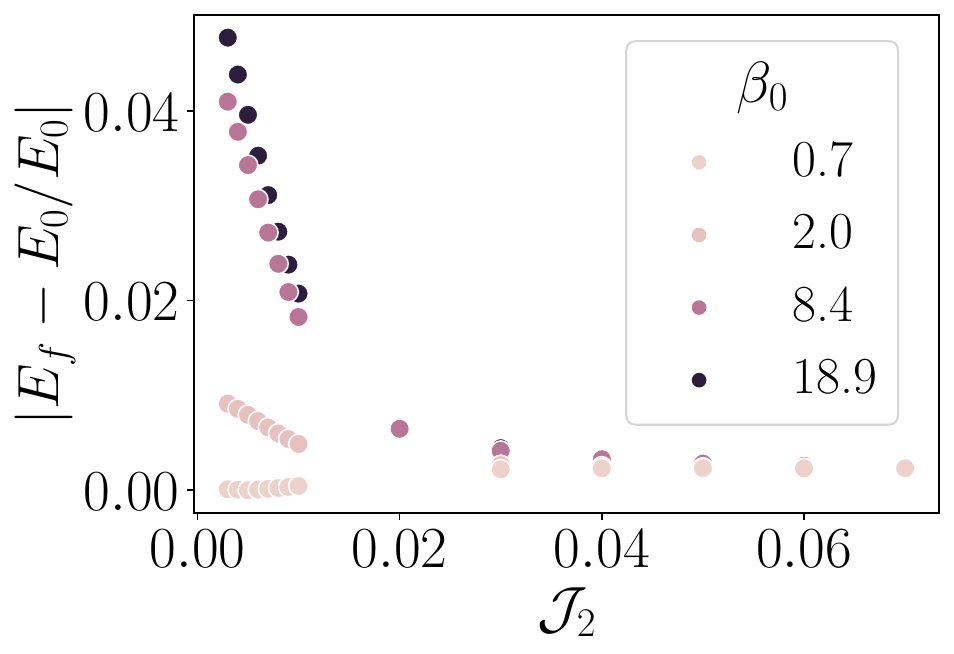}
    \caption{Relative error between the initial energy $E_0$ and the energy at the final time $E_f$ as a function of the quench strength, in the mixed quench. The colors of the markers represent different initial temperatures. Here, all calculations were obtained with $\Delta t = 0.06$ and $t_{\text{max}}=1700$. As explained in the main text, the errors are somewhat larger for smaller quench parameter $\Jj_2$ and larger initial inverse temperature $\beta_0$.}
    \label{fig:energy_errors}
\end{figure}

\begin{figure}
\centering
    \begin{subfigure}{0.44\linewidth}
        \caption{}
        \includegraphics[width=\linewidth]{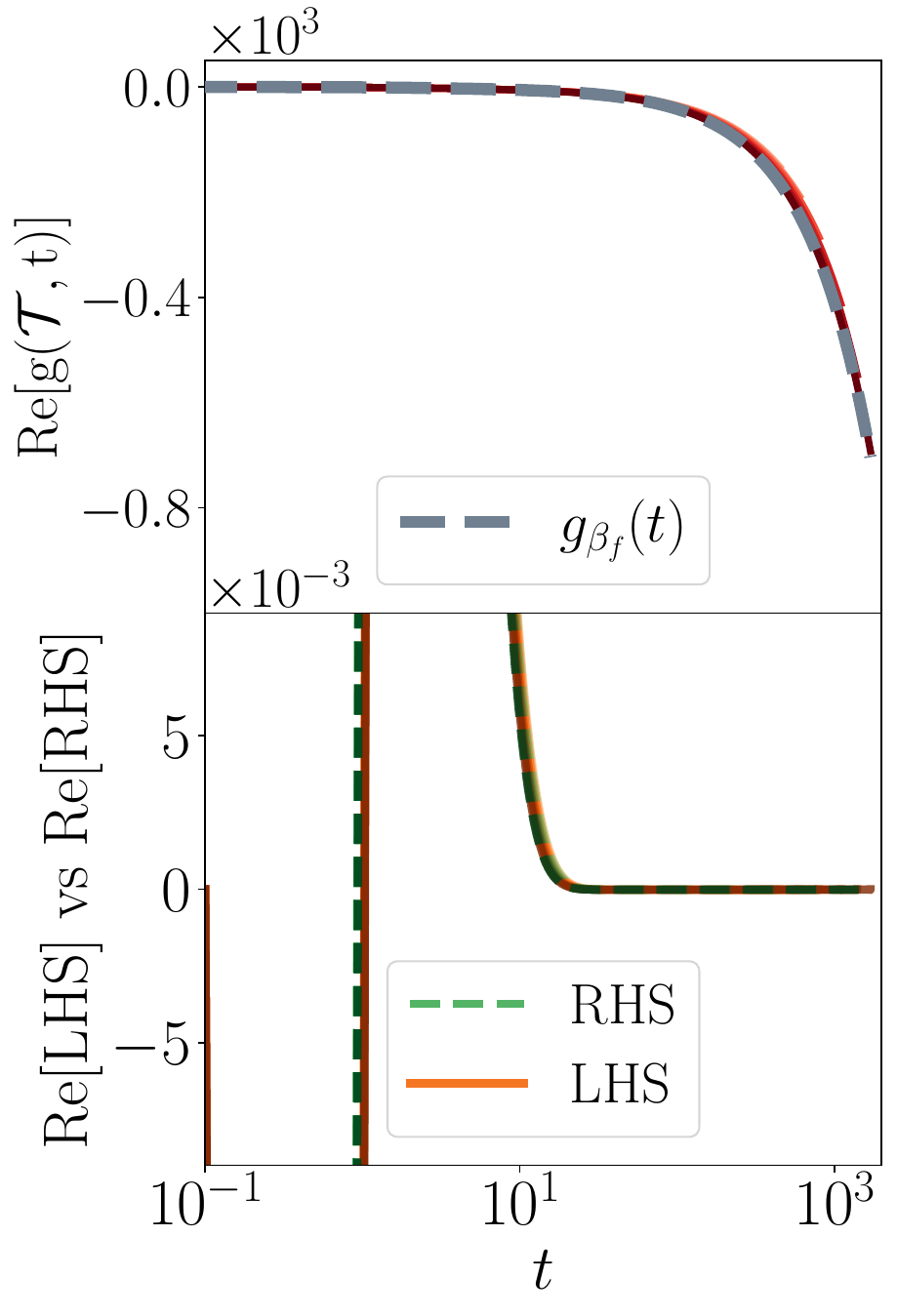}
    \end{subfigure}
    \begin{subfigure}{0.49\linewidth}
        \caption{}
        \includegraphics[width=\linewidth]{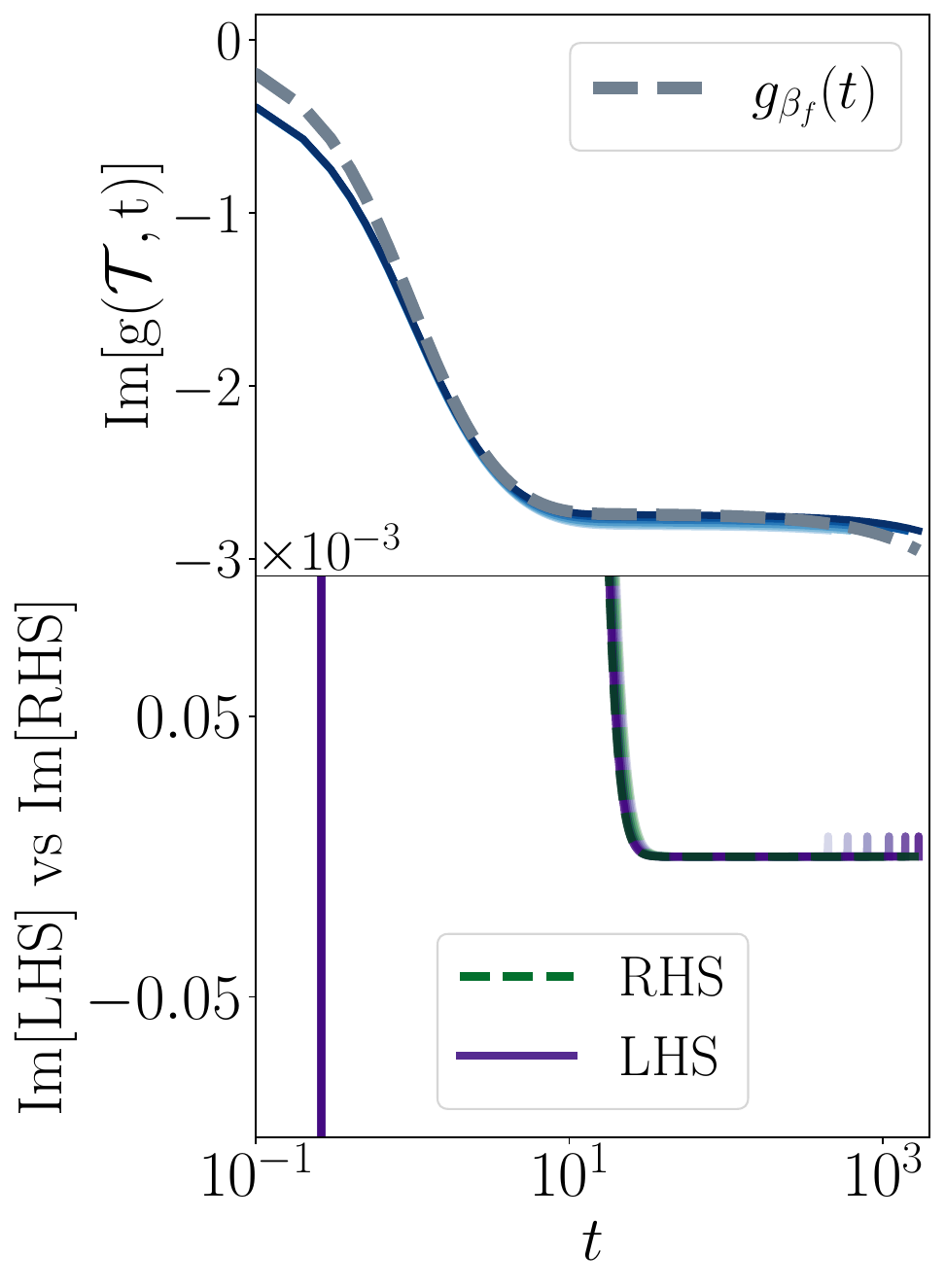}
    \end{subfigure}
\caption{Comparison of the stationary limit of the Kadanoff-Baym solution to the corresponding thermal system for $\cj_2 = 0.003$ and $\beta_0=18.9$. The color coding in (a) and (b) follows the same convention as explained in Fig. \ref{fig:sationary_limit}.}
    \label{fig:sationary_limit_Jtwo0.003}
    \end{figure}

    \begin{figure}
\centering
    \begin{subfigure}{0.44\linewidth}
        \caption{}
        \includegraphics[width=\linewidth]{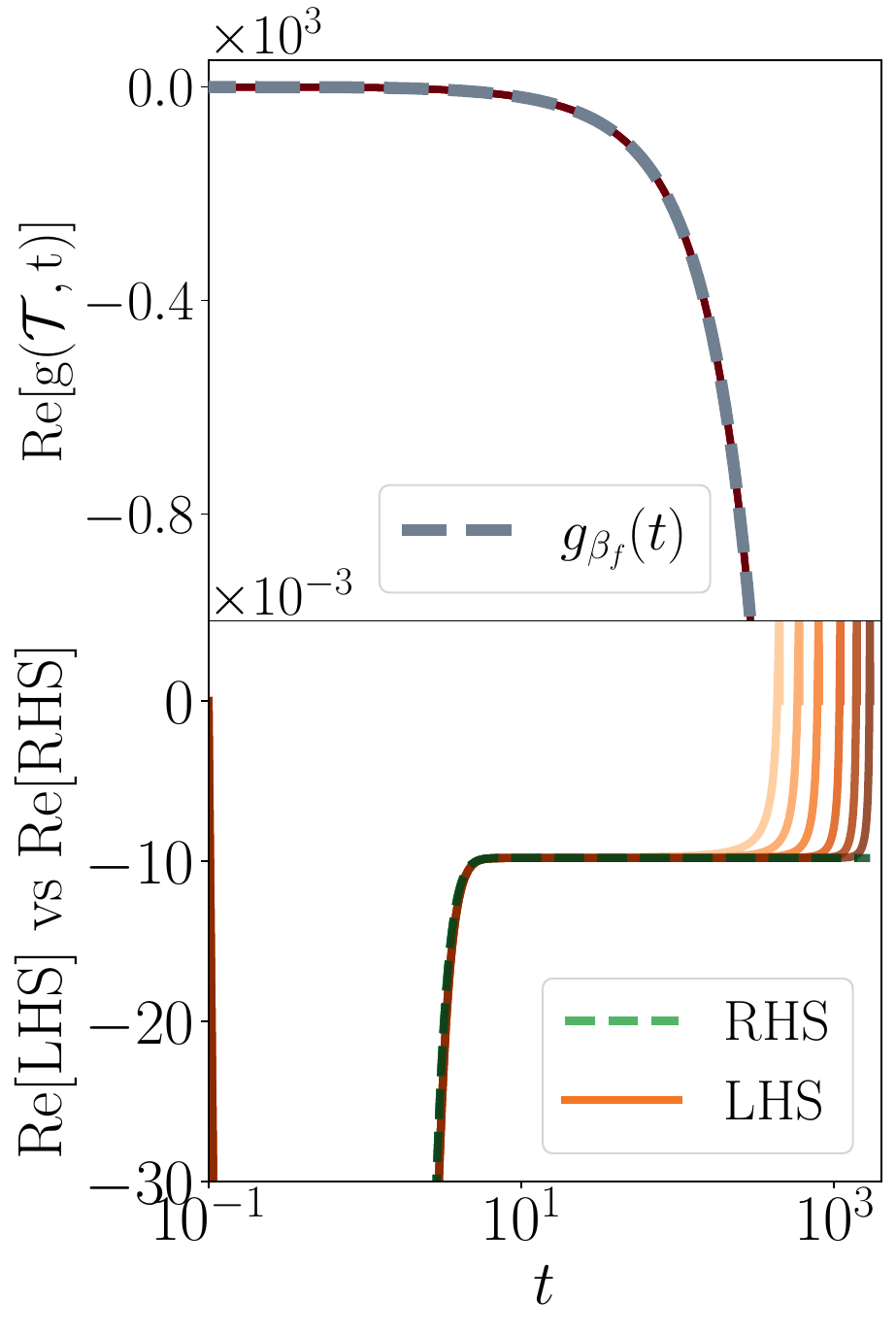}
    \end{subfigure}
\hfil
    \begin{subfigure}{0.49\linewidth}
        \caption{}
        \includegraphics[width=\linewidth]{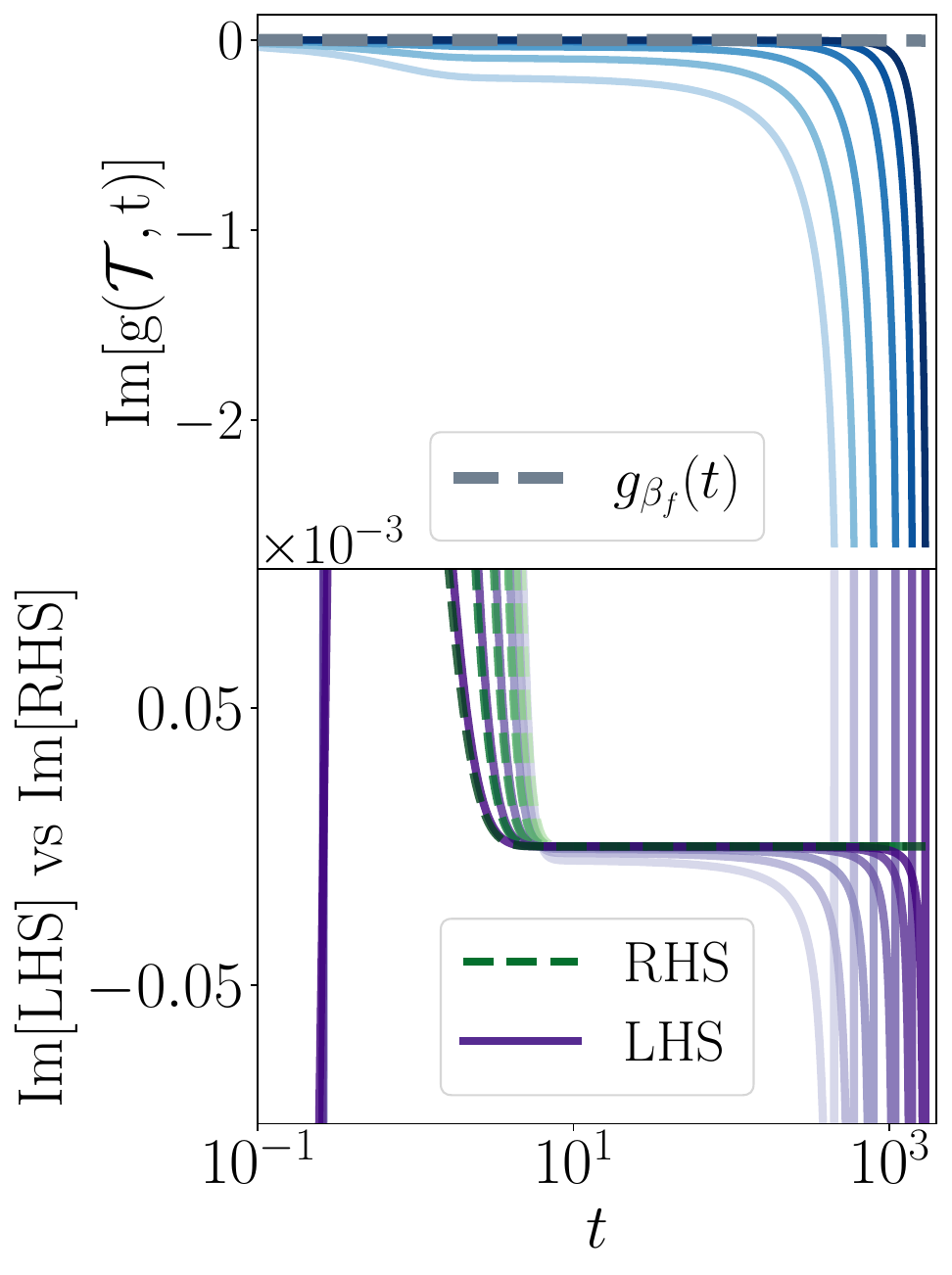}
    \end{subfigure}
\caption{Comparison of the stationary limit of the Kadanoff-Baym solution to the corresponding thermal system for $\cj_2 = 0.07$ and $\beta_0=18.9$. The color coding in (a) and (b) follows the same convention as explained in Fig. \ref{fig:sationary_limit}.}
    \label{fig:sationary_limit_Jtwo0.07}
    \end{figure}

\begin{figure}
    \centering
    \includegraphics[width=0.95\linewidth]{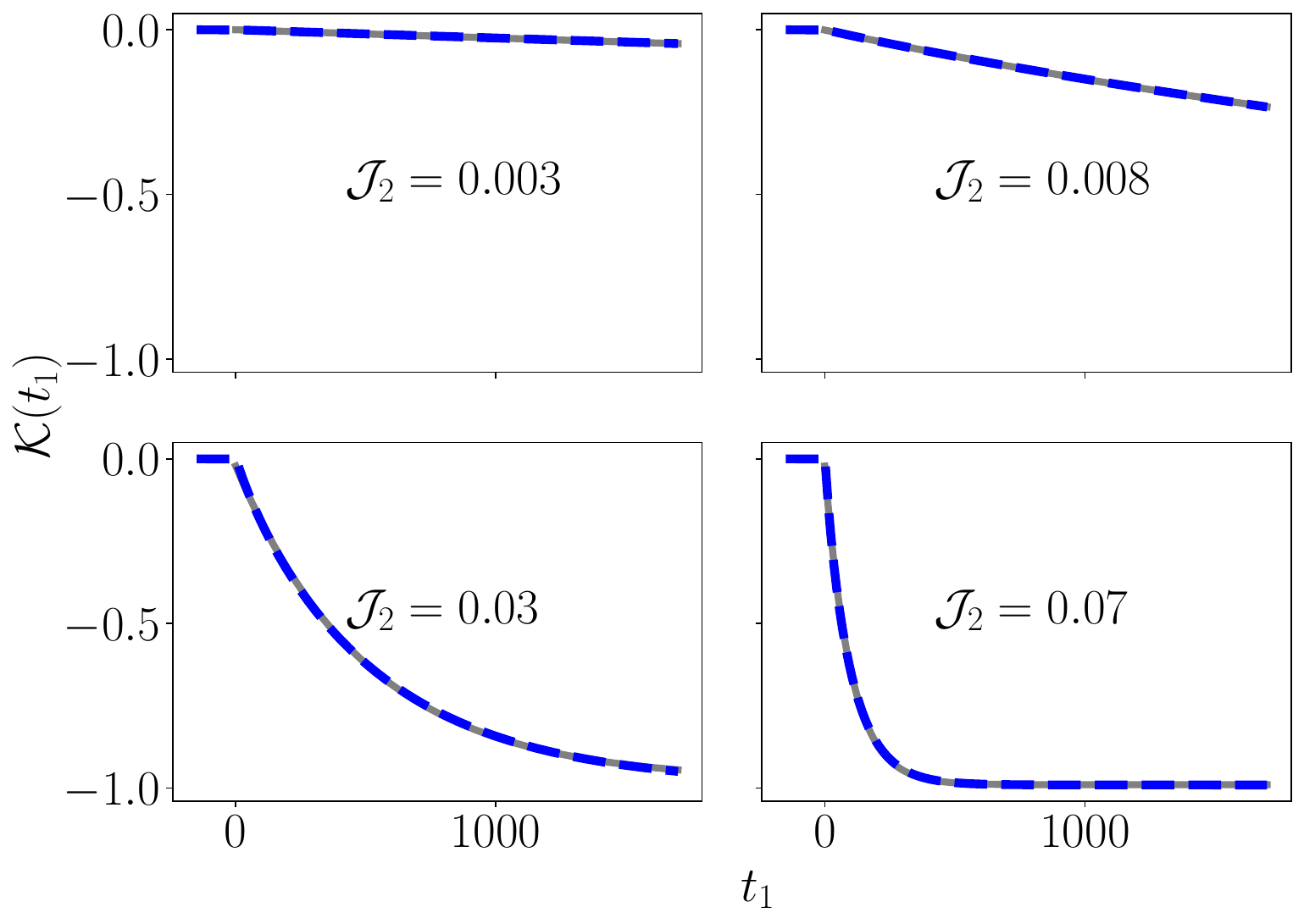}
    \caption{Goodness of fit for $\cK(t_1)$ using the exponential ansatz in Eq. \eqref{eq:expo_ansatz}. The dashed line corresponds $\cK(t_1)$, while the solid one depicts the fitted curve. All four instances shown here have $\beta_0=18.9$ as the initial inverse temperature.}
    \label{fig:fit_accuracy}
\end{figure}

\section{Additional figures}\label{app. additional figures}

The following figures are referenced in the main text and serve as supplementary material. Fig. \ref{fig:energy_errors} shows the relative error between the initial and final values of the total energy in the mixed-quench for different temperatures and kinetic couplings. As pointed out in the main text, higher errors are found at small $cj_2$ and large $\beta_0$ since they represent systems with low temperatures.
    
Figs. \ref{fig:sationary_limit_Jtwo0.003} and \ref{fig:sationary_limit_Jtwo0.07} depict the analysis of the stationary limit of the Kadanoff-Baym solution as presented in Fig. \ref{fig:sationary_limit} for $\cj_2=0.003$ and $\cj_2 = 0.07$, which represent the weakest and strongest quenches considered in this study respectively. The figures indicate that, as in the case of $\cj_2 = 0.02$, both cases eventually reach stationary behavior corresponding to thermal equilibrium at different relaxation times. These are additional evidences and, accordingly, all conclusions of Section \ref{sec. thermalization} hold true.

Figures \ref{fig:sationary_limit_Jtwo0.003} and \ref{fig:sationary_limit_Jtwo0.07} depict the analysis of the stationary limit of the Kadanoff-Baym solution as presented in Fig. \ref{fig:sationary_limit} for $\mathcal{J}_2=0.003$ and $\mathcal{J}_2=0.07$, which represent the weakest and strongest quenches considered in this study, respectively. The figures indicate that, as in the case of $\mathcal{J}_2=0.02$, both cases eventually reach stationary behavior corresponding to thermal equilibrium at different relaxation times. Additionally, Fig. \ref{fig:fit_accuracy} provides a goodness of fit curve for the kinetic energy density (used to determine the thermalization rate in this work) where we can see that the actual curve and the fitted curve, via Eq. \eqref{eq:expo_ansatz}, has a remarkable overlap. These are additional evidences and, accordingly, all conclusions of Sec. \ref{sec. thermalization} hold true.

\end{document}